\documentclass[journal]{IEEEtran}
\usepackage{amsmath}
\usepackage{amsfonts}
\usepackage{amssymb}
\usepackage{graphicx}
\usepackage{latexsym}
\usepackage{epsfig}
\usepackage{textcomp}
\usepackage{url}

\begin{document}

\title{A mechanistic model for +1 frameshifts in eubacteria}
\author{
\medskip  
L. Ponnala$^{1}$$^{*}$, D. L. Bitzer$^{2}$, A. Stomp$^{3}$, M. A. Vouk$^{2}$ \\
\medskip
\small
$^{1}$Center for Advanced Computing, Cornell University, Ithaca NY 14853 \\ $^{2}$Department of Computer Science, North Carolina State University, Raleigh, NC 27695 \\ $^{3}$Department of Forestry, North Carolina State University, Raleigh, NC 27695 USA \\ 
\bigskip $^{*}
$To whom correspondence should be addressed. E-mail: lp257@cornell.edu
}
\maketitle

\begin{abstract}
This work applies the methods of signal processing and the concepts of control system design to model the maintenance and modulation of reading frame in the process of protein synthesis. The model shows how translational speed can modulate translational accuracy to accomplish programmed +1 frameshifts and could have implications for the regulation of translational efficiency.

A series of free energy estimates were calculated from the ribosome's interaction with mRNA sequences during the process of translation elongation in eubacteria. A sinusoidal pattern of roughly constant phase was detected in these free energy signals. Signal phase was identified as a useful parameter for locating programmed +1 frameshifts encoded in bacterial genes for release factor 2. A displacement model was developed that captures the mechanism of frameshift based on the information content of the signal parameters and the relative abundance of tRNA in the bacterial cell. Results are presented using experimentally verified frameshift genes across eubacteria.

A set of MATLAB\textregistered\ programs that implement our methods are available upon request from the corresponding author.


\end{abstract}

\section{Introduction}
In electrical devices, input signals control device states.  If the translating ribosome followed this design, its reading frame states, Frame 0, Frame +1 and Frame +2 (or -1), would be controlled by an input signal.  In electrical devices, control system design takes the form of a mathematical model of a control system algorithm which decodes input signals to determine device state.  The analytical tools of signal processing provide methods for detecting signals, extracting them from noise, characterizing signal parameters, and identifying the parameters and parameter behaviors that are predictive of device states.  To use these tools requires a mathematical model of the machine and an algorithm that simulates the machine process. \blfootnote{This paper is based on work done while Lalit Ponnala was a graduate student at North Carolina State University. The work was supported in part by NC State DURP funds.}

Our previous work \cite{ponn.jbsb} has shown that a free energy signal containing a periodic component of frequency $f=1/3$ can be extracted for each mRNA of a specific eubacterium. Signal extraction is done using an algorithm that creates successive alignments of the bacterium's 16S rRNA 3'-terminal nucleotide tail with the mRNA sequence. For each sequence alignment, a free energy of hybridization is calculated, the value of which is a function of the degree of complementarity. This algorithm simulates scanning of the mRNA by the 16S rRNA tail, as suggested by Weiss \emph{et al} \cite{weiss.embo}.

Our hypothesis is that the free energy signal arising from hybridization of the 16S rRNA tail with the mRNA is the input signal that controls reading frame.  Modulation of reading frame could be accomplished through this signal if it supplied a force that adjusted the position of the mRNA relative to the ribosome.   The first step towards validation of this hypothesis is the development of a mathematical model that defines ribosome position as a function of free energy signal parameters. The second step involves experimental testing of model predictions.  This paper presents the development of the mathematical model describing control system design.

\section{Signal characterization and extraction}
Our previous work \cite{ponn.jbsb} has shown that the free energy
signal contains a periodic $f=1/3$ component embedded in noise. A suitable
model for the free energy signal is
\begin{equation}
    y_{n} = \mu + Asin\left(2\pi \frac{1}{3} n + \phi \right) + z_{n} \ , \ \ \ n = 0 \ldots (L-1)
    \label{estsignalmodel2}
\end{equation}
where $L$ is the number of nucleotides in the mRNA sequence, and $z_{n}$ is additive IID noise with mean $0$ and variance
$\sigma^{2}$. Estimates of signal amplitude $A$ and phase $\phi$ were obtained using
a regression procedure. We found that genes belonging to a
specific organism had a roughly constant phase $\phi$ in their free energy
signals and that the mean phase angle of all genes in the species ($\theta_{sp}$) varied linearly with species (G+C) content \cite{ponn.jbsb}. However, the statistical error associated with these estimates was large.

The free energy signal is noisy, resulting in a low
signal-to-noise ratio (SNR). The signal periodicity of three nucleotides can be used to improve the signal to noise ratio.  The noise component of the signal can be reduced by calculating nucleotide-based averages of free energy triplets.  This approach will result in the SNR growing linearly with the number of codons.

\subsection{Method of accumulation}
    A hypothetical memory for the ribosome system can be created consisting of a stack of 3 registers. The memory system maintains updates of the free energy released due to the interaction between the 16S rRNA tail and the mRNA sequence. As the energy values accumulate in the memory registers, information pertaining to the reading frame gets updated.

    We denote the register contents by the vector $\mathbf{R}^{(k)}, k = 1 \ldots \frac{L}{3}$, where $\frac{L}{3}$ is the number of codons in an mRNA sequence. We store the first three energy values (computed from alignments of the 16S rRNA tail with the first 3 bases of the mRNA sequence, i.e. the first codon) in consecutive registers i.e.
    \[ \textbf{R}^{(1)} = \left[ \begin{array}{c} y_{0} \\ y_{1} \\ y_{2} \end{array} \right] \]

    We then accumulate, or update, the free energies from the first codon by adding to them the free energy values corresponding to the second codon position, resulting in
    \[ \textbf{R}^{(2)} = \left[ \begin{array}{c} y_{0}+y_{3} \\ y_{1}+y_{4} \\ y_{2}+y_{5} \end{array} \right] \]

    After accumulating the signal for a length of $k$ codons, the register contents will be
    \[ \textbf{R}^{(k)} = \left[ \begin{array}{c} R^{(k)}_{1} \\ R^{(k)}_{2} \\ R^{(k)}_{3} \end{array} \right] = \left[ \begin{array}{c} \displaystyle\sum_{n=0}^{k-1}y_{3n} \\ \displaystyle\sum_{n=0}^{k-1}y_{3n+1} \\ \displaystyle\sum_{n=0}^{k-1}y_{3n+2} \end{array} \right] \]

    This procedure is repeated until the last mRNA codon is reached, i.e., until $k = \frac{L}{3}$.

\subsection{Cumulative magnitude and phase}
The register contents $\textbf{R}^{(k)}$ represent a snapshot of the free energy signal pattern. The three points have a sinusoidal nature due to the dominant periodicity of the energy pattern. This allows us to calculate the cumulative magnitude $M_{k}$ and phase $\theta_{k}$ by interpolation. As a result, $\textbf{R}^{(k)}$ can be represented as a phasor $M_{k}e^{j\theta_{k}}$ \cite{giancoli}. We equate the contents of the registers, after subtracting their mean, to points on a sine-wave and solve Equations \eqref{eq1}, \eqref{eq2} and \eqref{eq3} for $M_{k}$ and $\theta_{k}$.

\begin{equation}
  r^{(k)}_{1} = R^{(k)}_{1} - \left( \frac{\displaystyle\sum_{n=1}^{3}R^{(k)}_{n}}{3}\right) = M_{k}sin\left(\theta_{k}\right)
    \label{eq1}
\end{equation}

\begin{equation}
  r^{(k)}_{2} = R^{(k)}_{2} - \left( \frac{\displaystyle\sum_{n=1}^{3}R^{(k)}_{n}}{3} \right) = M_{k}sin\left(\theta_{k}+\frac{2\pi}{3}\right)
    \label{eq2}
\end{equation}

\begin{equation}
  r^{(k)}_{3} = R^{(k)}_{3} - \left( \frac{\displaystyle\sum_{n=1}^{3}R^{(k)}_{n}}{3} \right) = M_{k}sin\left(\theta_{k}+\frac{4\pi}{3}\right)
    \label{eq3}
\end{equation}

\subsection{Signal-to-Noise Ratio}
Based on our free energy signal model (Equation \eqref{estsignalmodel2}), we have
\begin{equation}
    r^{(k)}_{1} = \left(kA\right)sin\left(\phi \right) + \left(\displaystyle\sum_{j=0}^{k-1} z_{3j} \right) - \frac{1}{3} \displaystyle\sum_{j=0}^{3k-1} z_{j}
\end{equation}

\begin{equation}
    r^{(k)}_{2} = \left(kA\right)sin\left(\frac{2\pi}{3}+\phi \right) + \left(\displaystyle\sum_{j=0}^{k-1} z_{3j+1} \right) - \frac{1}{3} \displaystyle\sum_{j=0}^{3k-1} z_{j}
\end{equation}

\begin{equation}
    r^{(k)}_{3} = \left(kA\right)sin\left(\frac{4\pi}{3}+\phi \right) + \left(\displaystyle\sum_{j=0}^{k-1} z_{3j+2} \right) - \frac{1}{3} \displaystyle\sum_{j=0}^{3k-1} z_{j}
\end{equation}

Therefore, 
    \[ M_{k} = kA \] and 
    \[ \sigma^{2}_{k} = \left(\frac{2k}{3}\right)\sigma^2 \]
    where $\sigma^{2}_{k}$ is the noise variance of the contents of the memory register $\textbf{R}^{(k)}$. The SNR of the register contents is given by
    \[ \Gamma_{k} = \frac{M^{2}_{k}}{2\sigma^{2}_{k}} = \frac{3k}{2}\left(\frac{A^2}{2\sigma^{2}}\right) \]

    Thus, the accumulation of points corresponding to the same sinusoidal pattern causes the SNR to grow linearly with the number of codons.

\subsection{Visualization using polar plots}
\begin{figure}
    \centering
  \includegraphics[height=2.0in,width=2.0in]{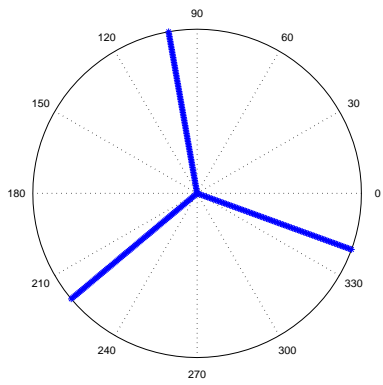}
    \caption{Thick lines indicate phase boundaries for each reading frame, relative to an initial signal phase of -20\textdegree}
    \label{fig:nb}
\end{figure}

The magnitude $M_{k}$ and phase $\theta_{k}$ of the register contents can be visualized on a polar plot, with the radial coordinate representing magnitude and the angular coordinate representing phase. Because the free energy signal frequency equals 1/3 cycles/nucleotide, each 120\textdegree\ sector of the polar plot represents one nucleotide (see Figure \ref{fig:nb}).  For the free energy signal to play a role in reading frame determination, it would be expected that variation in $M_{k}$ and/or $\theta_{k}$ would correlate with shifts in reading frame.  To determine if such a correlation might exist, two genes were selected: \emph{aceF}, a gene which does not encode a frameshift, and \emph{prfB}, a well-studied gene whose mRNA sequence is known to encode a programmed frameshift at codon 26 \cite{far.mr}.

\begin{figure}
    \centering
  \includegraphics[height=2.0in,width=2.0in]{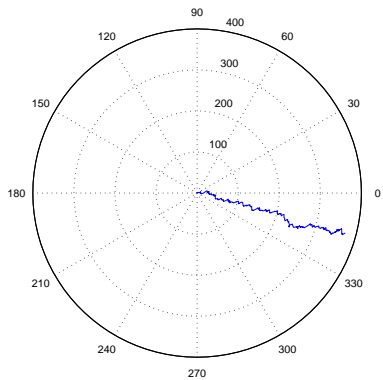}
    \caption{Polar plot for gene \emph{aceF} in \emph{E. coli}}
    \label{fig:aceF_p}
\end{figure}

\begin{figure}
    \centering
  \includegraphics[height=2.0in,width=2.0in]{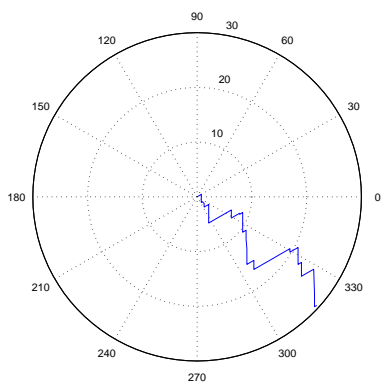}
    \caption{Polar plot for gene \emph{tsf} in \emph{E. coli}}
    \label{fig:tsf_p}
\end{figure}

\begin{figure}
    \centering
  \includegraphics[height=2.0in,width=2.0in]{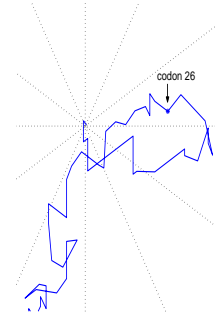}
    \caption{Partial polar plot for gene \emph{prfB} in \emph{E. coli}: arrow points to the location of frameshift, marked by a *}
    \label{fig:prfB_pz}
\end{figure}

\begin{figure}
    \centering
  \includegraphics[height=2.0in,width=2.0in]{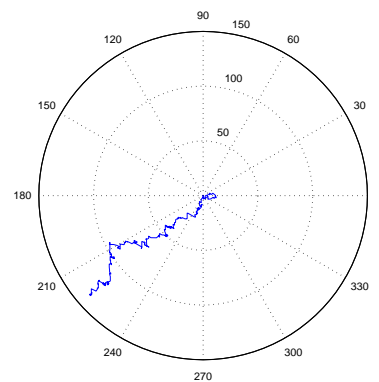}
    \caption{Polar plot for gene \emph{prfB} in \emph{E. coli}}
    \label{fig:prfB_p}
\end{figure}

Although the polar plot for \emph{aceF} (Figure \ref{fig:aceF_p})
shows some variation, the cumulative phase stays roughly constant at
about -15\textdegree, within the sector of one nucleotide. Similar 
phase constancy was observed in all the 1673 verified genes in \emph{E. coli} of
length 200 codons or greater \cite{ponn.embs06}. However, considerable variation in track within the nucleotide
sector can occur (see Figure \ref{fig:tsf_p}). By comparison, the
polar plots of \emph{prfB} (Figures \ref{fig:prfB_pz} and
\ref{fig:prfB_p}) are quite different.  The plot starts in the same
nucleotide sector as that for \emph{aceF}, but around codon 26 it
swings through approximately 240\textdegree.  When the phase change
is complete, the plot re-establishes itself within a different
nucleotide sector and remains there, with small variation, to the
end of the gene. Although provocative and consistent with our
hypothesis, analysis of other genes known to encode frameshifts
would strengthen the correlation.

RECODE\footnote{http://recode.genetics.utah.edu/} is a database of non-canonical translational events such as frameshifts, ribosomal hops and codon redefinition \cite{recode01}\cite{recode03}. Experimentally verified \emph{prfB} gene sequences for twelve prokaryotes other than \emph{E. coli} were obtained and their free energy signals were calculated using the corresponding species' 16S tail, and signal parameters were generated using the cumulative method. The \emph{prfB} polar plots for all the examined species are shown in the Appendix. A significant phase change is observed around the frameshift location in all these genes, consistent with the results obtained using the \emph{prfB} gene in \emph{E. coli}.

\subsection{Drawbacks}
Our cumulative model of signal phase, although useful for revealing frameshift sites encoded in gene sequences, has one significant drawback.  For every additional codon, a greater perturbation of the free energy signal will be needed to shift the cumulative phase.  This means that the model will have difficulty identifying frameshifts if they occur towards the end of a long gene sequence. Also, there is no experimental evidence that indicates that the entire gene sequence upstream of a frameshift site has a controlling influence on the frameshift.  The sequence elements that result in a shift in reading frame during translation are small and can be localized in a short sequence within the coding region \cite{far.mr}.  To accommodate these concerns we developed a new model that estimates instantaneous signal phase at each codon.

\section{Displacement Model}
\subsection{Calculation of displacement}

For a gene without a frameshift, the polar plot would lengthen itself radially (due to growth in magnitude) but stay at a roughly constant phase angle ($\theta_{k} \approx \theta_{sp}$). When a +1 frameshift happens, the phase moves to a new nucleotide sector, +240\textdegree\ or -120\textdegree\ away.  From the \emph{prfB} polar plot, we see that the phase shifts about 60\textdegree\ before it gets to the frameshift location (from approximately -20\textdegree\ to approximately +40\textdegree), the equivalent of one-half of a nucleotide. Then it begins its track at the angle that reestablishes it in the new nucleotide sector, +240\textdegree\ from where it originated.  We designate $x=0$ as the initial state, i.e., reading frame 0, as one of the two stable states of the ribosome-mRNA system.  We assign unit increments in $x$ for every 60\textdegree\ increment in phase, i.e. for every $\frac{1}{2}$ nucleotide-shift in the mRNA sequence.  If the ribosome shifts a whole nucleotide, as it does in the +1 frameshift, we have $x=2$.  So a +1 frameshift can be modeled as a state transition from $x=0$ to $x=2$. The intermediate value $x=1$ can be thought of as a \emph{boundary} point, where there is equal likelihood of picking either the codon in Frame 0 or the codon in Frame +1.

As stated earlier, the cumulative energy signal, owing to its sinusoidal nature, can be represented as $\mathbf{V}_{k} = M_{k}e^{j\theta_{k}}$. We will refer to $\mathbf{V}_{k}$ as the \emph{cumulative vector}. The contents of $\mathbf{V}_{k}$ contain a summation of the entire free energy signal up to codon $k$. The derivative of $\mathbf{V}_{k}$ with respect to codon position $k$ gives the instantaneous energy available at codon $k$.
\begin{equation}
    \textbf{D}_{k} = \frac{d}{dk}\left(M_{k}e^{j\theta_{k}}\right) = M_{k}\frac{d}{dk}\left(e^{j\theta_{k}}\right) + e^{j\theta_{k}}\frac{dM_{k}}{dk}
    \label{diffvec}
\end{equation}

The magnitude and phase of the differential vector $\textbf{D}_{k}$, referred to as differential magnitude and differential phase, are given by Equation \eqref{diffmag} and Equation \eqref{diffph} respectively.

\begin{equation}
\left|\textbf{D}_{k}\right| = \sqrt{\left(\frac{dM_{k}}{dk}\right)^{2} + \left(M_{k}\frac{d\theta_{k}}{dk} \right)^{2}}
\label{diffmag}
\end{equation}
\begin{equation}
\angle \textbf{D}_{k} = \theta_{k} + \arctan\left(\frac{M_{k}\frac{d\theta_{k}}{dk}}{\frac{dM_{k}}{dk}}\right)
\label{diffph}
\end{equation}

To calculate $\left|\textbf{D}_{k}\right|$ and $\angle \textbf{D}_{k}$, we will need the derivatives ($\frac{dM_{k}}{dk}$ and $\frac{d\theta_{k}}{dk}$), which can be evaluated using function approximation techniques \cite{chekin}. A second order polynomial can be fitted to a window of points centered around $M_{k}$, to evaluate its derivative, $\frac{dM_{k}}{dk}$. An identical procedure is followed for computing $\frac{d\theta_{k}}{dk}$.


We observe that for a signal that stays roughly in phase, $\frac{d\theta_{k}}{dk} \approx 0$, and so, $\left|\textbf{D}_{k}\right| \approx \frac{dM_{k}}{dk}$ and $\angle \textbf{D}_{k} \approx \theta_{k}$. We know, from previous work that the free energy signals in a given eubacterium have a roughly constant phase \cite{ponn.jbsb}. For \emph{E. coli}, that angle is
$\theta_{sp} \approx -20$\textdegree. For a normal, non-frameshifting gene of length $L$ nucleotides in \emph{E. coli}, we see that $\theta_{k} \rightarrow \theta_{sp}$ as $k \rightarrow \frac{L}{3}$. Within the context of our hypothesis, the differential vector $\textbf{D}_{k}$ represents a force acting on the ribosome at codon $k$ that adjusts the position of the ribosome relative to the mRNA, i.e., that modulates reading frame.


Another element believed to play an integral part in programmed frameshifts is ribosomal pausing \cite{far.mr}. Sipley and Goldman \cite{sipgold} provide experimental evidence that supports a frameshift model in which ribosomal pause time is a major determinant of frameshift probability, with pause time a function of tRNA availability.  Therefore, we introduce the concept of \emph{wait-time}, a measure of how long the ribosome waits for the tRNA to associate with the ribosome A-site, into our displacement model.

\subsection{Estimating wait-time}
The actual availability of tRNA, estimated using two-dimensional polyacrylamide gel electrophoresis, was found to be proportional to codon frequency for moderately expressed genes \cite{ikem.mbe}. Using a set of mRNA sequences in \emph{E. coli} that have $N$ codons in all, the frequency of each codon (except the stop codons) can be calculated as
\begin{equation}
    f_{i} = \frac{N_{i}}{N} , \ \ \ i = 1 \ldots 61
    \label{freqcalc}
\end{equation}
where $N_{i}$ is the number of codons of type $i$. If a particular tRNA recognizes only one codon, then the codon frequency would be indicative of its availability. If there is more than one codon recognized by a tRNA isoacceptor, then the availability of that isoacceptor will be the sum of the individual codon frequencies. We estimate the availability of each tRNA isoacceptor using
\begin{equation}
    \gamma_{p} = \displaystyle\sum_{i=1}^{n_{p}} f_{i} , \ \ \ p = 1 \ldots 20
    \label{trnacalc}
\end{equation}
where $n_{p}$ is the number of codons that code for amino acid $p$.

Codons having abundant tRNAs would have short wait-times, and vice-versa. We assume a decreasing linear relationship between the wait-time $\tau$ and the tRNA availability $\gamma$, as shown in Equation \eqref{nloop}. The wait-time gives an approximate number of cycles for which the ribosome can adjust itself while waiting for the appropriate tRNA. The number of wait cycles for a few sample codons are shown in Table \ref{tab:wttab}.
\begin{equation}
    \tau_{p} = \frac{\max(\gamma)-\gamma_{p}}{\min(\gamma)}
    \label{nloop}
\end{equation}

\begin{table}
\begin{center}
\begin{tabular}{|c|c|c|}
\hline {\em Codon} & {\em Amino-acid} & {\em Number of wait-cycles} \\
\hline
aac &   Asn &   7   \\
ccu &   Pro &   16  \\
acg &   Thr &   13  \\
cuu &   Leu &   13  \\
uuc &   Phe &   7   \\
gca &   Ala &   2   \\
\hline
\end{tabular}
\caption{Wait-times for a few sample codons in \emph{E. coli}}
\label{tab:wttab}
\end{center}
\end{table}

\subsection{The complete model}
The vector $\textbf{D}_{k}$ represents a force that could produce a linear movement of the ribsome one way or the other until the corresponding tRNA is found for the codon in the A-site. The displacement at each codon position is calculated incrementally ($\Delta x$), with the sign of $\Delta x$ indicating the direction of movement (+ = downstream, - = upstream). The total displacement $x_{k}$ is obtained by accumulating $\Delta x$ for the corresponding number of wait cycles. When the ribosome is in reading frame 0, we define $x=0$ and when it moves into the +1 frame, we define $x=2$. We claim that the following equation captures the behavior in both reading frame states:
\begin{equation}
    \Delta x_{k} = -C \left|\textbf{D}_{k}\right| \sin\left(\angle \textbf{D}_{k} + \frac{\pi x_{k}}{3} - \theta_{sp}\right)
    \label{dxcalc}
\end{equation}
The argument of the sine function contains the instantaneous measurement of phase:
\begin{equation}
    \theta_{\Delta x} = \frac{\pi x_{k}}{3} - \theta_{sp}
    \label{phasedx}
\end{equation}

Observe that when $x=0$, the cumulative phase is at the species angle i.e., $\angle \textbf{D}_{k} = \theta_{sp}$, leading to $\Delta x=0$. When $x=2$, we have $\angle \textbf{D}_{k} = \theta_{sp} + \frac{4\pi}{3}$, again leading to $\Delta x=0$. To calculate $\Delta x$, we introduce a constant of proportionality $C$, and calibrate it using the \emph{prfB} signal. Mathematically, \emph{C} measures the rate at which the ribosome adjusts itself to perturbations in $x$. For each unit of wait-time (also referred to as a \emph{wait-cycle}), the incremental displacement $\Delta x^{j}_{k}$ gets added onto the current position $x^{j}_{k}$. The total displacement is then assigned to the next codon $k+1$. Note that we are using the superscript $j$ to index increments made during the wait-time of the ribosome. If the ribosome waits for $\tau$ cycles at codon $k$, the total initial displacement at codon $k+1$ would be assigned as

\begin{equation}
    x^{0}_{k+1} = \displaystyle\sum_{j=1}^{\tau} \Delta x^{j}_{k}
    \label{xtotal}
\end{equation}

\subsection{Stability}
In practice, all the above equations hold approximately, so it is important to establish stability of the ribosome-mRNA system in a rigorous manner \cite{strogatz}. Equation \eqref{dxcalc} can be written as a recursive relation
\begin{equation}
    x^{j+1}_{k}=x^{j}_{k}-C \left|\textbf{D}_{k}\right| \sin\left(\angle \textbf{D}_{k} + \frac{\pi x^{j}_{k}}{3} - \theta_{sp}\right)
    \label{recrel}
\end{equation}

\begin{figure}
    \centering
  \includegraphics[height=2.0in,width=2.0in]{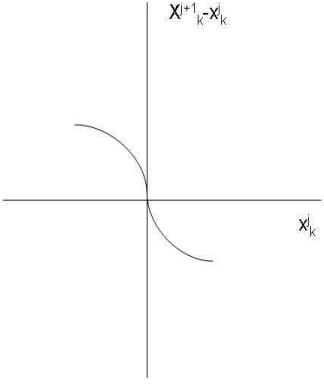}
    \caption{Vector field generated by Equation \eqref{dxcalc}}
    \label{fig:stab2}
\end{figure}

\subsubsection{Stability of $x^{*}=0$}
When the ribosome is in reading frame 0, $x^{j}_{k}=0$ and $\angle \textbf{D}_{k} = \theta_{sp}$. Substituting $x^{j}_{k}=0$ into Equation \eqref{recrel} leads to $x^{j+1}_{k}=x^{j}_{k}$, and hence, $x^{*}=0$ is a fixed point. Let $\eta_{j}=x^{j}_{k}-x^{*}$ be a small perturbation away from $x^{*}$. To see whether the perturbation grows or decays, we substitute $x^{j}_{k}=\eta_{j}+x^{*}$ into Equation \eqref{recrel}. The recursive relation can now be written as
    \[ x^{*}+\eta_{j+1} = x^{*}+\eta_{j} - C\left|\textbf{D}_{k}\right| \sin\left(\angle \textbf{D}_{k} + \frac{\pi (x^{*}+\eta_{j})}{3} - \theta_{sp}\right) \]
Substituting $x^{*}=0$, we get
\begin{equation}
    \eta_{j+1} = \eta_{j} - C\left|\textbf{D}_{k}\right| \sin\left(\frac{\pi \eta_{j}}{3}\right)
    \label{etaeqn}
\end{equation}
Since $\eta_{j}$ is small, we have
    \[ \eta_{j+1} \approx \eta_{j} - C\left|\textbf{D}_{k}\right| \frac{\pi \eta_{j}}{3} = \left(1 - C\frac{\pi \left|\textbf{D}_{k}\right|}{3}\right)\eta_{j} \]
By making $C$ fairly small, it can be ensured that $\left(C\frac{\pi \left|\textbf{D}_{k}\right|}{3}\right) < 1 \ \forall k$. This implies that $\eta_{j}$ decays to zero as $j$ gets large, since $\left(1 - \frac{\pi \left|\textbf{D}_{k}\right|}{3}\right) < 1$. Thus, small perturbations cause the displacement to converge to the fixed point $x^{*}=0$. The idea is illustrated in Figure \ref{fig:stab2}.

\subsubsection{Stability of $x^{*}=2$}
When the ribosome is in reading frame +1, $x^{j}_{k}=2$ and $\angle \textbf{D}_{k} = \theta_{sp} + \frac{4\pi}{3}$. Substituting these into Equation \eqref{recrel} yields $x^{j+1}_{k}=x^{j}_{k}$, so $x^{*}=2$ is a fixed point. For a nearby point $x^{j}_{k}=x^{*}+\eta_{j}$, the recursive relation takes the form
    \[ x^{*}+\eta_{j+1} = x^{*}+\eta_{j} - C\left|\textbf{D}_{k}\right| \sin\left(\angle \textbf{D}_{k} + \frac{\pi (x^{*}+\eta_{j})}{3} - \theta_{sp}\right) \]
Substituting $x^{*}=2$, we get an equation identical to Equation \eqref{etaeqn}. Following identical steps, we may establish the stability of the fixed point $x^{*}=2$.

The above arguments have established that the Equations \eqref{dxcalc} and \eqref{phasedx} are structured so that the states $x=0$ and $x=2$ represent stable fixed points of the ribosome-mRNA system. Transition between the states is governed by the differential vector $\textbf{D}_{k}$ and the time $\tau$ for which the ribsome waits at codon $k$.

\section{Results}
Two model parameters, the species phase angle, $\theta_{sp}$, and the constant, $C$, must be specified to generate displacement values. The species phase angle $\theta_{sp}$ is the mean phase angle estimated from the set of verified genes as annotated in GENBANK\footnote{http://www.ncbi.nlm.nih.gov/Genbank/}, using the method described in \cite{ponn.jbsb}. For \emph{E. coli}, the estimated value is $\theta_{sp} = -13$\textdegree. For gene \emph{prfB} in \emph{E. coli}, the value of $C=0.005$ gave the highest resolution of a jump in displacement at codon 26. These values of $\theta_{sp}$ and $C$ were used for subsequent analyses of other genes in \emph{E. coli}.  The values of these parameters for other bacteria are listed in the Appendix. At the first codon of a gene sequence, the ribosome is locked into Frame 0, so we use $x_{1}=0$. The stop codons are assigned a large number of wait-cycles, typically 1000.

\begin{figure}
    \centering
  \includegraphics[width=2.5in]{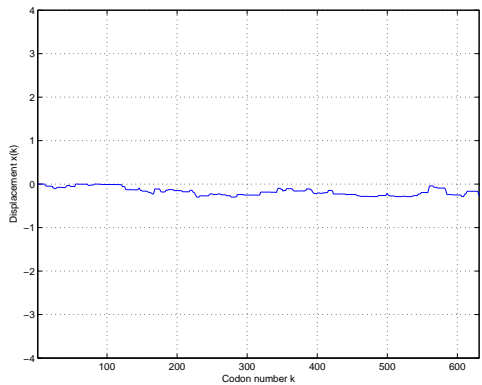}
    \caption{Displacement plot for gene \emph{aceF} in \emph{E. coli}}
    \label{fig:aceF_x}
\end{figure}

\begin{figure}
    \centering
  \includegraphics[width=2.5in]{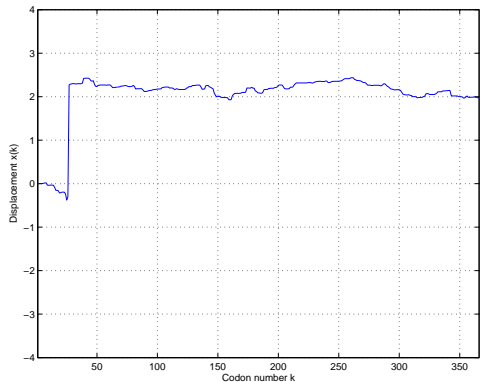}
    \caption{Displacement plot for gene \emph{prfB} in \emph{E. coli}}
    \label{fig:prfB_x}
\end{figure}

The displacement plots for the \emph{aceF} and \emph{prfB} genes of \emph{E. coli} are given in Figures \ref{fig:aceF_x} and \ref{fig:prfB_x}, respectively.  Several features of these plots are of note.  The displacement plot for \emph{aceF} (Figure \ref{fig:aceF_x}), a gene lacking a frameshift, shows that $x \approx 0$ for the entire length of the coding region. This behavior of $x$ indicates that our method does not detect a frameshift in this gene, the expected result.  In contrast, the displacement plot for the \emph{prfB} gene (Figure \ref{fig:prfB_x}) shows a sudden shift in $x$ at codon 26, the absolute value of which is slightly greater than 2 and it is in the positive direction.   Our algorithm is scaled such that a displacement value of $x=2$ indicates a shift of one nucleotide, so in this case, the displacement indicates a +1 nucleotide shift in reading frame.  This is also an expected result given that codon 26 is the location of a +1 frameshift in the \emph{prfB} gene.  For the remainder of the sequence, i.e., from codon 27 to the end of the gene, the value of $x$ remains roughly at $x=2$. This indicates that the gene stays in the new reading frame. The \emph{prfB} displacement plots for the remaining bacteria that we analyzed are given in the Appendix.

Link \emph{et al} \cite{link} assessed the \emph{in vivo} abundances of proteins in \emph{E. coli} using electrphoresis, and ranked the genes in decreasing order of yield. We calculated the free energy signals for 87 such genes in \emph{E. coli}, and analyzed them using our model. We found that for 86 of these genes, $-1 < x_{k} < 1$ for all values of $k$, indicating that the ribosome stays in frame for the entire length of each sequence. For the one remaining gene, we found slight deviation from the boundary value of $x_{k}=1$ at $k=70$, indicating a low probability of picking the in-frame codon at that location. The polar plots and displacement plots for 10 of these genes are included in the Appendix.

\section{Discussion}
Our previous work defined an algorithm that simulates possible hybridization between the 3'-terminal nucleotides of the 16S rRNA and the mRNA. The algorithm revealed a periodic, free energy signal in the coding regions of the genes in a number of bacterial species \cite{ponn.jbsb}.  Based on the ideas of Weiss \emph{et al} \cite{weiss.embo}, Trifonov \cite{trif.bch} and others, we hypothesized that this free energy signal could be supplying the information to modulate reading frame.

Using the free energy signal we developed a mathematical model optimized to precisely predict the codon location of the frameshift site within the \emph{prfB} coding sequence.  The model is an adaptive algorithm that estimates the displacement of the ribosome from its original reading frame (Frame 0).  This algorithm enables us to track the state of the ribosome-mRNA system.  The physical interpretation of the differential vector, $\textbf{D}_{k}$, in the model is that it represents the amount of force available at codon $k$ to adjust the position of the mRNA.  The amount of this adjustment potential that is actually realized is proportional to the time the ribosome waits for a tRNA to occupy the A-site.  If the tRNA is relatively abundant, little of the adjustment is realized; if the tRNA is rare implying a long pause before the A-site is occupied, more adjustment of the mRNA relative to the ribosome occurs.  The displacement $x$, captures the position adjustment.  In a recursive form, the model starts with the previous position, derived from the energy signal for all the codons up to but not including the current codon, and uses the new displacement value to update the position, or state, of the mRNA relative to the ribosome.

In the course of developing our model, we have made several approximations and assumptions. One model assumption is that the presence of rare codons is the only factor modulating elongation rate. This assumption is consistent with Spirin \cite{spirin} who asserts that the wait time due to the relative abundance of the tRNA can be assumed to be a dominating factor in inducing frameshifts.  Although mRNA secondary structure is believed to result in ribosomal pausing, its absence from our model is based on the observation that a strong correlation has not been observed in all cases between mRNA secondary structure and framshifting \cite{kontos}.

A second assumption concerns the proportionality between frequency of tRNA isoacceptor (calculated using Equation \eqref{trnacalc}) and actual tRNA availability. This proportionality is found to break down at low frequencies for genes encoding highly abundant proteins \cite{ikem.mbe}. The codon bias in such genes is extreme, and this implies that the actual tRNA availability may be more than that estimated using our simple frequency calculation. This introduces a small error into the wait-time estimated using Equation \eqref{nloop}. However, this small error would not significantly impact our overall results obtained by assuming that the wait-time is inversely proportional to our estimated tRNA availability. Another approximation involves the calculation of species mean phase angle $\theta_{sp}$. We have used \emph{all} the coding sequences annotated as ``verified'' in the GENBANK database, leading to a large variance in the estimate of $\theta_{sp}$. A more confident estimate may be obtained by using genes whose authenticity has a greater degree of certainity, such as the genes studied by Link \emph{et al} \cite{link}.

Our model has utility as both a tool that could be used for sequence annotation and for its implications as to the mechanism of reading frame maintenance and frameshifting.  Sequence annotation is an early objective for genome sequencing projects.  Frameshift sites are difficult to recognize \cite{moon.fs} for current gene annotation programs such as GENMARK \cite{genmark} and GLIMMER \cite{glimmer}. Our model implies that a free energy signal that is used to maintain reading frame is encoded in the coding regions of authentic genes.  The existence of this signal can be visualized using either polar plots of signal phase and magnitude or in displacement plots.  We are currently exploring this approach with the objective of developing an annotation program that can identify authentic coding regions and frameshift locations.

The utility of this model from the mechanistic perspective is that it suggests how both reading frame maintenance and reading frame shifts could be encoded in mRNA sequences using translational speed to modulate positional accuracy.  The model captures the idea that the instantaneous component of hybridization energy, $\textbf{D}_{k}$ (whose amount is a function of the mRNA sequence), is available to the ribosomal complex to adjust the position of the mRNA relative to the ribosomal decoding center by an amount that is proportional to the time required for a tRNA or release factor to fully occupy the A-site.  The model implies that the codon bias of mRNAs could reflect the existence of a position-adjusting mechanism to maintain reading frame. Through codon selection, each mRNA sequence carries the information to fine-tune the position of each codon in the decoding center taking into consideration variable translational speed.

One consequence of our interpretation of the functional significance of codon bias is that it could give insight into the empirically demonstrated connection between native and recombinant protein yields and codon bias.  Using the free energy signal parameters as indicators of elongation accuracy, one way to think about our model is that it yields a qualitative estimate of the frameshift tendency within a coding sequence.  To the degree that protein yield losses are determined by elongation errors, such as incorrect recruitment of tRNA, our model can show where such errors are most likely to occur in the coding sequence.  Our model can also determine which possible sequence modifications would reduce the likelihood of such errors.  By fitting a likelihood function to the displacement data $x_{k}$, we could quantify the ``correctness'' of a coding sequence for translation.  These predictions would then need to be experimentally tested.

Our model also illustrates the value of applying engineering concepts to biological systems.  The translation process operates with high reliability in potentially variable environments.  As such, it can be considered a dynamic process in which the existence of a control system for reading frame maintenance is a reasonable engineering assumption.  Mathematical modeling of control systems for dynamic processes has been the subject of considerable research \cite{maybeck79}.  Signal processing techniques have been used with considerable success to estimate the various states of a dynamic process using noisy measurements.  The Kalman filter \cite{kalman60}\cite{brown92} is one of the most useful control system models.  This filter uses recursive updating of the process state based on discrete sampling of input signal information. One example application is maintaining a ship's geographical position despite drift, a problem that bears some similarity to the problem faced by the ribosomal complex in maintaining reading frame.

Each cycle of translation elongation requires the ribosomal complex to return to the same ``position'', i.e., the positioning of the tRNA carrying the nascent polypeptide chain in the P-site.  The precision of this position is critical as the P-site tRNA spatially defines the A-site boundary in the ribosomal complex \cite{bv.rna}.  The translational process must accomplish precise positioning of the P-site tRNA in the face of considerable process variation, including potentially changing environmental conditions of salt concentration, temperature, pH, and variable process components such as tRNAs and mRNA sequences.  The requirement for the ribosomal complex to return to position in the face of environmental perturbations is analogous to the drift problem encountered in the ship example.  In our model the equation for calculating instantaneous phase
(Equation \eqref{phasedx}) is analogous to the \emph{measurement equation} of a Kalman filter, and the recursive relation (Equation \eqref{recrel}) is analogous to its \emph{state update} equation. We have identified two states $x=0$ and $x=2$ corresponding to reading frames 0 and +1, respectively.  The ribosome-mRNA system is shown to be stable in each of these two states, i.e., small perturbations to the state $x_{k}$ arising from minor signal deviations will die out eventually.  Our algorithm lays the ground work for using adaptive filtering techniques to detect frameshifts in coding sequences.  The logical next step is to design an algorithm that describes the transition into the -1 frame, and thereby develop a generalized model of reading frame maintenance in bacteria.


\nocite{moon.fs}
\nocite{shah.fs}

\bibliographystyle{IEEEtran}
\bibliography{RefBioinf}

\begin{thebibliography}{10}
\providecommand{\url}[1]{#1}
\csname url@rmstyle\endcsname
\providecommand{\newblock}{\relax}
\providecommand{\bibinfo}[2]{#2}
\providecommand\BIBentrySTDinterwordspacing{\spaceskip=0pt\relax}
\providecommand\BIBentryALTinterwordstretchfactor{4}
\providecommand\BIBentryALTinterwordspacing{\spaceskip=\fontdimen2\font plus
\BIBentryALTinterwordstretchfactor\fontdimen3\font minus
  \fontdimen4\font\relax}
\providecommand\BIBforeignlanguage[2]{{%
\expandafter\ifx\csname l@#1\endcsname\relax
\typeout{** WARNING: IEEEtran.bst: No hyphenation pattern has been}%
\typeout{** loaded for the language `#1'. Using the pattern for}%
\typeout{** the default language instead.}%
\else
\language=\csname l@#1\endcsname
\fi
#2}}

\bibitem{ponn.jbsb}
L.~Ponnala, A.-M. Stomp, D.~L. Bitzer, and M.~A. Vouk, ``Analysis of free
  energy signals arising from nucleotide hybridization between rrna and mrna
  sequences during translation in eubacteria,'' \emph{EURASIP Journal on
  Bioinformatics and Systems Biology}, vol. 2006, pp. Article ID 23\,613, 9
  pages, 2006, doi:10.1155/BSB/2006/23613.

\bibitem{weiss.embo}
R.~B. Weiss, D.~M. Dunn, A.~E. Dahlberg, J.~F. Atkins, and R.~F. Gesteland,
  ``Reading frame switch caused by base-pair formation between the 3' end of
  16{S} r{RNA} and the m{RNA} during elongation of protein synthesis in
  {E}scherichia coli,'' \emph{EMBO J}, vol.~7, no.~5, pp. 1503--1507, 1988.

\bibitem{giancoli}
D.~C. Giancoli, \emph{Physics for Scientists and Engineers}.\hskip 1em plus
  0.5em minus 0.4em\relax Prentice Hall, 1989.

\bibitem{far.mr}
P.~J. Farabaugh, ``Programmed translational frameshifting,'' \emph{Microbiol
  Rev}, vol.~60, no.~1, pp. 103--134, Mar 1996.

\bibitem{ponn.embs06}
L.~Ponnala, D.~L. Bitzer, A.~Stomp, and M.~A. Vouk, ``A computational model for
  reading frame maintenance,'' in \emph{Proceedings of the 28th IEEE EMBS
  Annual International Conference}.\hskip 1em plus 0.5em minus 0.4em\relax
  IEEE, Aug 30-Sep 3, New York City, USA 2006, pp. 4540--4543, iSBN:
  14244-0033-3.

\bibitem{recode01}
P.~V. Baranov, O.~L. Gurvich, O.~Fayet, M.~F. Prere, W.~A. Miller, R.~F.
  Gesteland, J.~F. Atkins, and M.~C. Giddings, ``{RECODE}: a database of
  frameshifting, bypassing and codon redefinition utilized for gene
  expression,'' \emph{Nucleic Acids Res}, vol.~29, no.~1, pp. 264--267, 2001.

\bibitem{recode03}
P.~V. Baranov, O.~L. Gurvich, A.~W. Hammer, R.~F. Gesteland, and J.~F. Atkins,
  ``{RECODE} 2003,'' \emph{Nucleic Acids Res}, vol.~31, no.~1, pp. 87--89,
  2003.

\bibitem{chekin}
W.~Cheney and D.~Kincaid, \emph{Numerical {M}athematics and {C}omputing},
  4th~ed.\hskip 1em plus 0.5em minus 0.4em\relax Brooks/Cole Publishing
  Company, 1999.

\bibitem{sipgold}
J.~Sipley and E.~Goldman, ``Increased ribosomal accuracy increases a programmed
  translational frameshift in {E}scherichia coli.'' \emph{Proc Natl Acad Sci
  USA}, vol.~90, no.~6, p. 2315–2319, March 15 1993.

\bibitem{ikem.mbe}
T.~Ikemura, ``Codon usage and t{RNA} content in unicellular and multicellular
  organisms,'' \emph{Mol Biol Evol}, vol.~2, no.~1, pp. 13--34, Jan 1985.

\bibitem{strogatz}
S.~H. Strogatz, \emph{Nonlinear dynamics and chaos: With applications to
  physics, biology, chemistry, and engineering}.\hskip 1em plus 0.5em minus
  0.4em\relax Perseus Books, Cambridge MA, 1994.

\bibitem{link}
A.~J. Link, K.~Robison, and G.~Church, ``Comparing the predicted and observed
  properties of proteins encoded in the genome of {E}scherichia coli,''
  \emph{Electrophoresis}, vol.~18, pp. 1259--1313, 1997.

\bibitem{trif.bch}
E.~N. Trifonov, ``Recognition of correct reading frame by the ribosome,''
  \emph{Biochimie}, vol.~74, no.~4, pp. 357--362, Apr 1992.

\bibitem{spirin}
A.~S. Spirin, \emph{Ribosomes}.\hskip 1em plus 0.5em minus 0.4em\relax
  Springer, 1999.

\bibitem{kontos}
H.~Kontos, S.~Napthine, and I.~Brierley, ``Ribosomal pausing at a frameshifter
  {RNA} pseudoknot is sensitive to reading phase but shows little correlation
  with frameshift efficiency,'' \emph{Mol Cell Biol}, vol.~21, no.~24, pp.
  8657--8670, Dec 2001.

\bibitem{moon.fs}
S.~Moon, Y.~Byun, H.-J. Kim, S.~Jeong, and K.~Han, ``Predicting genes expressed
  via -1 and +1 frameshifts,'' \emph{Nucleic Acids Res}, vol.~32, no.~16, pp.
  4884--4892, 2004.

\bibitem{genmark}
M.~Borodovsky and J.~McIninch, ``{GENMARK}: Parallel gene recognition for both
  dna strands,'' \emph{Computers Chem.}, vol.~17, no.~19, pp. 123--133, 1993.

\bibitem{glimmer}
A.~L. Delcher, D.~Harmon, S.~Kasif, O.~White, and S.~L. Salzberg, ``Improved
  microbial gene identification with {GLIMMER},'' \emph{Nucleic Acids
  Research}, vol.~27, no.~23, pp. 4636--4641, 1999.

\bibitem{maybeck79}
P.~S. Maybeck, \emph{Stochastic models, estimation, and control}, ser.
  Mathematics in Science and Engineering, 1979, vol. 141.

\bibitem{kalman60}
R.~E. Kalman, ``A new approach to linear filtering and prediction problems,''
  \emph{Transactions of the ASME - Journal of Basic Engineering}, vol. 82
  (Series D), pp. 35--45, March 1960.

\bibitem{brown92}
R.~G. Brown and P.~Y.~C. Hwang, \emph{Introduction to Random Signals and
  Applied Kalman Filtering}, 2nd~ed.\hskip 1em plus 0.5em minus 0.4em\relax
  John Wiley and Sons, Inc., 1992.

\bibitem{bv.rna}
P.~V. Baranov, R.~F. Gesteland, and J.~F. Atkins, ``{P}-site t{RNA} is a
  crucial initiator of ribosomal frameshifting,'' \emph{RNA}, vol.~10, pp.
  221--230, 2004.

\bibitem{shah.fs}
A.~A. Shah, M.~C. Giddings, J.~B. Parvaz, R.~F. Gesteland, J.~F. Atkins, and
  I.~P. Ivanov, ``Computational identification of putative programmed
  translational frameshift sites,'' \emph{Bioinformatics}, vol.~18, no.~8, pp.
  1046--1053, 2002.

\end{thebibliography}

\appendix

\subsection{Selected eubacteria}
A set of 12 eubacteria (apart from \emph{E. coli} shown in the paper) have been selected for analysis, based on the following factors:
\begin{itemize}
	\item Matching of accession number between RECODE (\url{http://recode.genetics.utah.edu/}) and GENBANK (\url{http://www.ncbi.nlm.nih.gov/genomes/lproks.cgi})
	\item Availability of a consensus sequence for the last 13 bases of the 16S rRNA, also referred to as the \emph{16S tail}
\end{itemize}

For each species, Table \ref{tab:sptab} indicates
\begin{itemize}
	\item its name
	\item its GENBANK accession number
	\item the 13 base-long \emph{16S tail}
	\item the GC-content of the species, expressed as a percentage
	\item the mean species phase angle, $\theta_{sp}$, in degrees
	\item the value of the parameter $C$, as defined in the model
	\item the number of the codon at which frameshift (FS) occurs, according to the RECODE database (following the convention that the first codon in the sequence, i.e. the start codon is numbered 1)
\end{itemize}

\clearpage
\begin{table}
\begin{center}
\begin{tabular}{|c|c|c|c|c|c|c|}
\hline 
{\emph{Name}} & {\emph{Genbank Acc}} & {\emph{16S tail}} & {\emph{(G+C)}} & {$\theta_{sp}$} & {\emph{C}} & {\emph{FS codon}} \\ \hline
Borrelia burgdorferi  & NC\_001318 & uuuccuccacuag & $28.2$ & $-63$ & $0.005$ & $20$ \\ \hline
Bacillus halodurans & NC\_002570 & uuuccuccacuag & $43.7$ & $-23$ & $0.005$ & $25$ \\ \hline
Bacillus subtilis & NC\_000964 & uuuccuccacuag & $43.5$ & $-24$ & $0.01$ & $25$ \\ \hline
Chlamydia muridarum & NC\_002620 & uuuccuccacuag & $40.3$ & $-55$ & $0.005$ & $24$ \\ \hline
Chlamydophila pneumoniae & NC\_000922 & uuuccuccacuag & $40.6$ & $-54$ & $0.005$ & $24$ \\ \hline
Chlamydia trachomatis & NC\_000117 & uuuccuccacuag & $41.3$ & $-55$ & $0.005$ & $24$ \\ \hline
Haemophilus influenzae & NC\_000907 & auuccuccacuag & $38.1$ & $-58$ & $0.005$ & $26$ \\ \hline
Pasteurella multocida & NC\_002663 & auuccuccacuag & $40.4$ & $-48$ & $0.01$ & $26$ \\ \hline   
Streptococcus mutans & NC\_004350 & uuuccuccacuag & $36.8$ & $-57$ & $0.005$ & $28$ \\ \hline
Salmonella typhimurium & NC\_003197 & auuccuccacuag & $52.2$ & $3$ & $0.005$ & $26$ \\ \hline
Treponema pallidum & NC\_000919 & uuuccuccacuag & $52.8$ & $-8$ & $0.005$ & $25$ \\ \hline
Xylella fastidiosa & NC\_002488 & uuuccuccacuag & $52.6$ & $-15$ & $0.005$ & $26$ \\ \hline    
\end{tabular}
\caption{Table of selected eubacteria}
\label{tab:sptab}
\end{center}
\end{table}

\clearpage
\subsection{+1 Frameshift genes}
\textbf{Borrelia burgdorferi}
\begin{figure}[h]
  \hfill
  \begin{minipage}[t]{.45\textwidth}
    \begin{center}  
      \epsfig{file=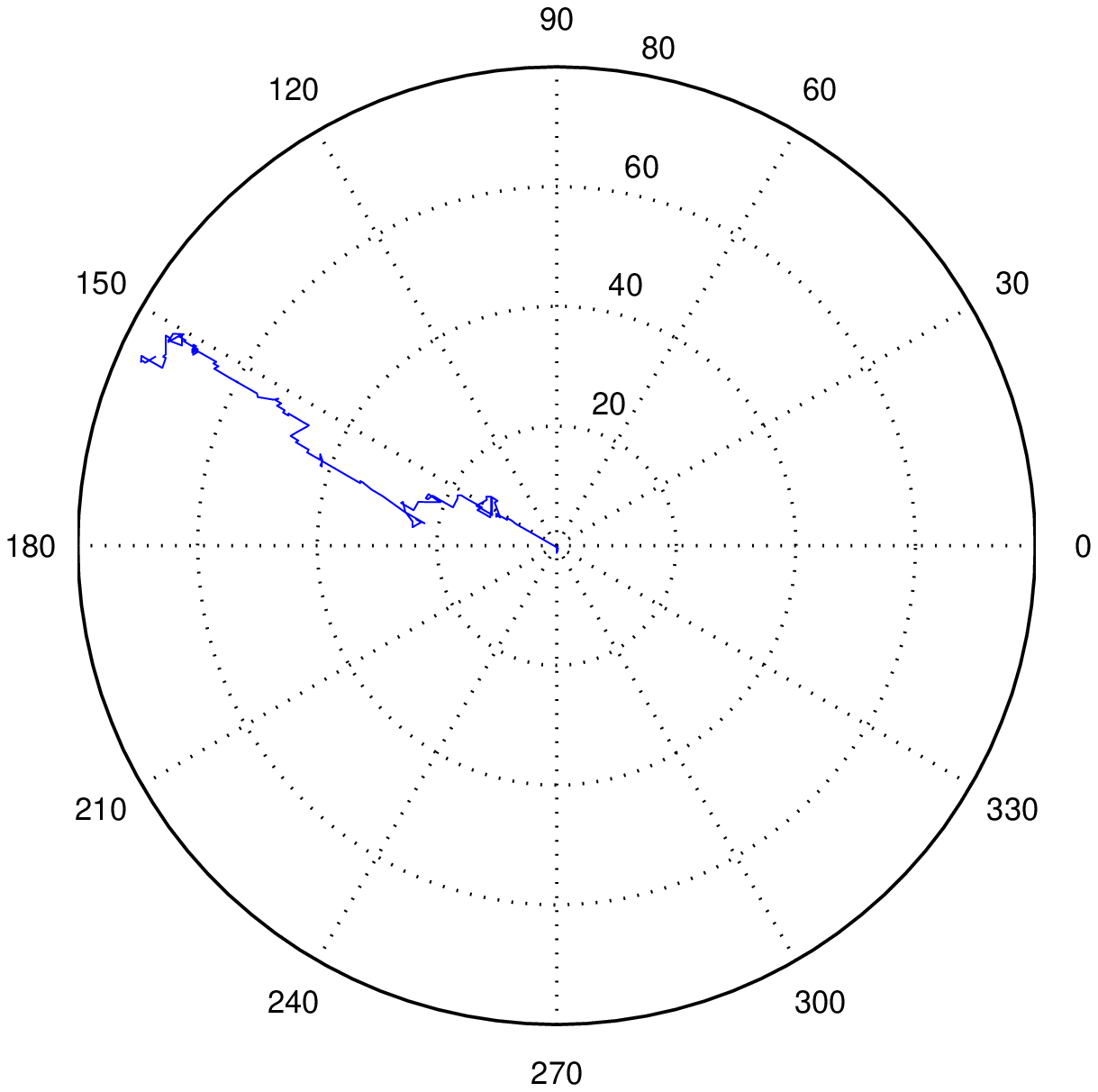, scale=0.5}
      \caption{Polar plot}
    \end{center}
  \end{minipage}
  \hfill
  \begin{minipage}[t]{.45\textwidth}
    \begin{center}  
      \epsfig{file=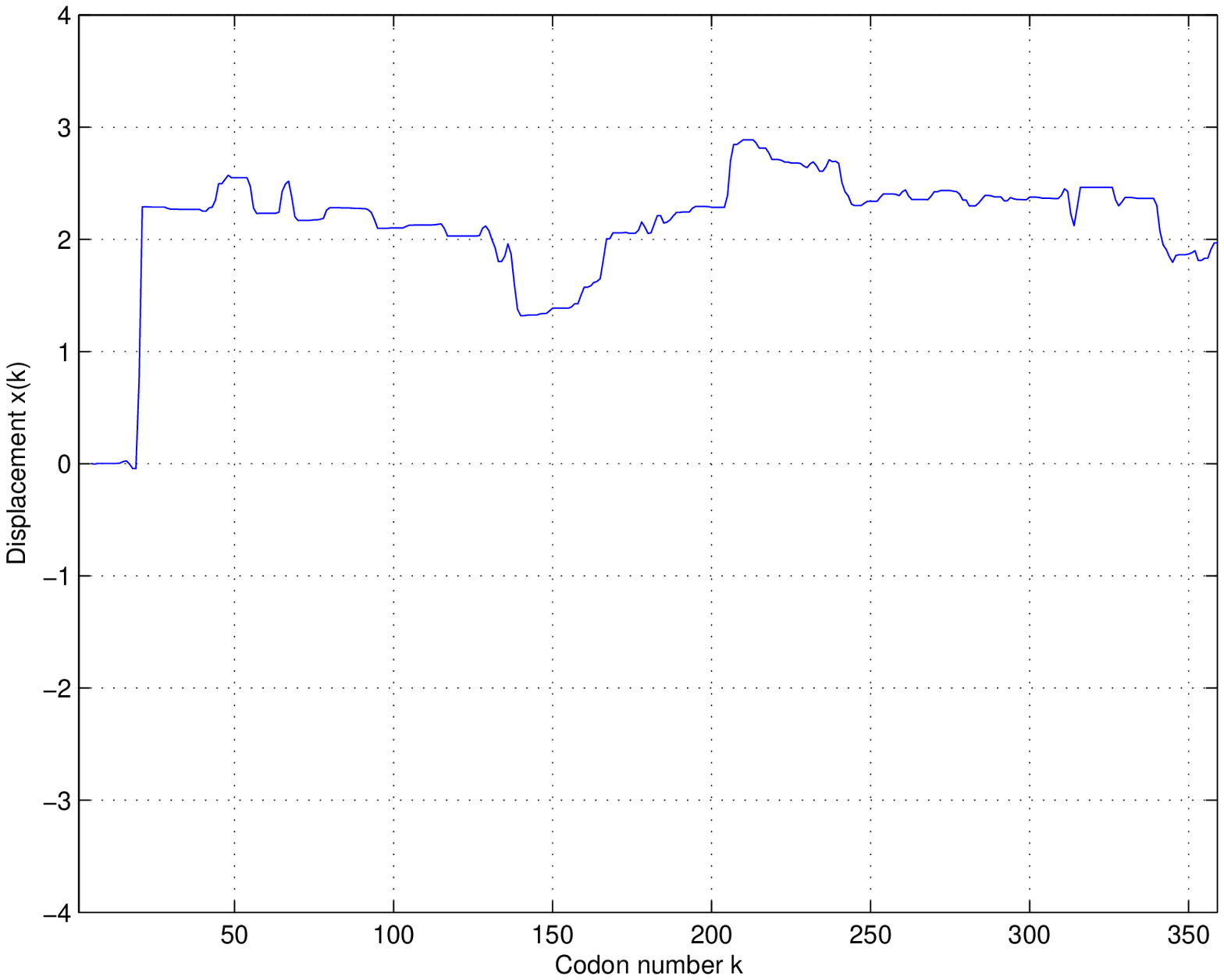, scale=0.5}
      \caption{Displacement plot}
    \end{center}
  \end{minipage}
  \hfill
\end{figure}

\clearpage
\textbf{Bacillus halodurans}
\begin{figure}[h]
  \hfill
  \begin{minipage}[t]{.45\textwidth}
    \begin{center}  
      \epsfig{file=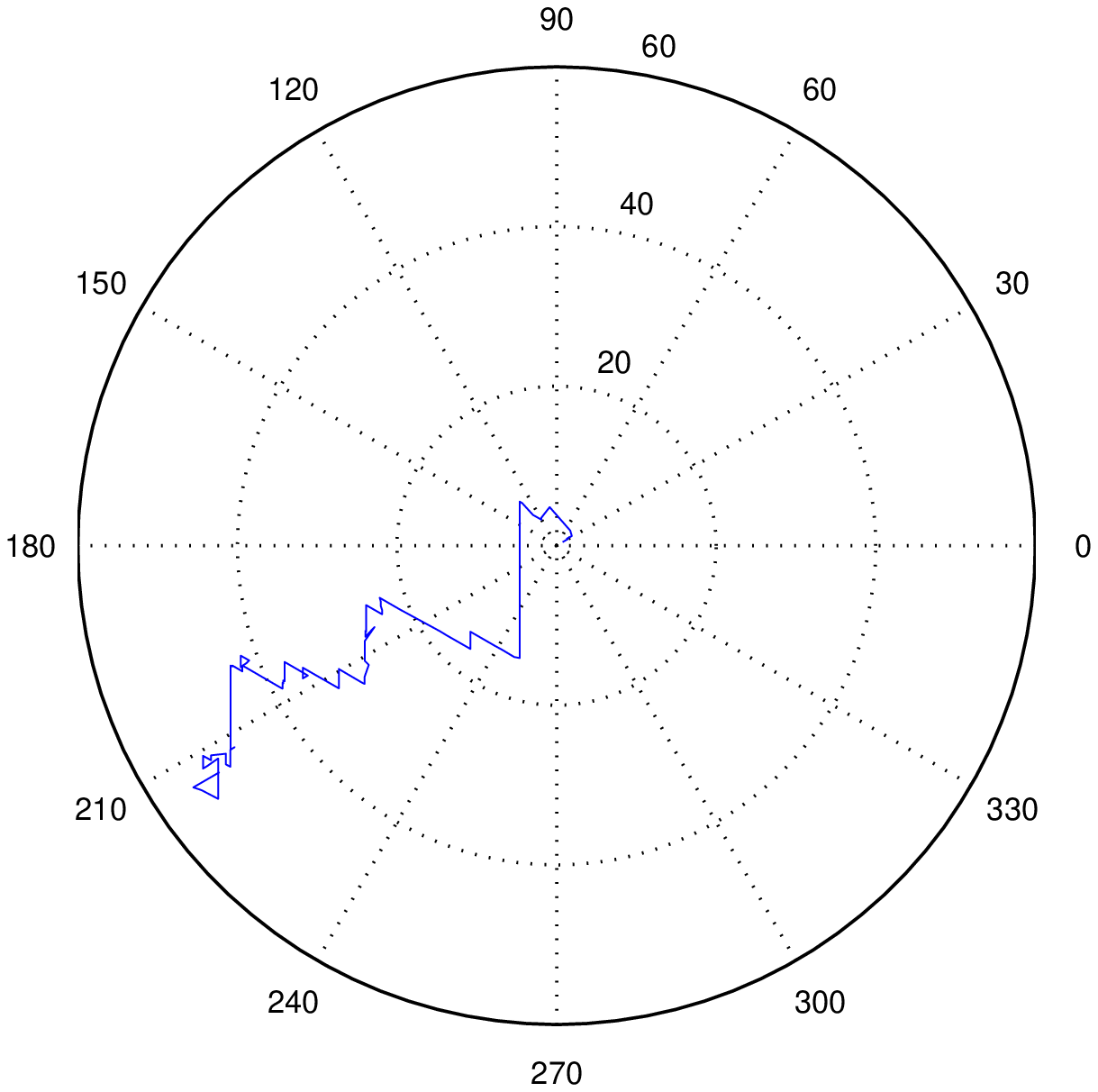, scale=0.5}
      \caption{Polar plot}
    \end{center}
  \end{minipage}
  \hfill
  \begin{minipage}[t]{.45\textwidth}
    \begin{center}  
      \epsfig{file=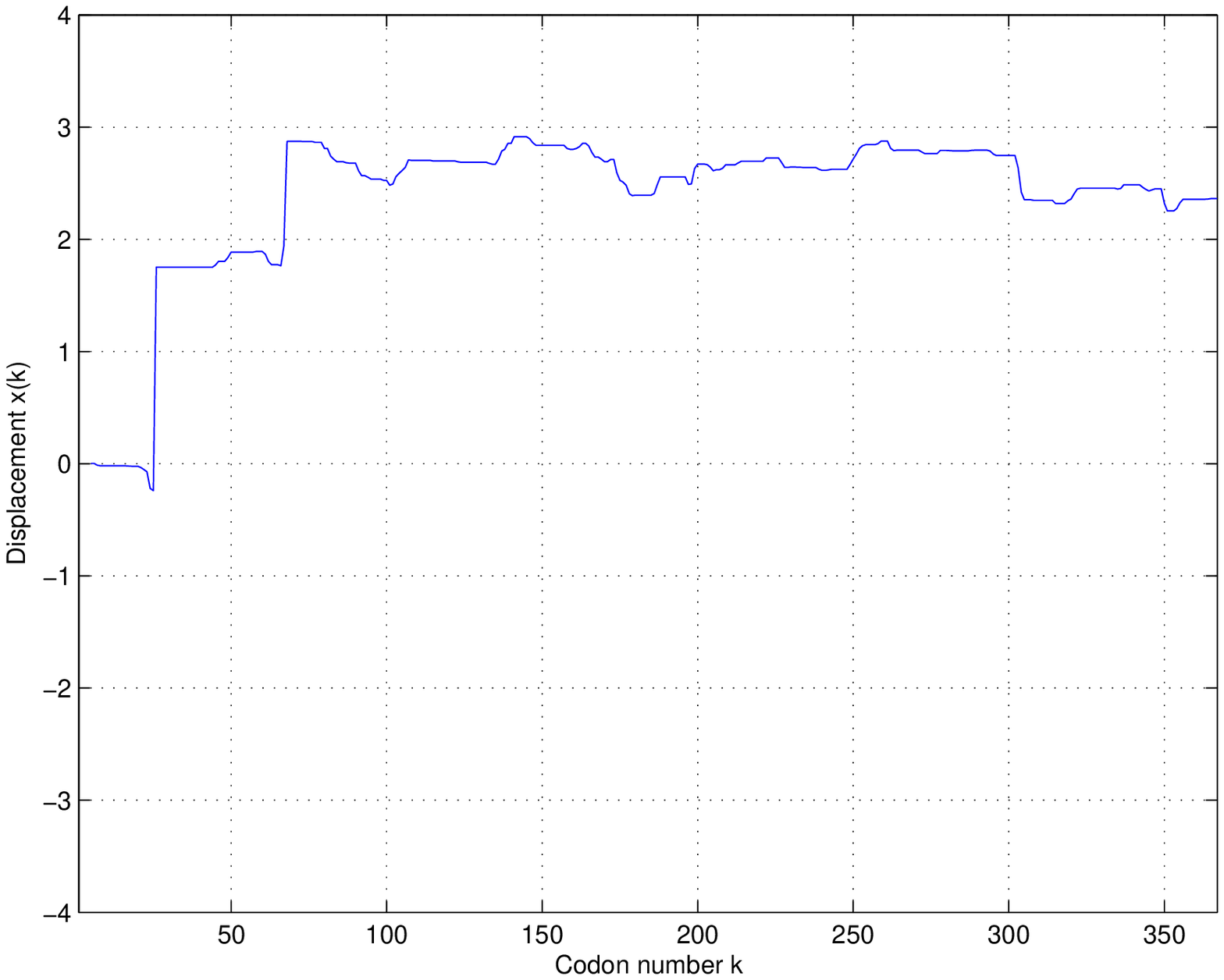, scale=0.5}
      \caption{Displacement plot}
    \end{center}
  \end{minipage}
  \hfill
\end{figure}

\clearpage
\textbf{Bacillus subtilis}
\begin{figure}[h]
  \hfill
  \begin{minipage}[t]{.45\textwidth}
    \begin{center}  
      \epsfig{file=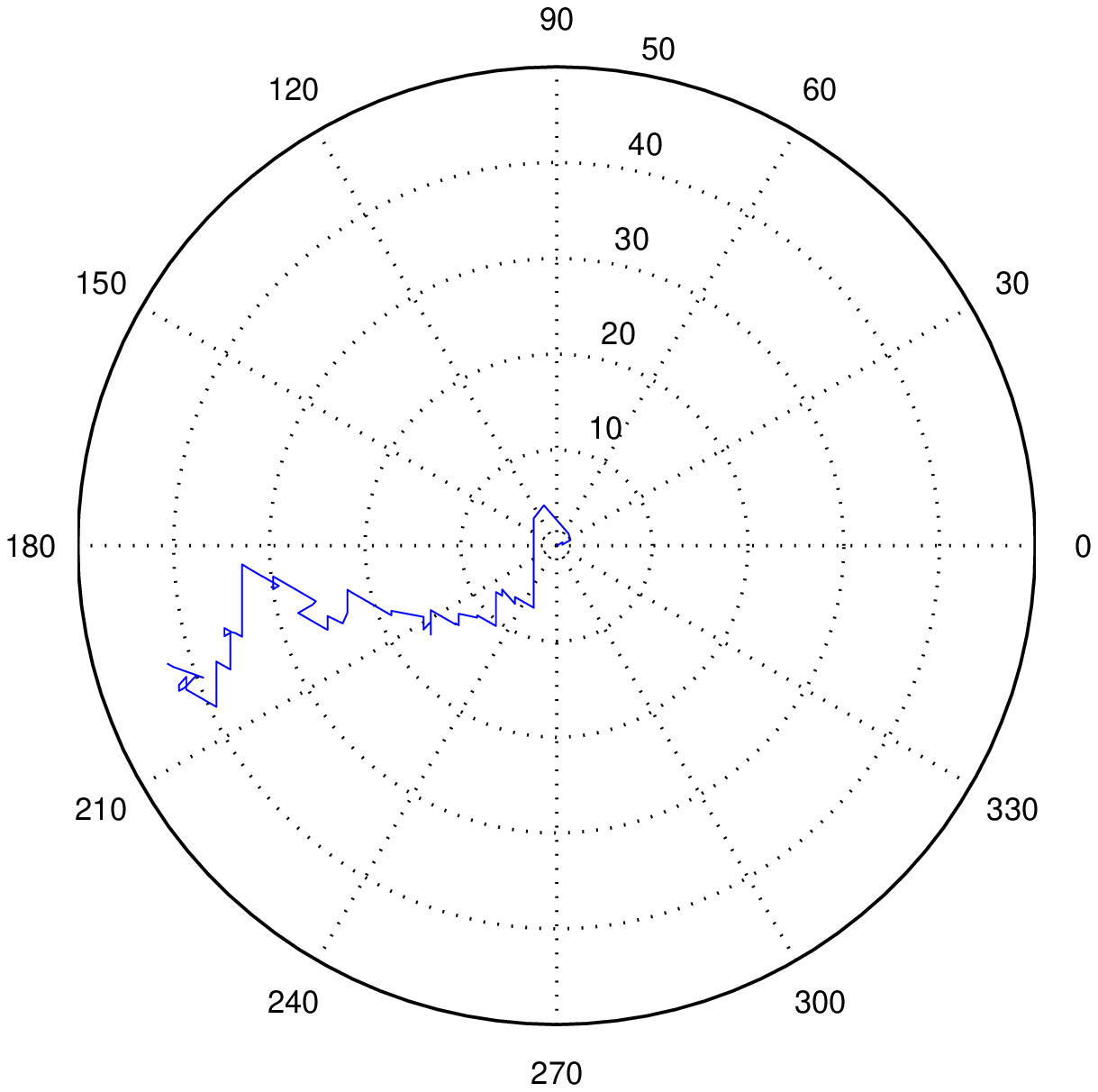, scale=0.5}
      \caption{Polar plot}
    \end{center}
  \end{minipage}
  \hfill
  \begin{minipage}[t]{.45\textwidth}
    \begin{center}  
      \epsfig{file=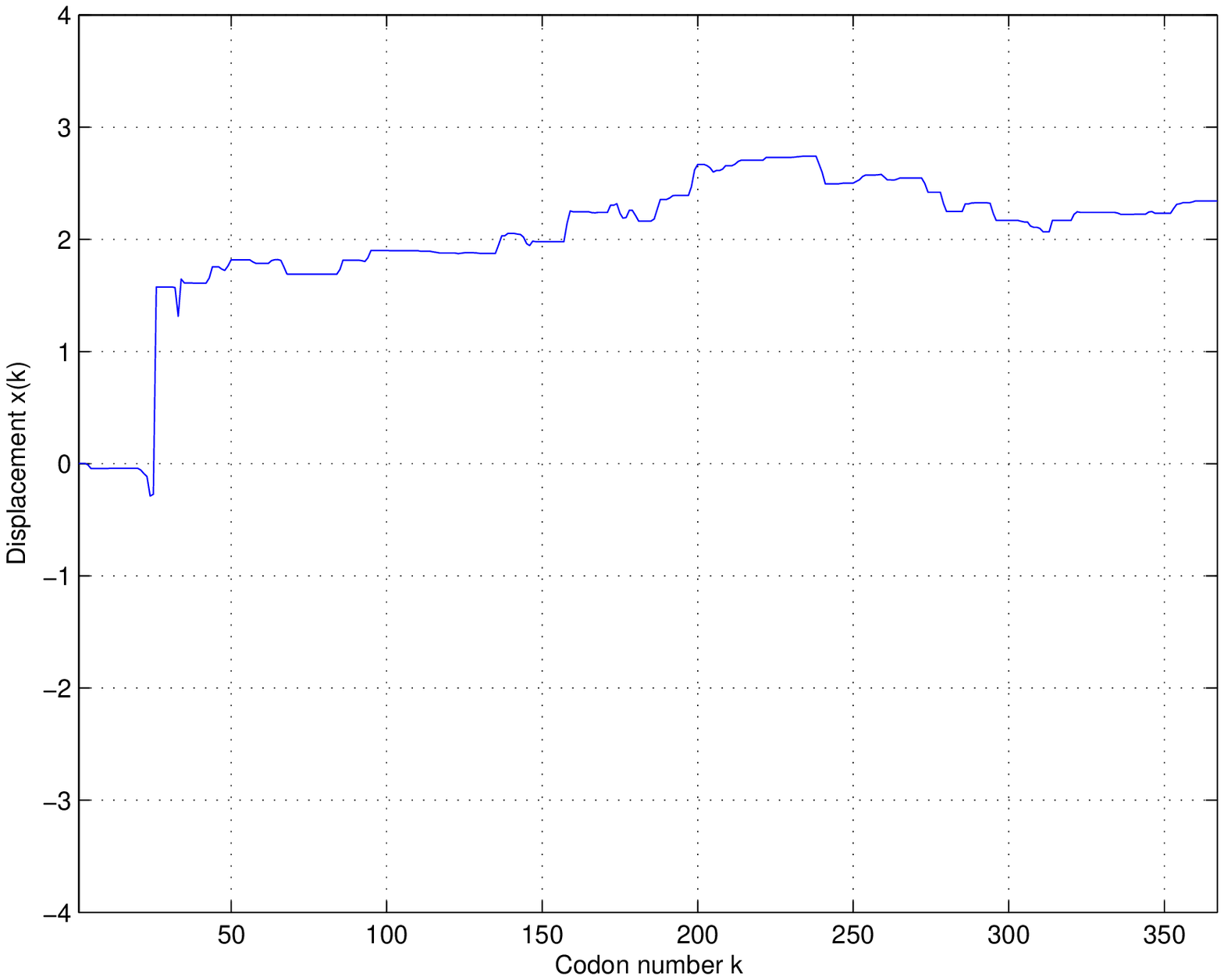, scale=0.5}
      \caption{Displacement plot}
    \end{center}
  \end{minipage}
  \hfill
\end{figure}

\clearpage
\textbf{Chlamydia muridarum}
\begin{figure}[h]
  \hfill
  \begin{minipage}[t]{.45\textwidth}
    \begin{center}  
      \epsfig{file=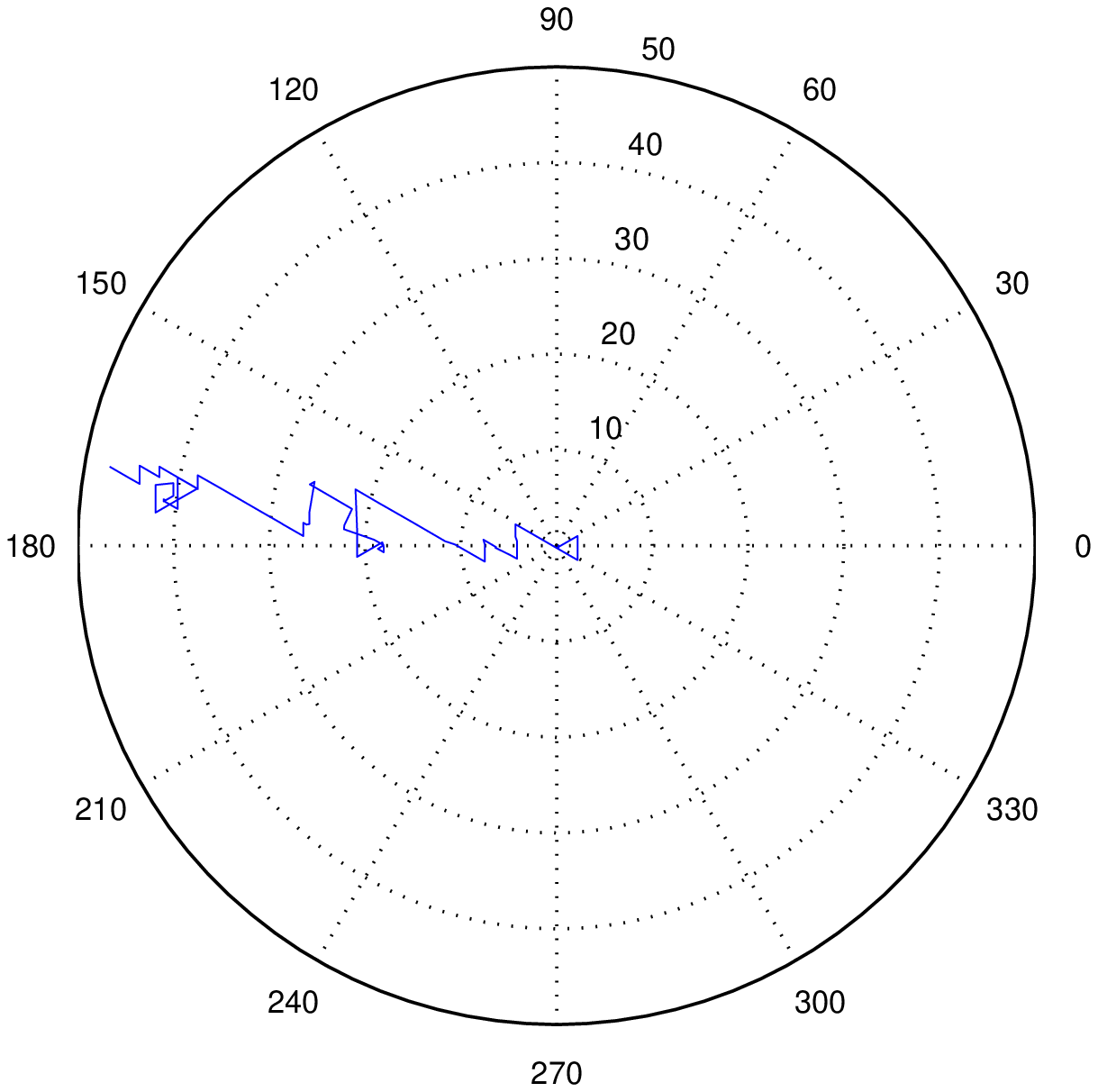, scale=0.5}
      \caption{Polar plot}
    \end{center}
  \end{minipage}
  \hfill
  \begin{minipage}[t]{.45\textwidth}
    \begin{center}  
      \epsfig{file=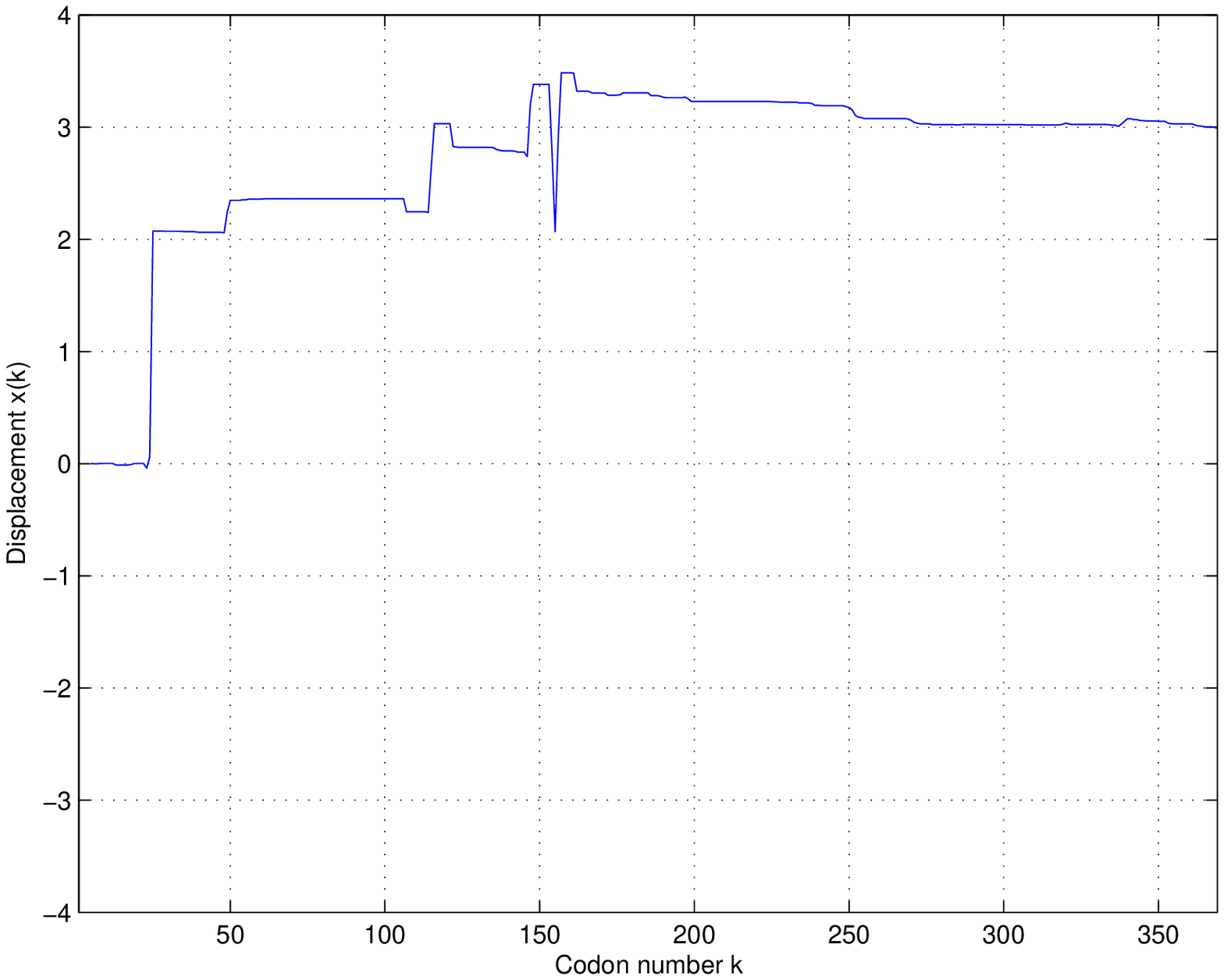, scale=0.5}
      \caption{Displacement plot}
    \end{center}
  \end{minipage}
  \hfill
\end{figure}

\clearpage
\textbf{Chlamydophila pneumoniae}
\begin{figure}[h]
  \hfill
  \begin{minipage}[t]{.45\textwidth}
    \begin{center}  
      \epsfig{file=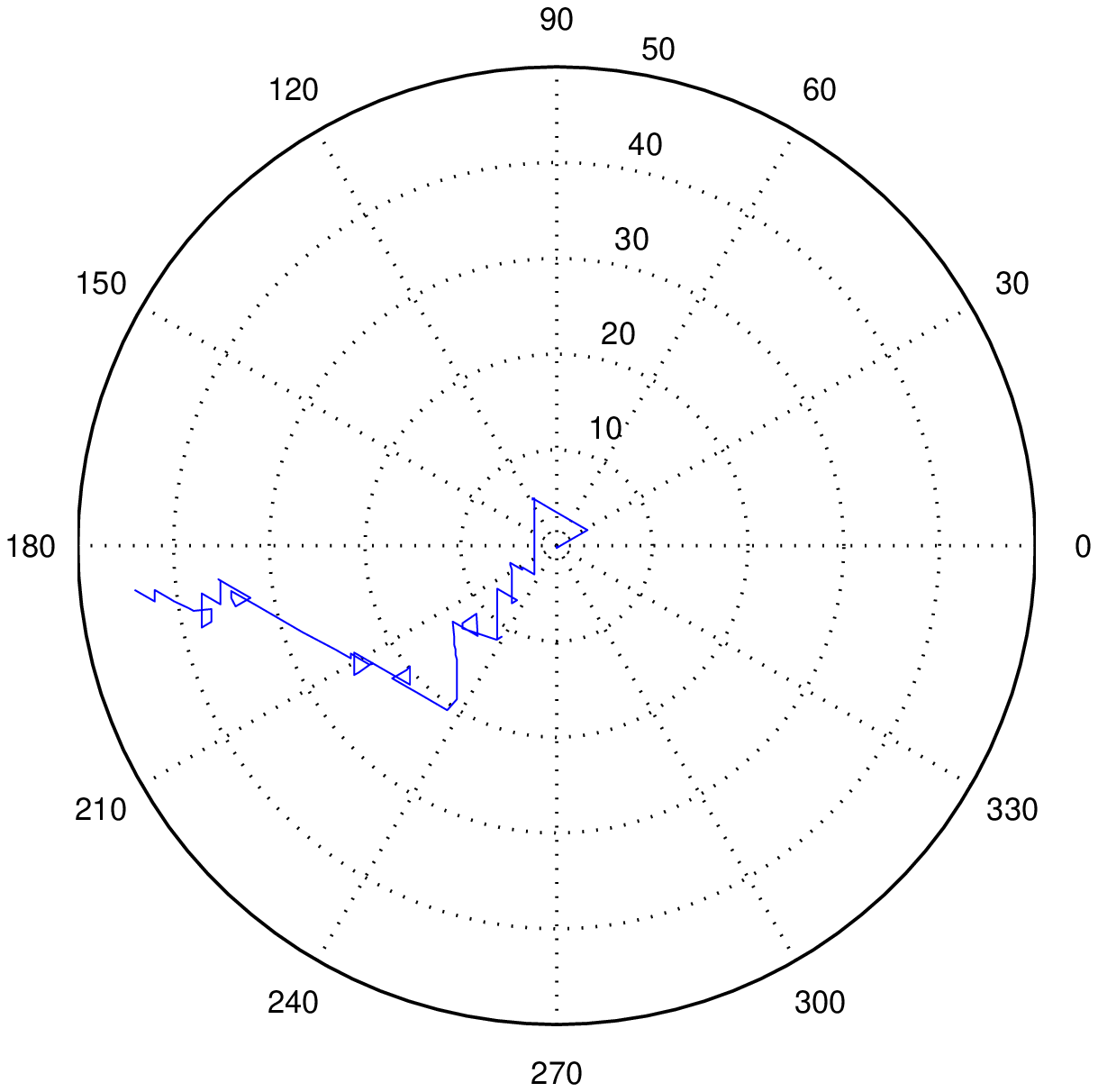, scale=0.5}
      \caption{Polar plot}
    \end{center}
  \end{minipage}
  \hfill
  \begin{minipage}[t]{.45\textwidth}
    \begin{center}  
      \epsfig{file=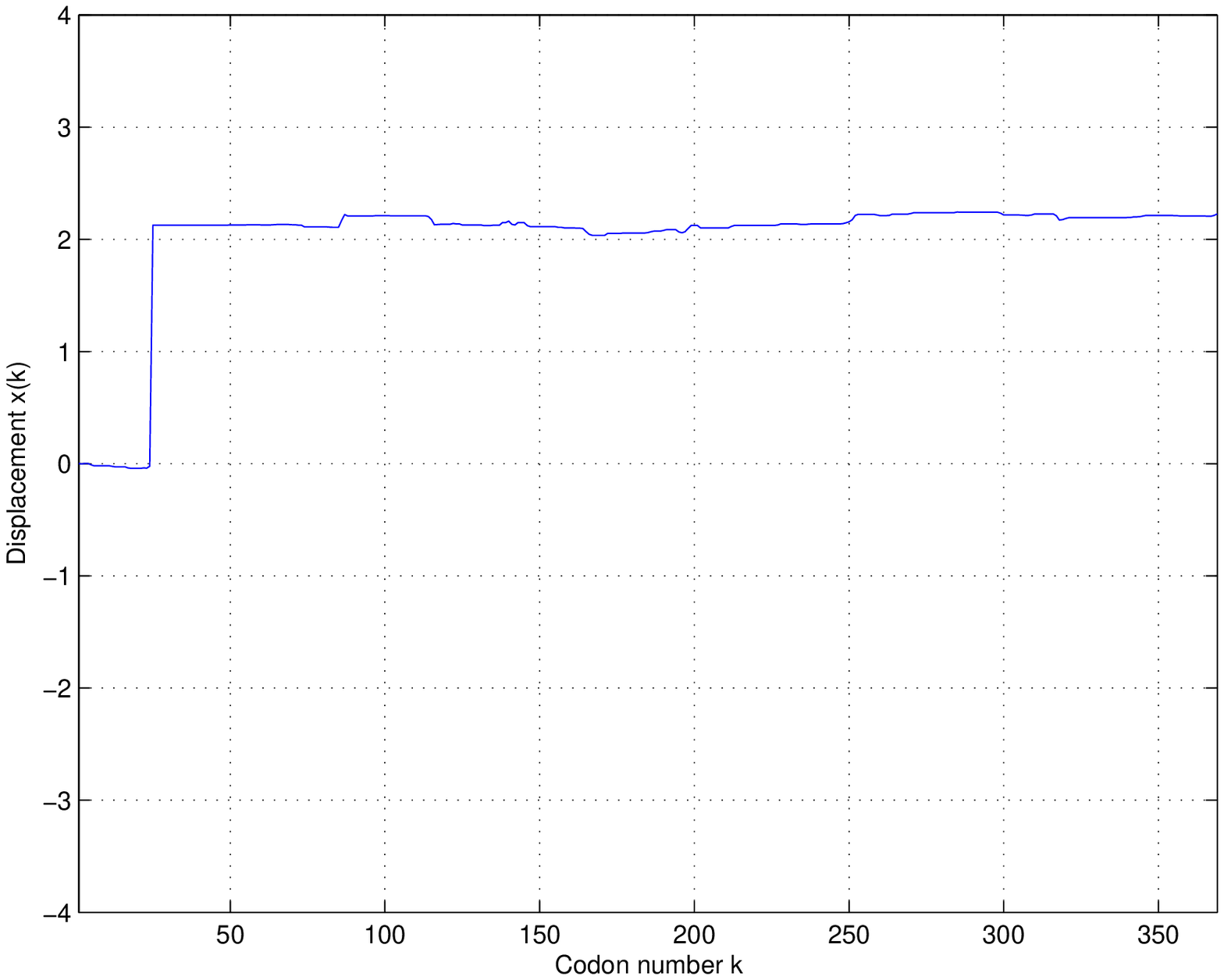, scale=0.5}
      \caption{Displacement plot}
    \end{center}
  \end{minipage}
  \hfill
\end{figure}

\clearpage
\textbf{Chlamydia trachomatis}
\begin{figure}[h]
  \hfill
  \begin{minipage}[t]{.45\textwidth}
    \begin{center}  
      \epsfig{file=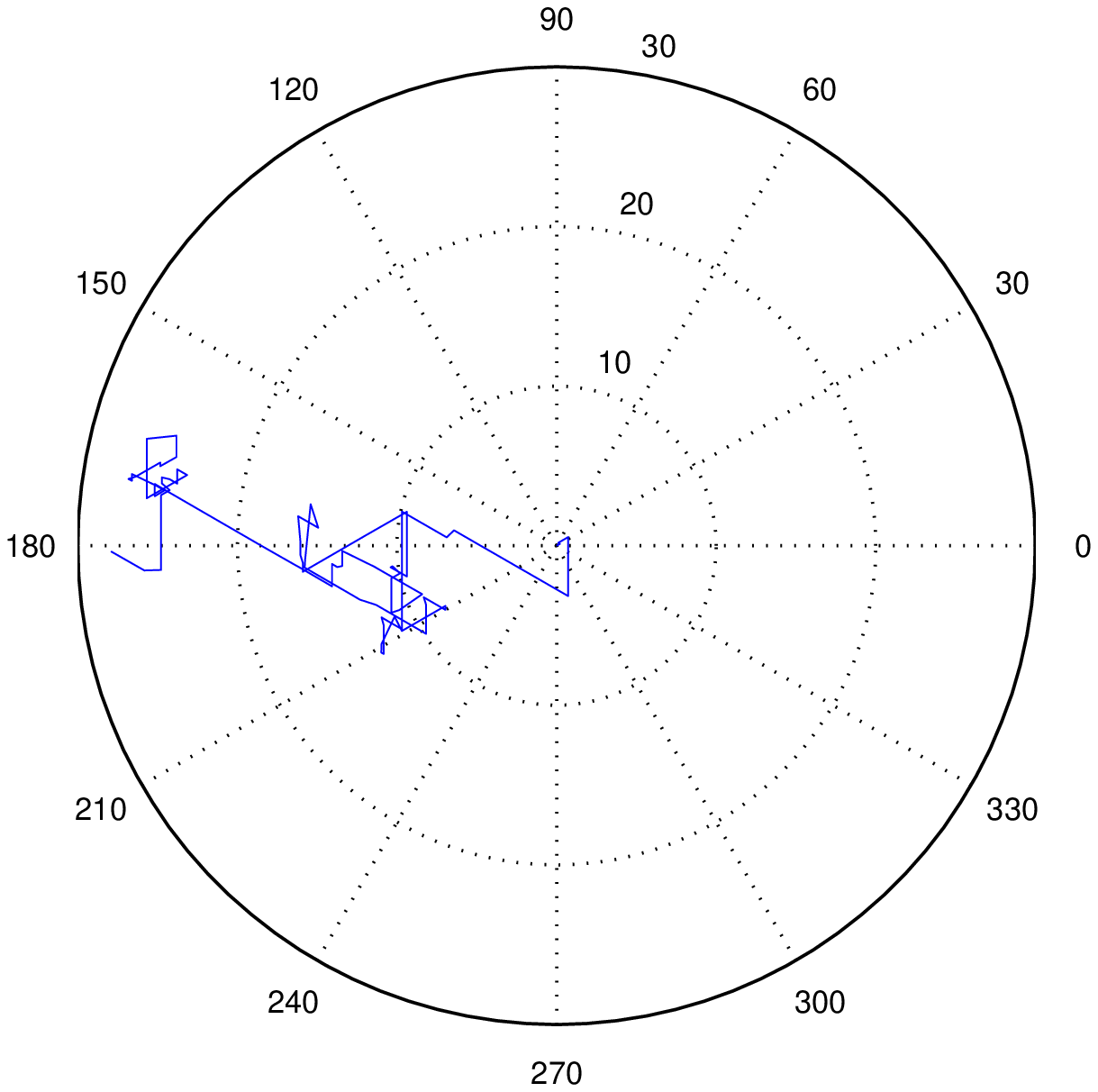, scale=0.5}
      \caption{Polar plot}
    \end{center}
  \end{minipage}
  \hfill
  \begin{minipage}[t]{.45\textwidth}
    \begin{center}  
      \epsfig{file=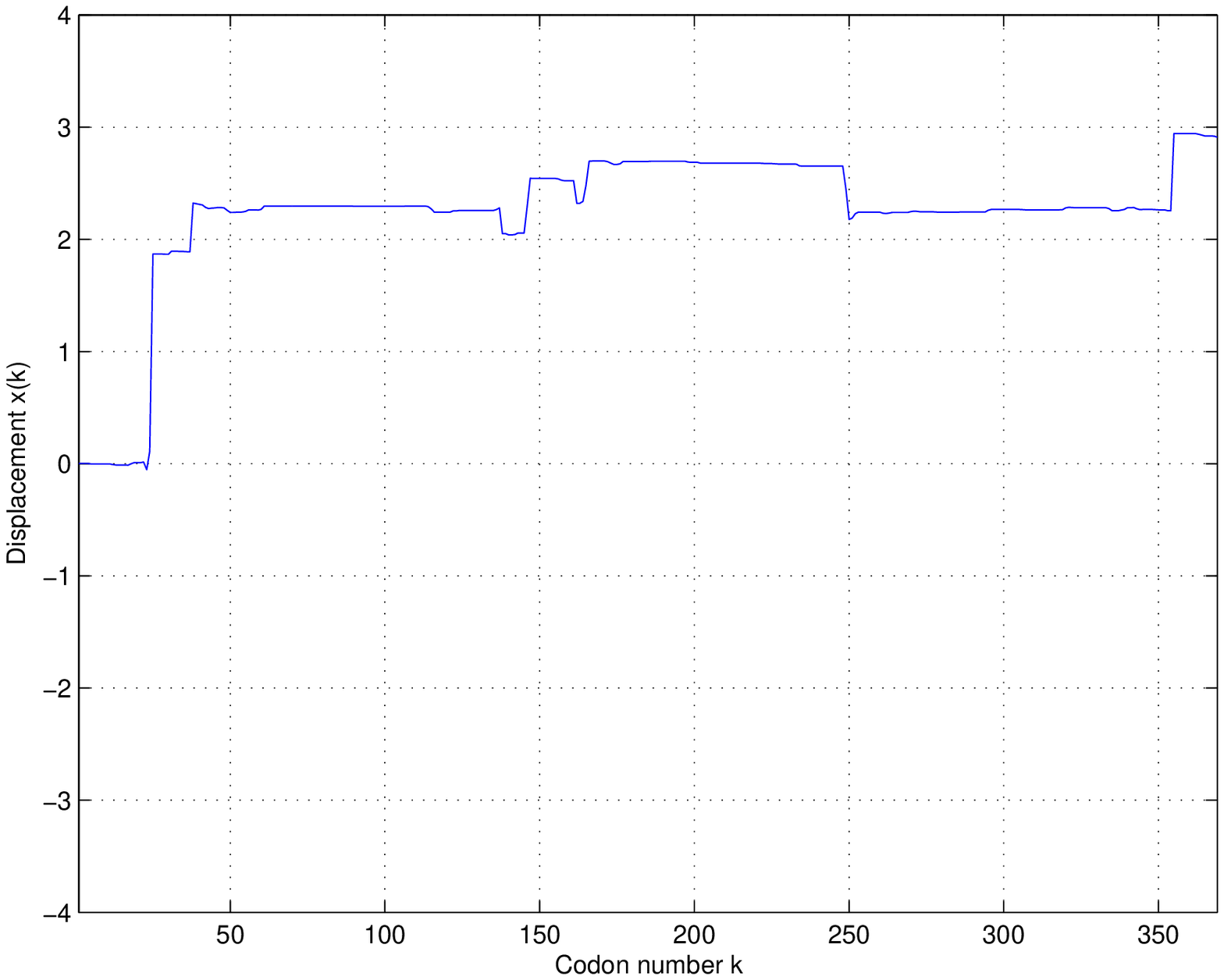, scale=0.5}
      \caption{Displacement plot}
    \end{center}
  \end{minipage}
  \hfill
\end{figure}

\clearpage
\textbf{Haemophilus influenzae}
\begin{figure}[h]
  \hfill
  \begin{minipage}[t]{.45\textwidth}
    \begin{center}  
      \epsfig{file=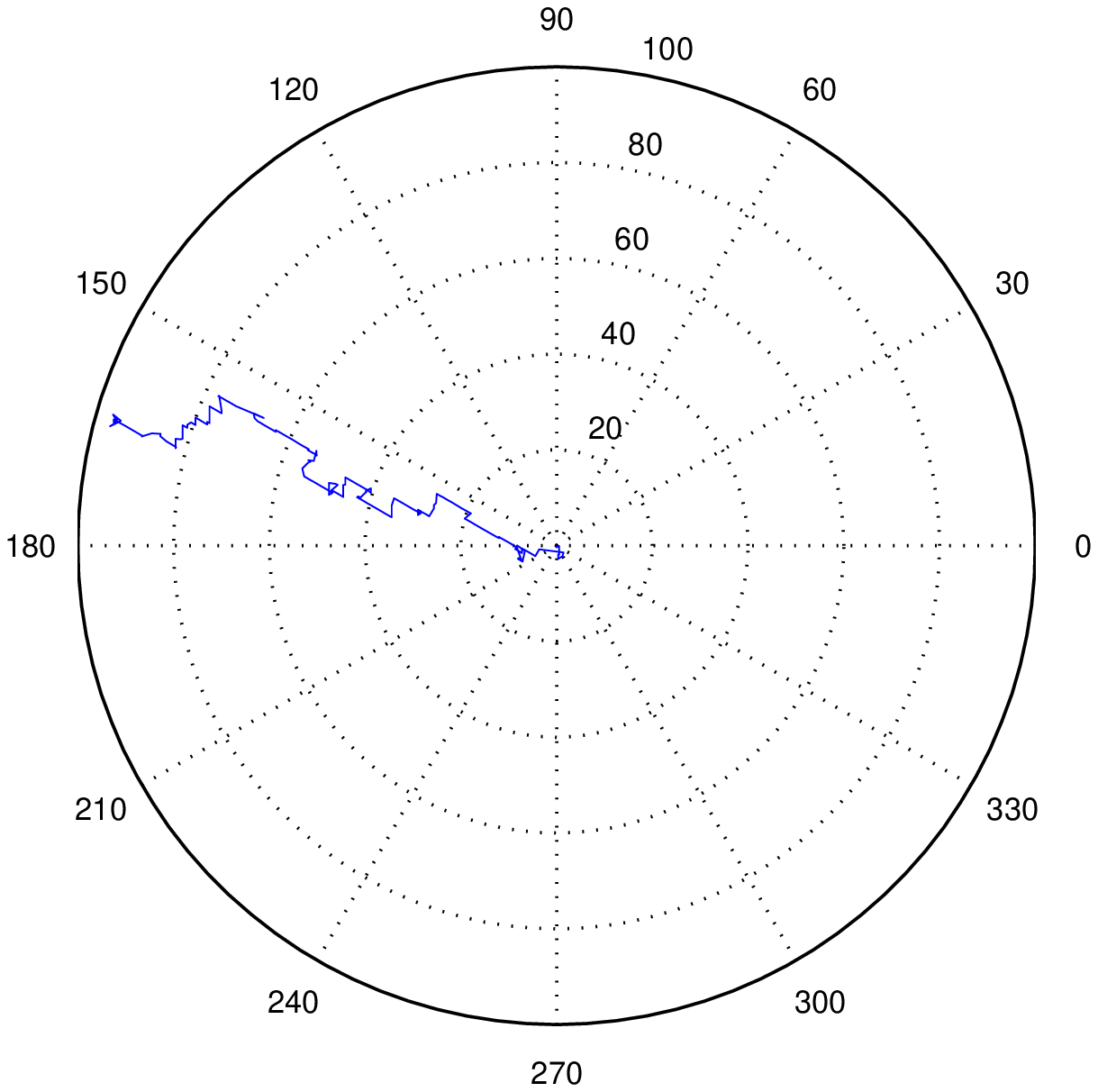, scale=0.5}
      \caption{Polar plot}
    \end{center}
  \end{minipage}
  \hfill
  \begin{minipage}[t]{.45\textwidth}
    \begin{center}  
      \epsfig{file=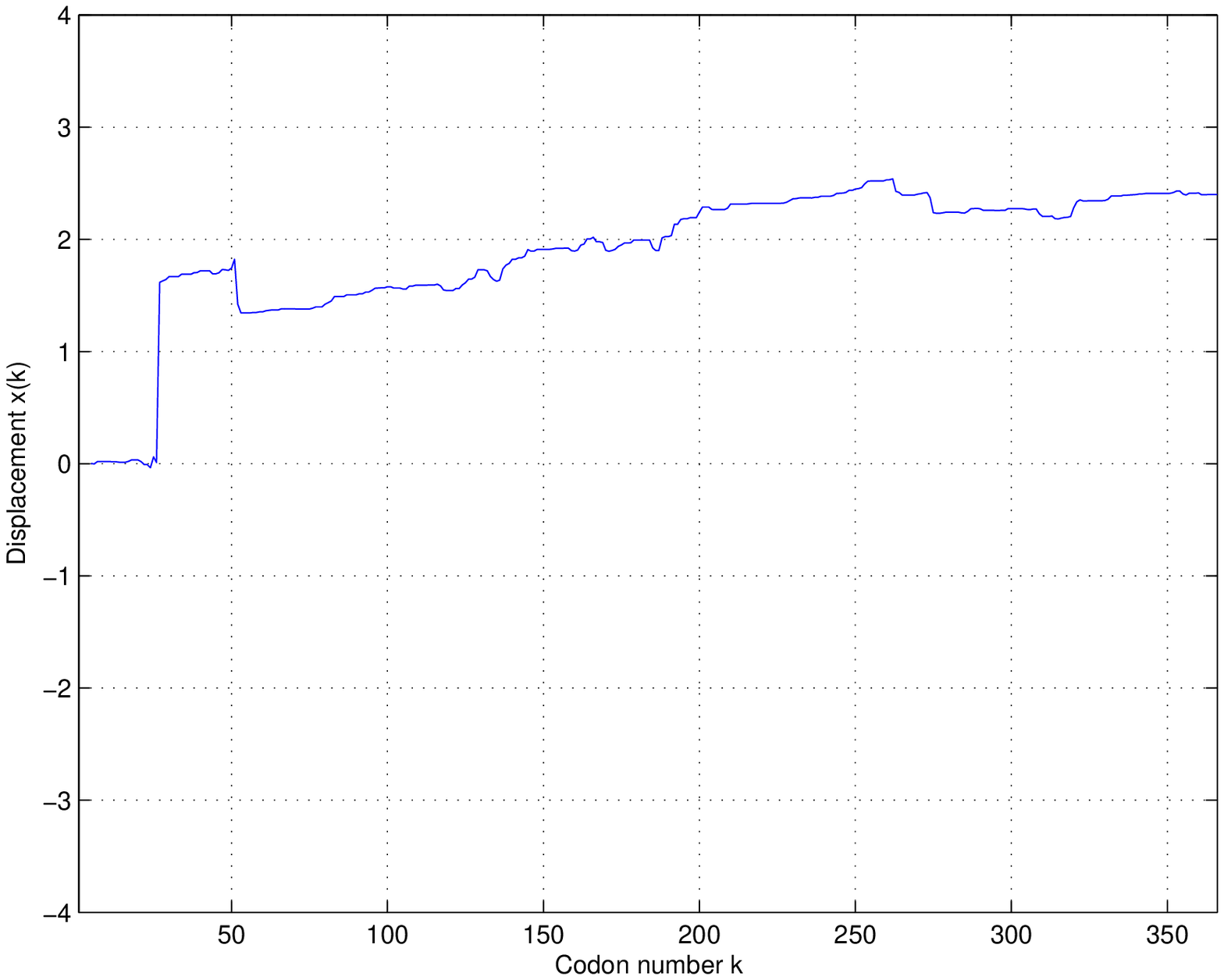, scale=0.5}
      \caption{Displacement plot}
    \end{center}
  \end{minipage}
  \hfill
\end{figure}

\clearpage
\textbf{Pasteurella multocida}
\begin{figure}[h]
  \hfill
  \begin{minipage}[t]{.45\textwidth}
    \begin{center}  
      \epsfig{file=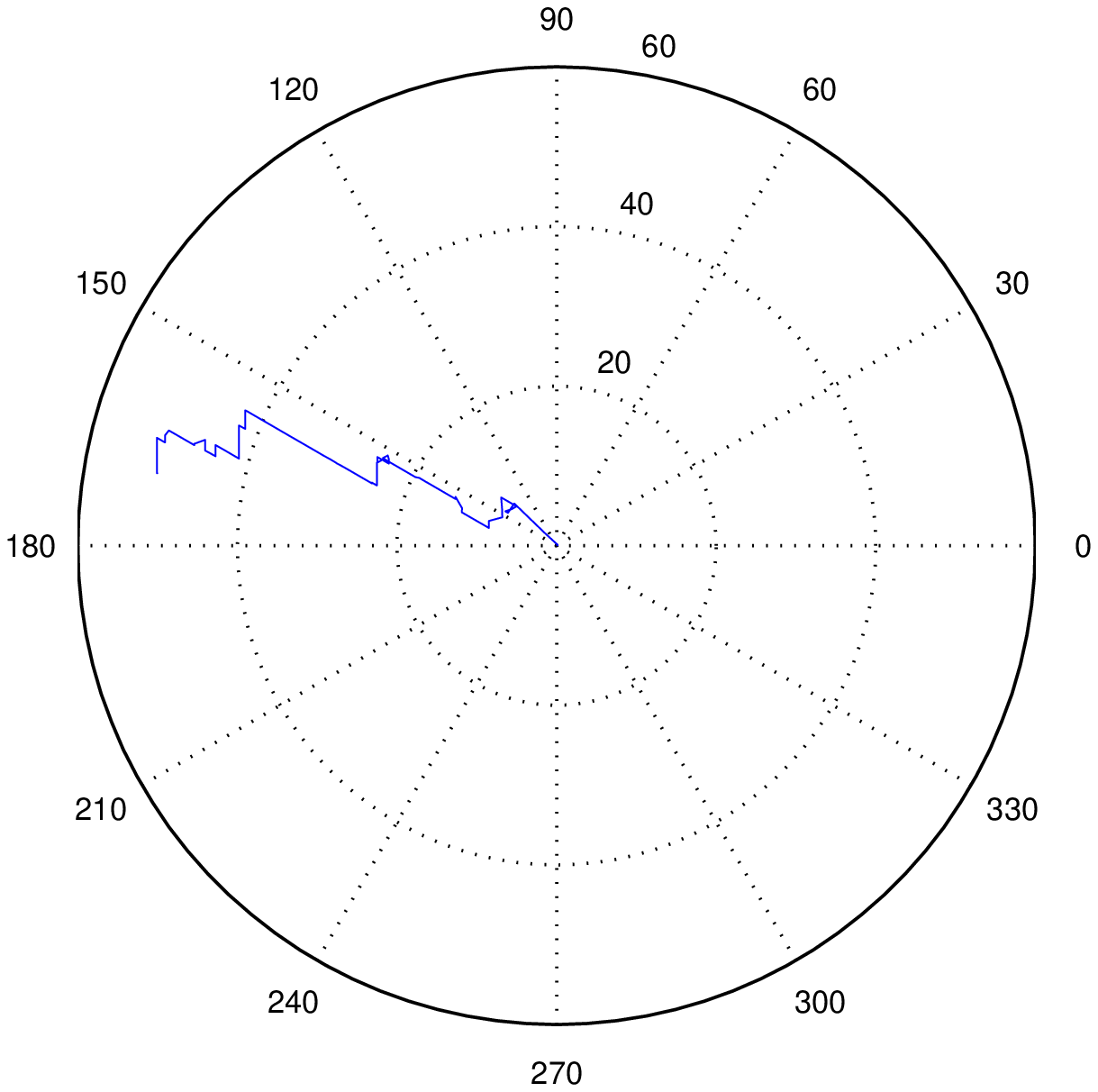, scale=0.5}
      \caption{Polar plot}
    \end{center}
  \end{minipage}
  \hfill
  \begin{minipage}[t]{.45\textwidth}
    \begin{center}  
      \epsfig{file=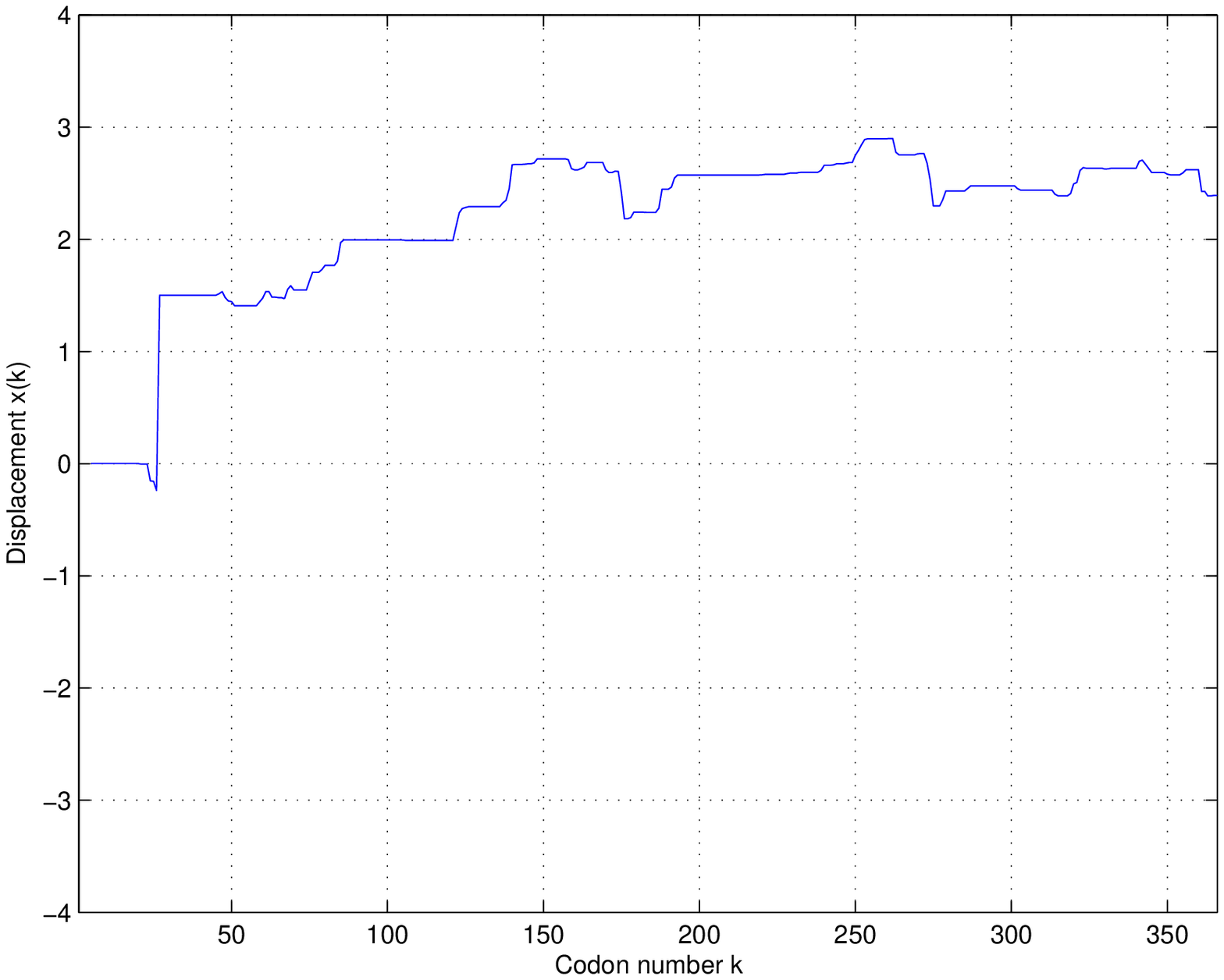, scale=0.5}
      \caption{Displacement plot}
    \end{center}
  \end{minipage}
  \hfill
\end{figure}

\clearpage
\textbf{Streptococcus mutans}
\begin{figure}[h]
  \hfill
  \begin{minipage}[t]{.45\textwidth}
    \begin{center}  
      \epsfig{file=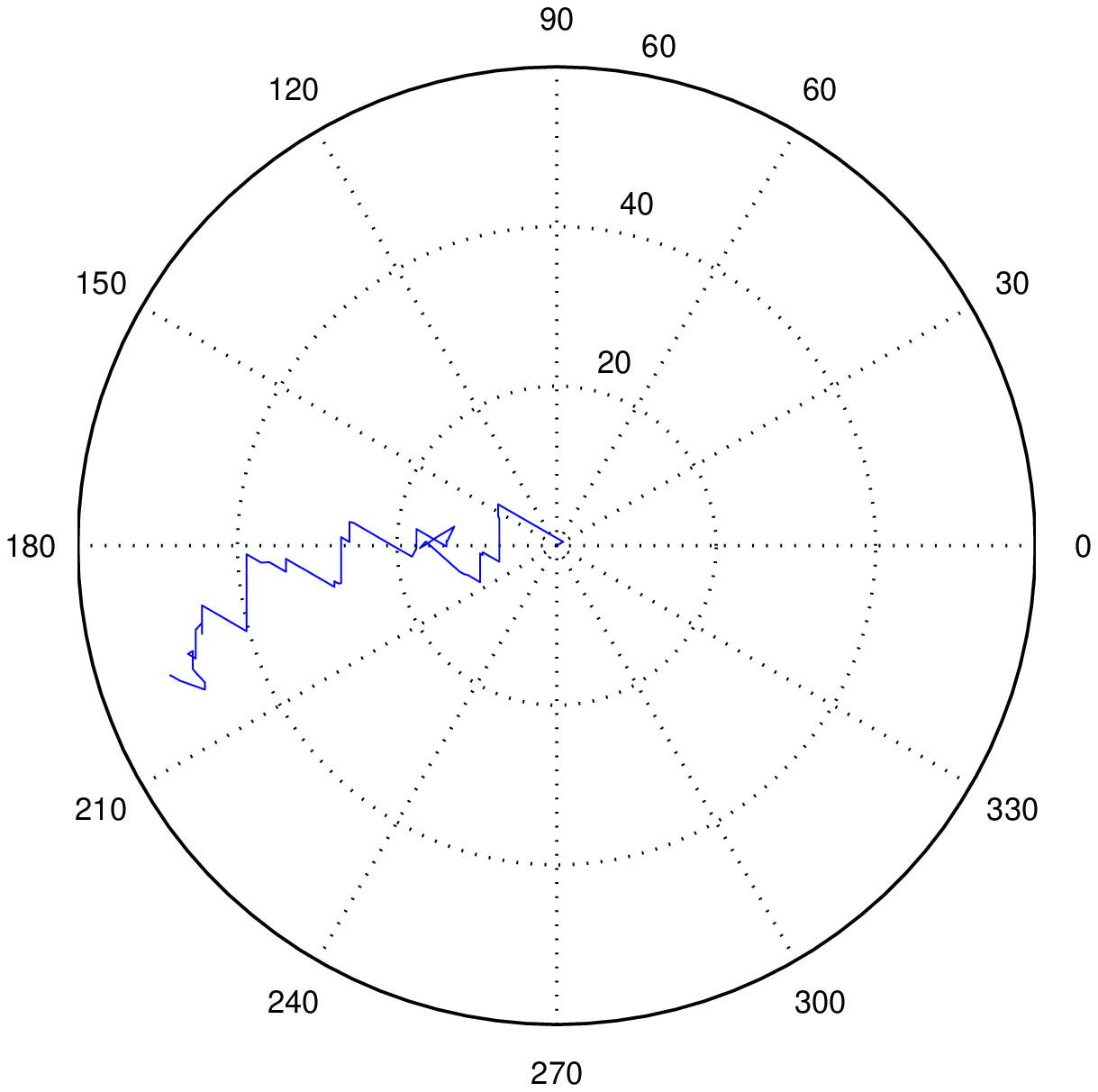, scale=0.5}
      \caption{Polar plot}
    \end{center}
  \end{minipage}
  \hfill
  \begin{minipage}[t]{.45\textwidth}
    \begin{center}  
      \epsfig{file=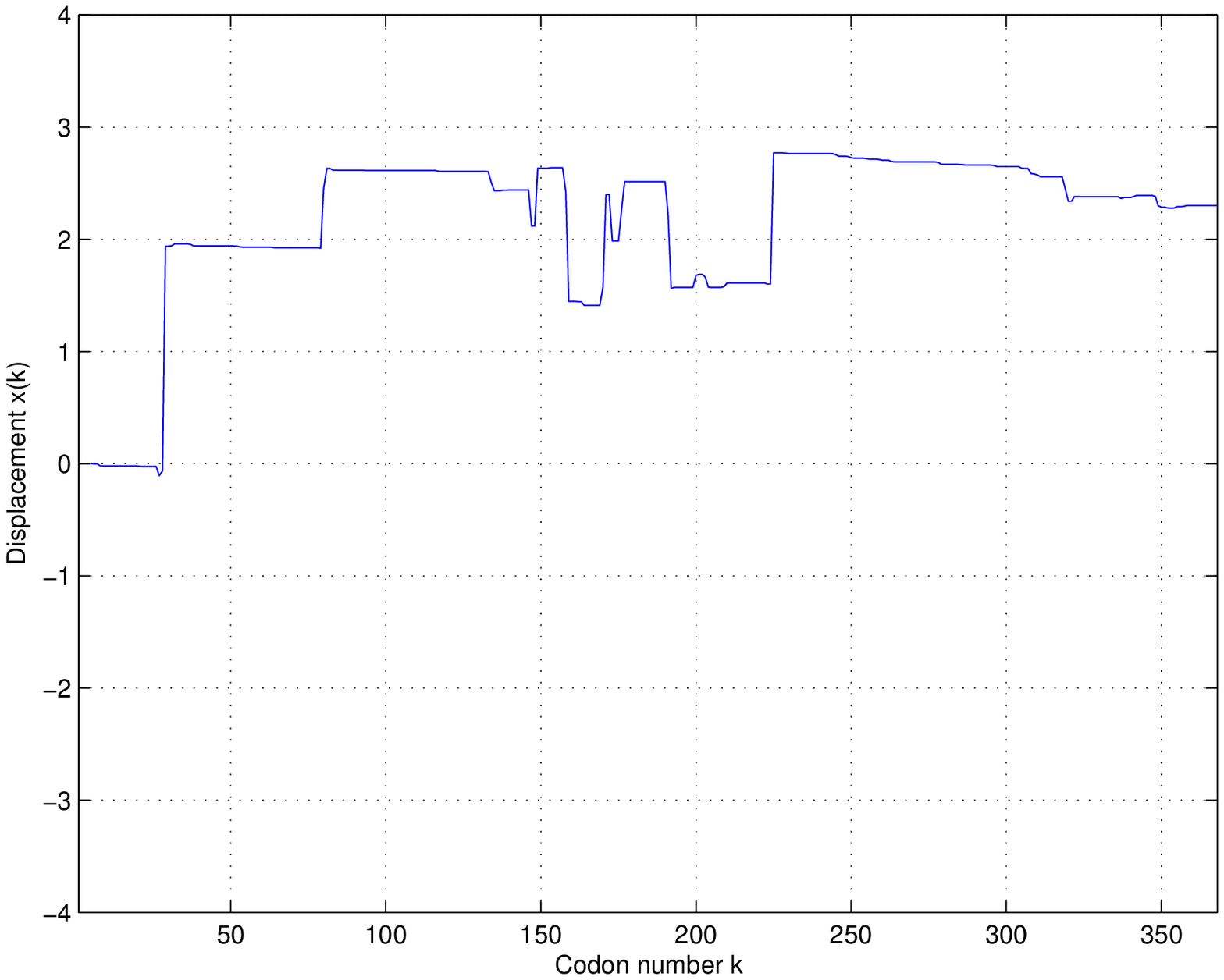, scale=0.5}
      \caption{Displacement plot}
    \end{center}
  \end{minipage}
  \hfill
\end{figure}

\clearpage
\textbf{Salmonella typhimurium}
\begin{figure}[h]
  \hfill
  \begin{minipage}[t]{.45\textwidth}
    \begin{center}  
      \epsfig{file=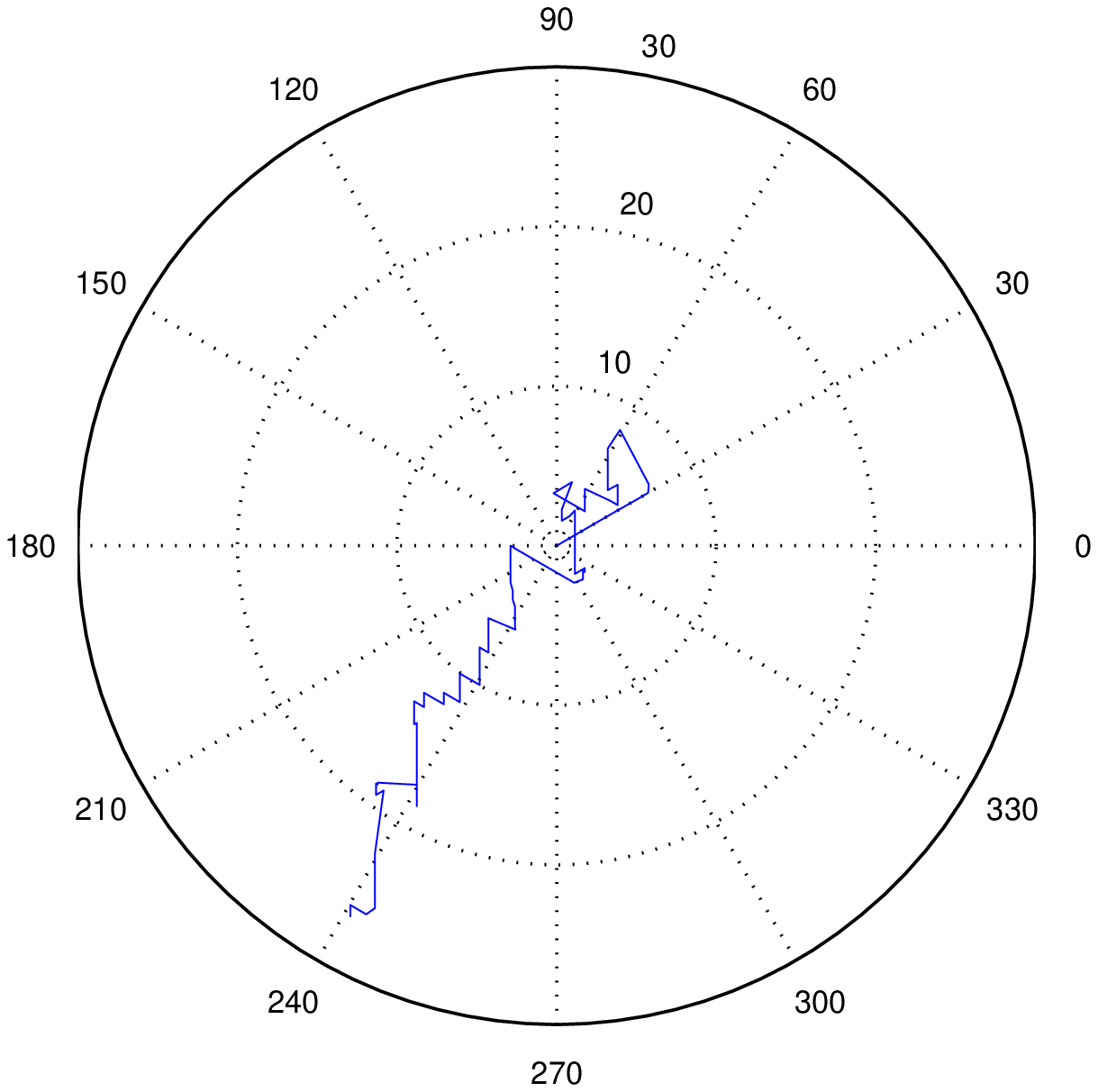, scale=0.5}
      \caption{Polar plot}
    \end{center}
  \end{minipage}
  \hfill
  \begin{minipage}[t]{.45\textwidth}
    \begin{center}  
      \epsfig{file=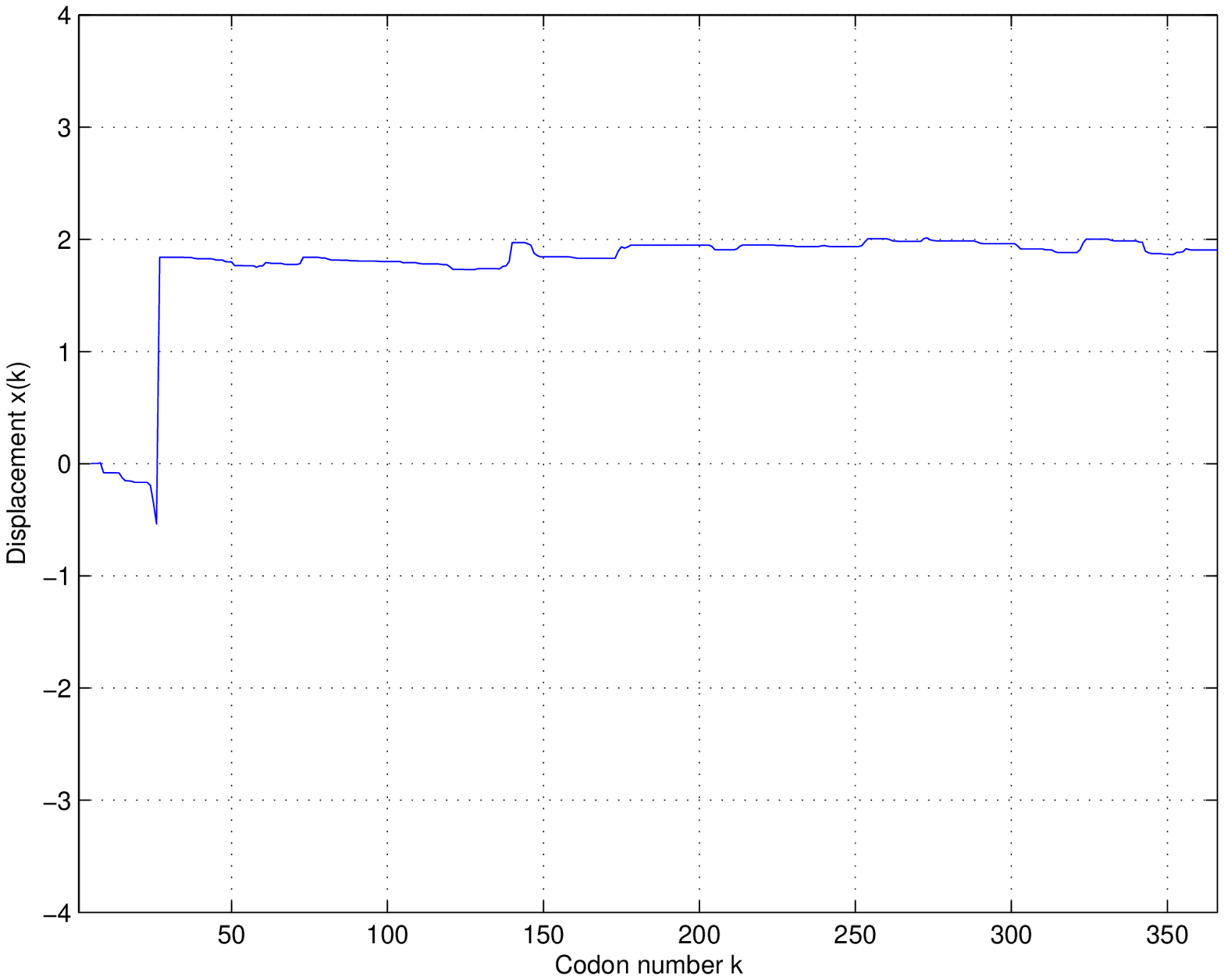, scale=0.5}
      \caption{Displacement plot}
    \end{center}
  \end{minipage}
  \hfill
\end{figure}

\clearpage
\textbf{Treponema pallidum}
\begin{figure}[h]
  \hfill
  \begin{minipage}[t]{.45\textwidth}
    \begin{center}  
      \epsfig{file=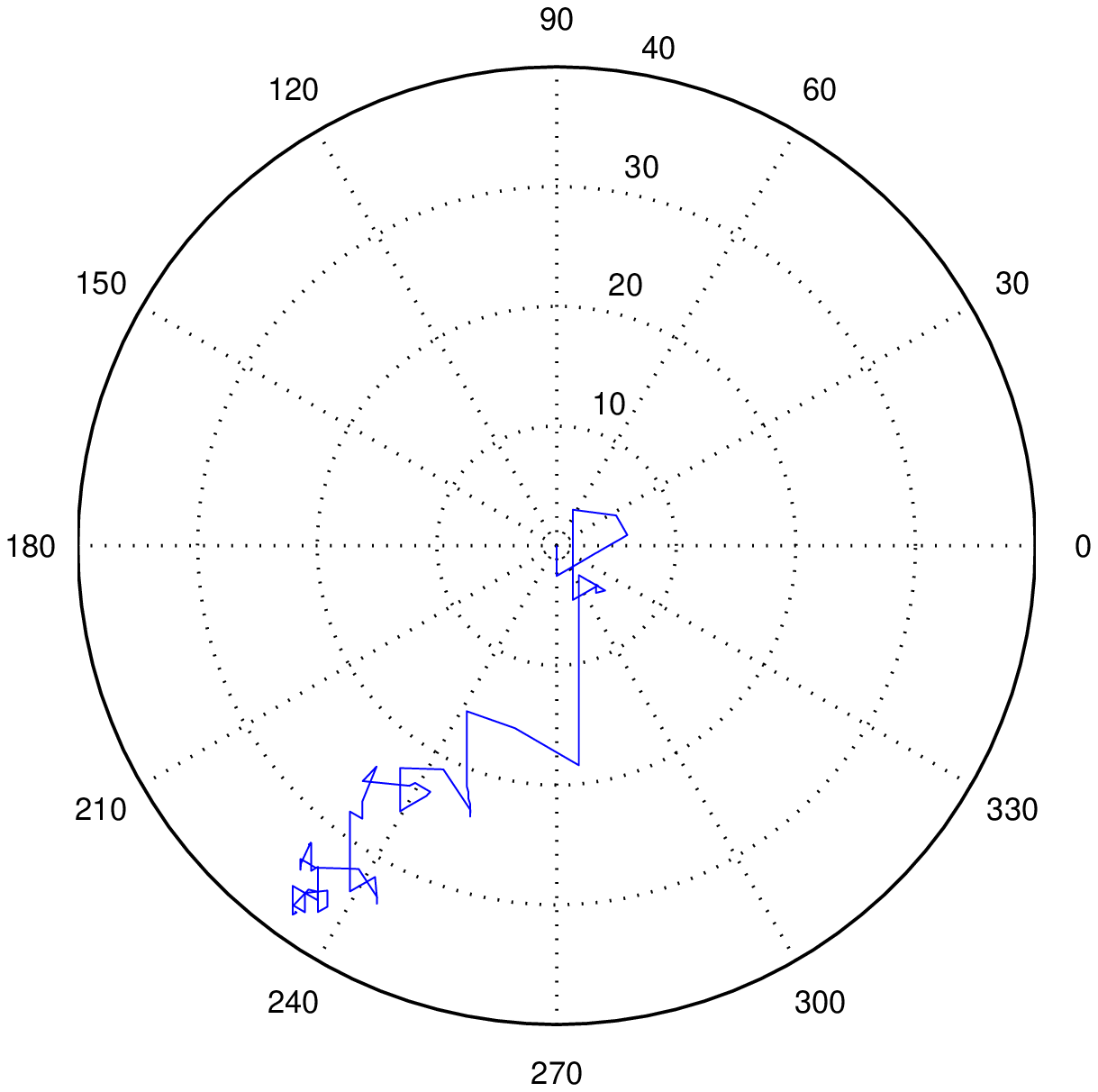, scale=0.5}
      \caption{Polar plot}
    \end{center}
  \end{minipage}
  \hfill
  \begin{minipage}[t]{.45\textwidth}
    \begin{center}  
      \epsfig{file=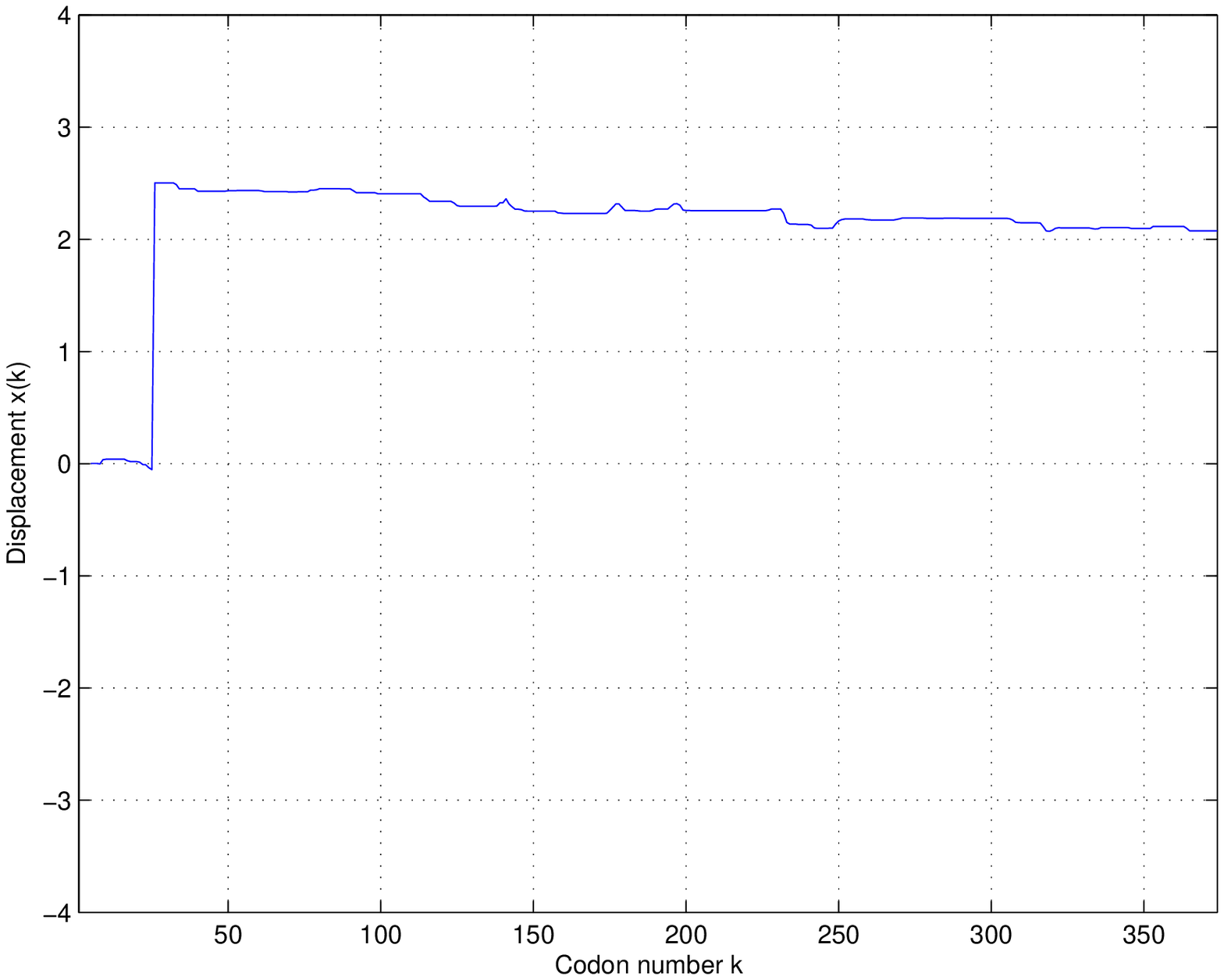, scale=0.5}
      \caption{Displacement plot}
    \end{center}
  \end{minipage}
  \hfill
\end{figure}

\clearpage
\textbf{Xylella fastidiosa}
\begin{figure}[h]
  \hfill
  \begin{minipage}[t]{.45\textwidth}
    \begin{center}  
      \epsfig{file=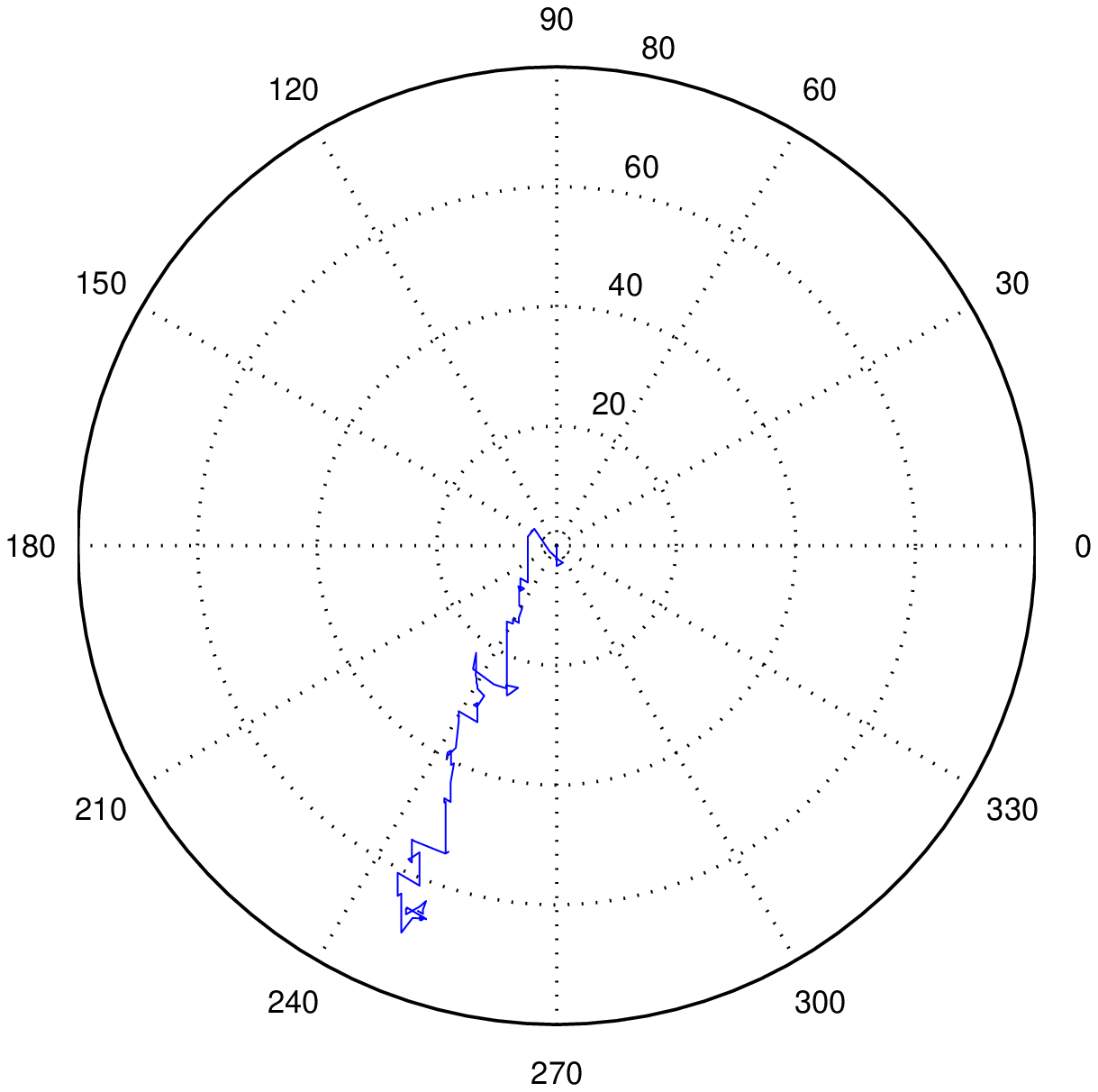, scale=0.5}
      \caption{Polar plot}
    \end{center}
  \end{minipage}
  \hfill
  \begin{minipage}[t]{.45\textwidth}
    \begin{center}  
      \epsfig{file=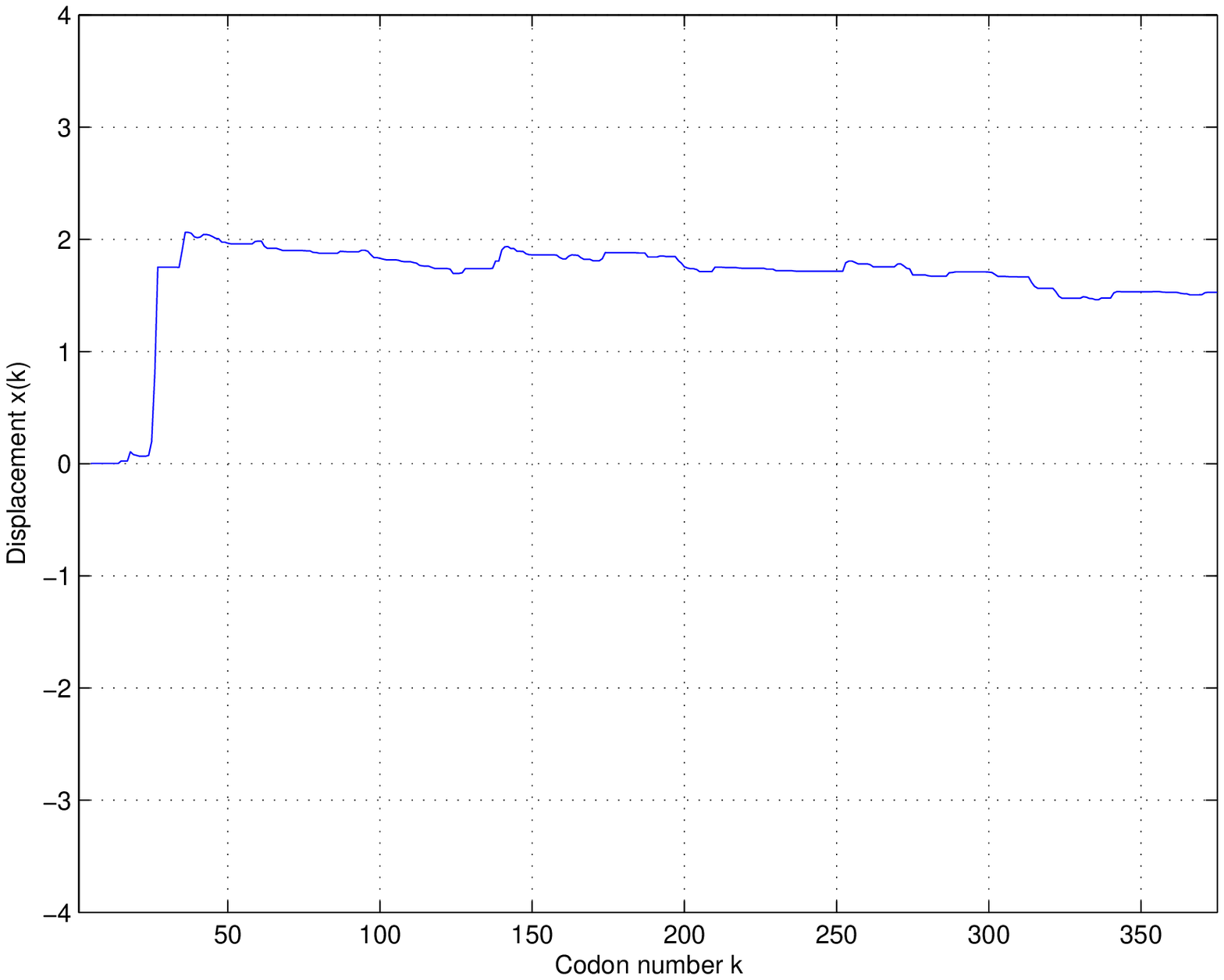, scale=0.5}
      \caption{Displacement plot}
    \end{center}
  \end{minipage}
  \hfill
\end{figure}

\clearpage
\subsection{Link Genes}
High-yield, non-frameshift genes in \emph{E. coli}. \\

\textbf{atpD}
\begin{figure}[h]
  \hfill
  \begin{minipage}[t]{.45\textwidth}
    \begin{center}  
      \epsfig{file=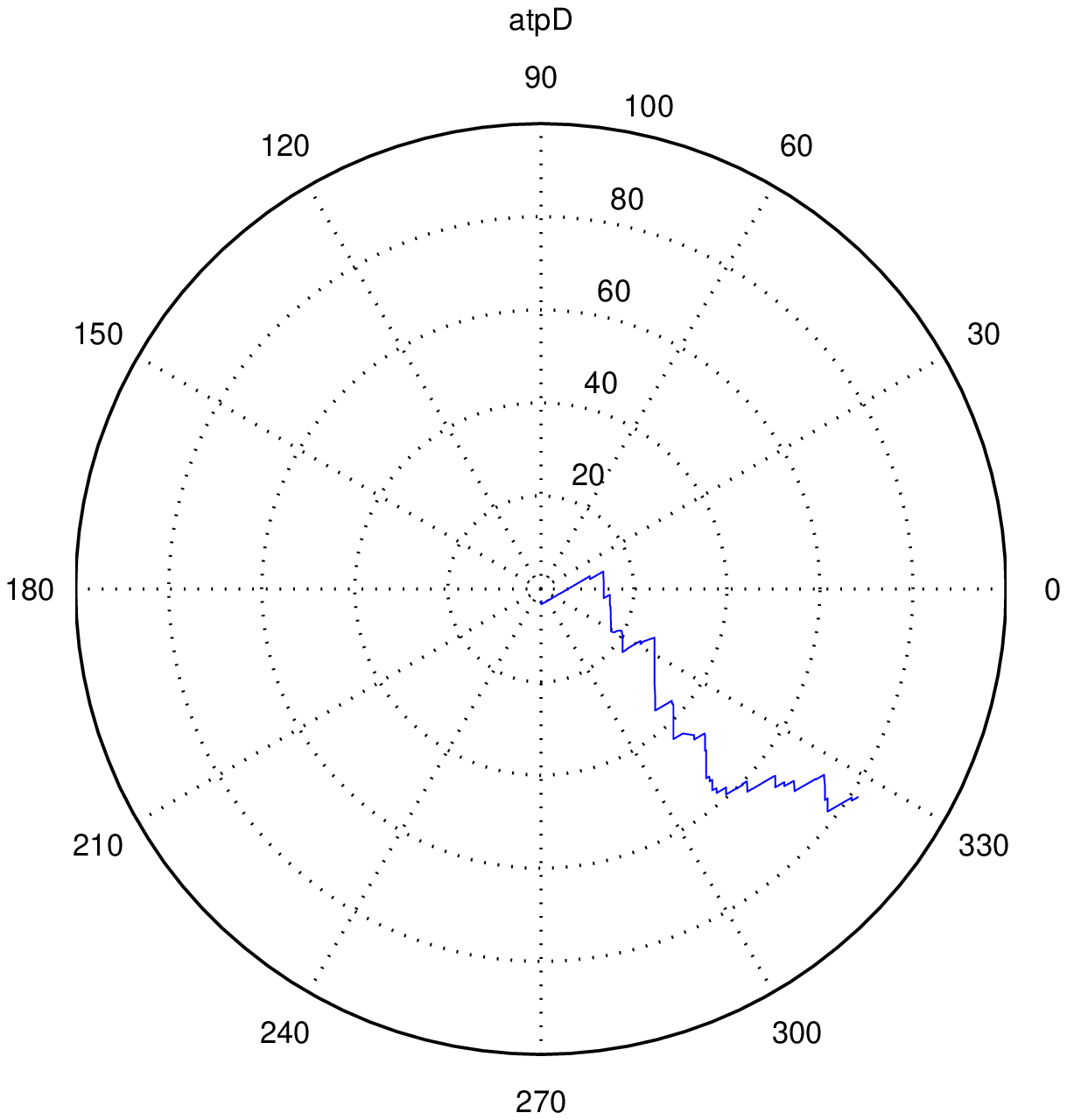, scale=0.5}
      \caption{Polar plot}
    \end{center}
  \end{minipage}
  \hfill
  \begin{minipage}[t]{.45\textwidth}
    \begin{center}  
      \epsfig{file=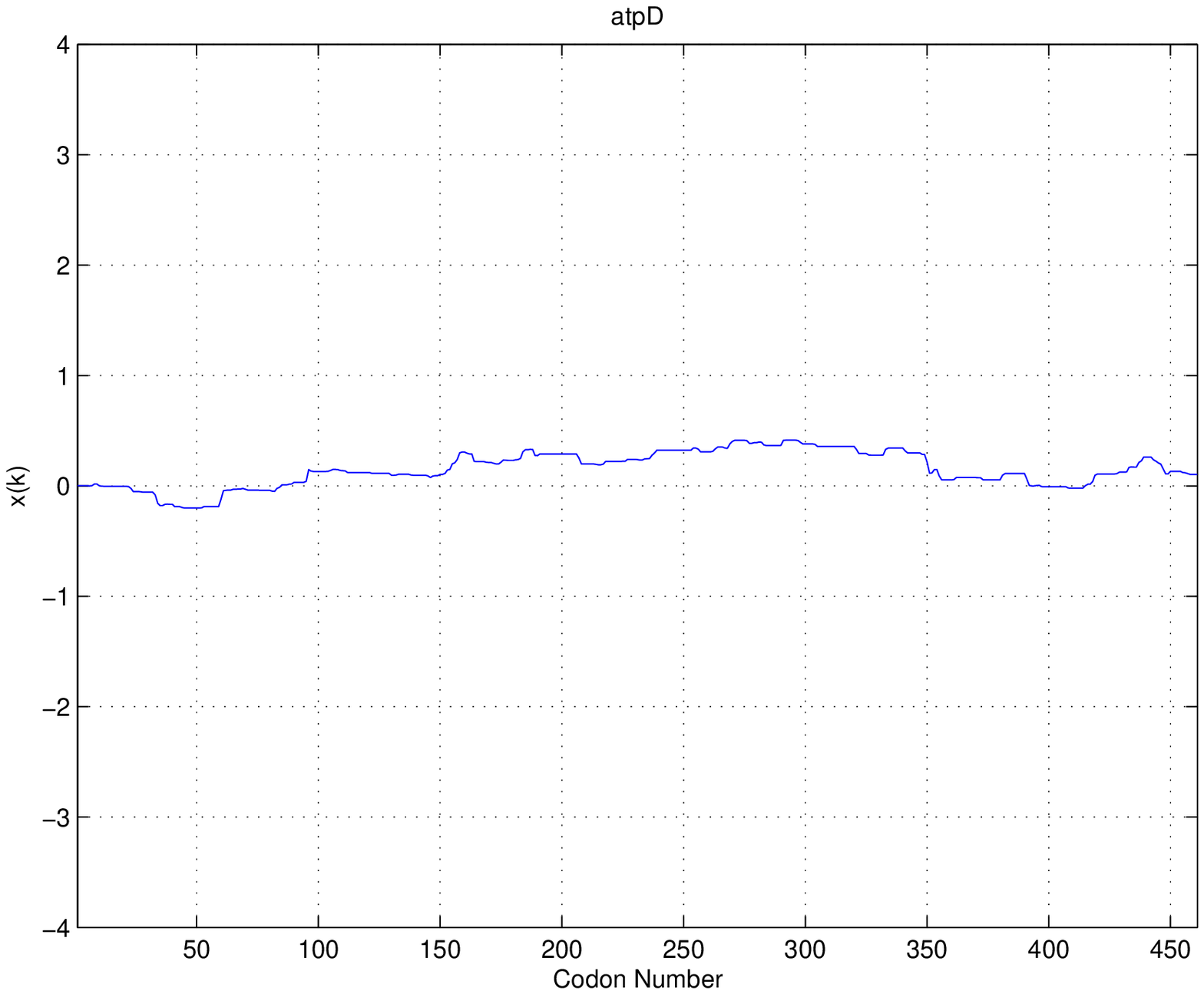, scale=0.5}
      \caption{Displacement plot}
    \end{center}
  \end{minipage}
  \hfill
\end{figure}

\clearpage
\textbf{malE}
\begin{figure}[h]
  \hfill
  \begin{minipage}[t]{.45\textwidth}
    \begin{center}  
      \epsfig{file=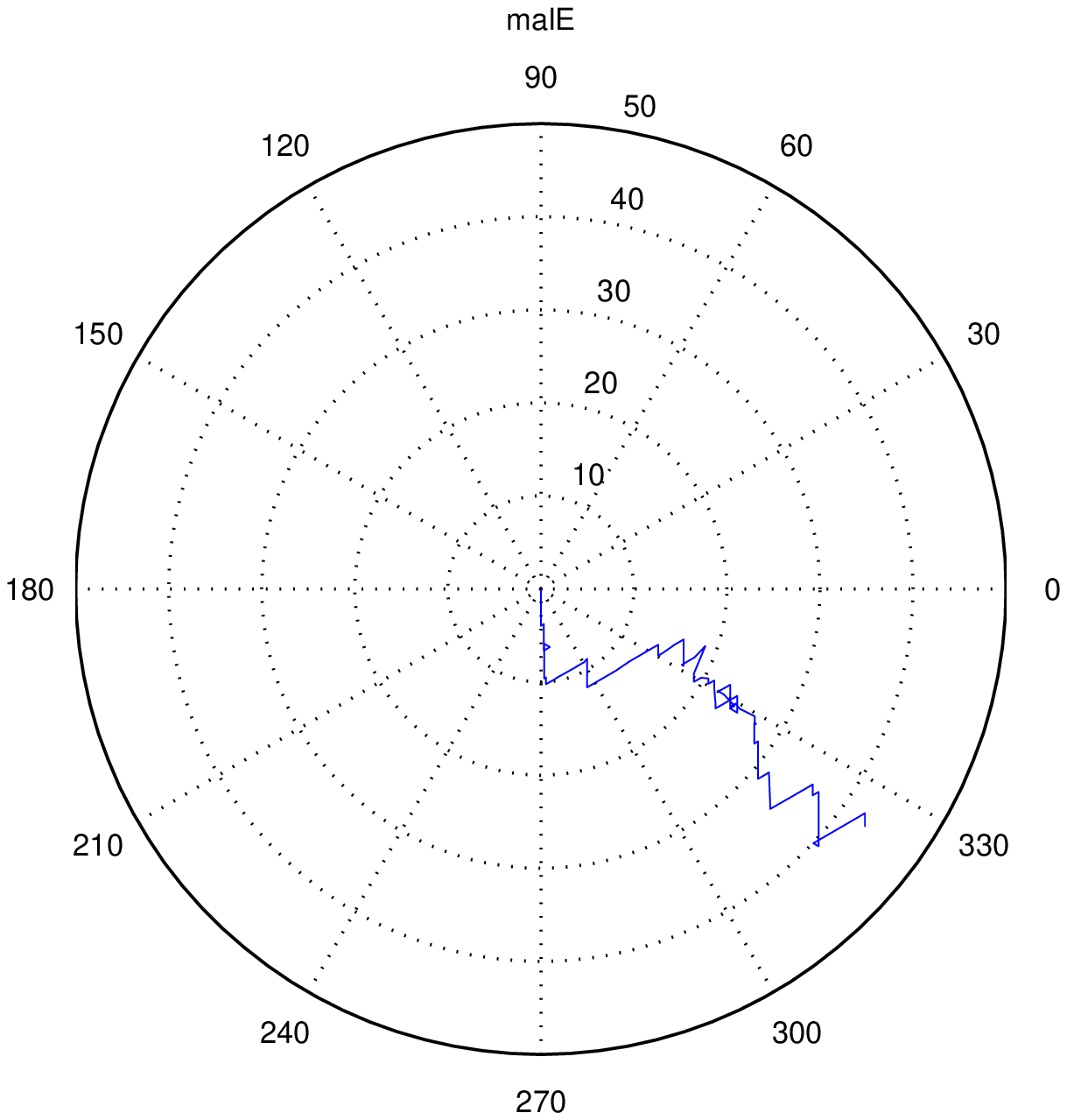, scale=0.5}
      \caption{Polar plot}
    \end{center}
  \end{minipage}
  \hfill
  \begin{minipage}[t]{.45\textwidth}
    \begin{center}  
      \epsfig{file=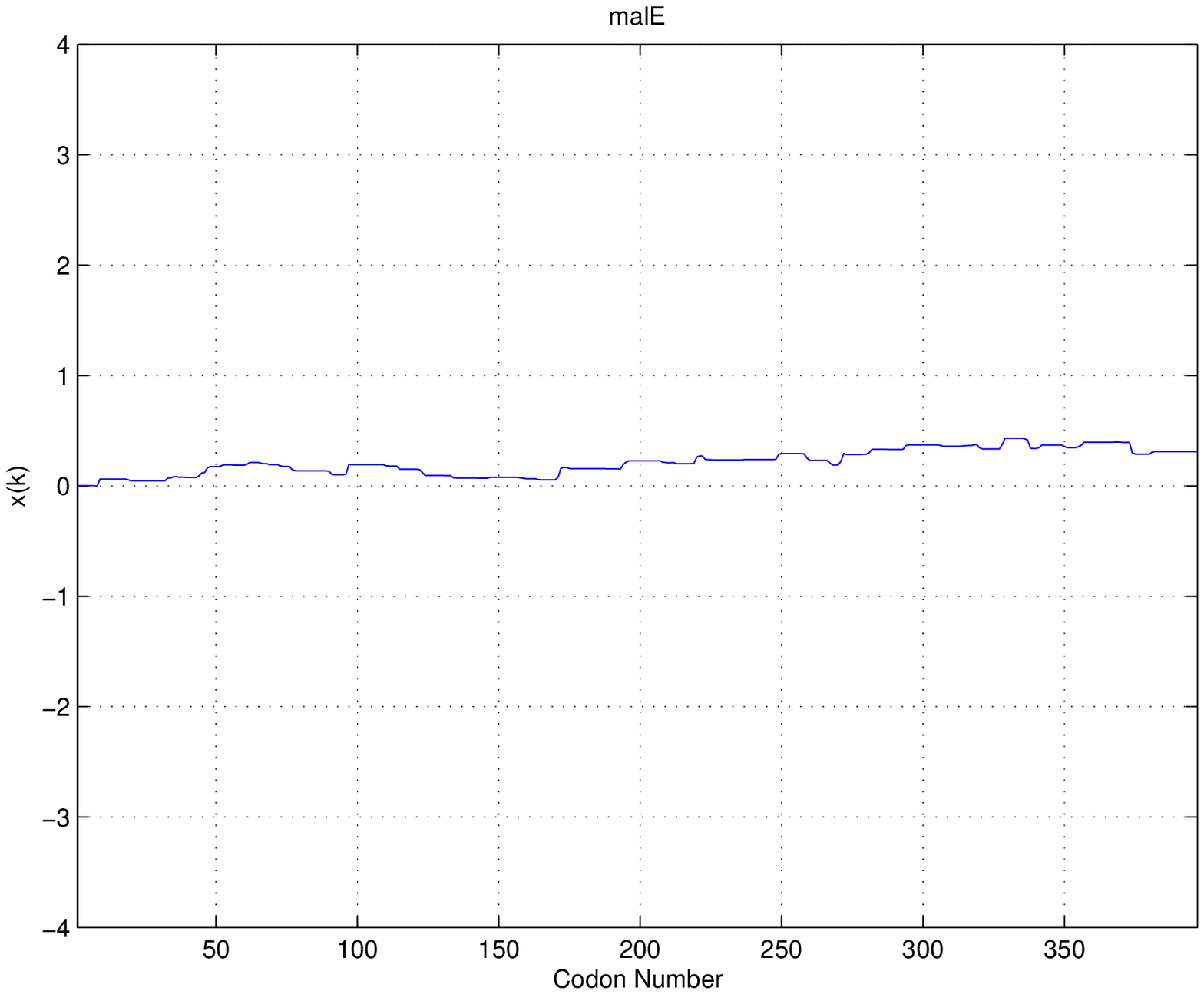, scale=0.5}
      \caption{Displacement plot}
    \end{center}
  \end{minipage}
  \hfill
\end{figure}

\clearpage
\textbf{manX}
\begin{figure}[h]
  \hfill
  \begin{minipage}[t]{.45\textwidth}
    \begin{center}  
      \epsfig{file=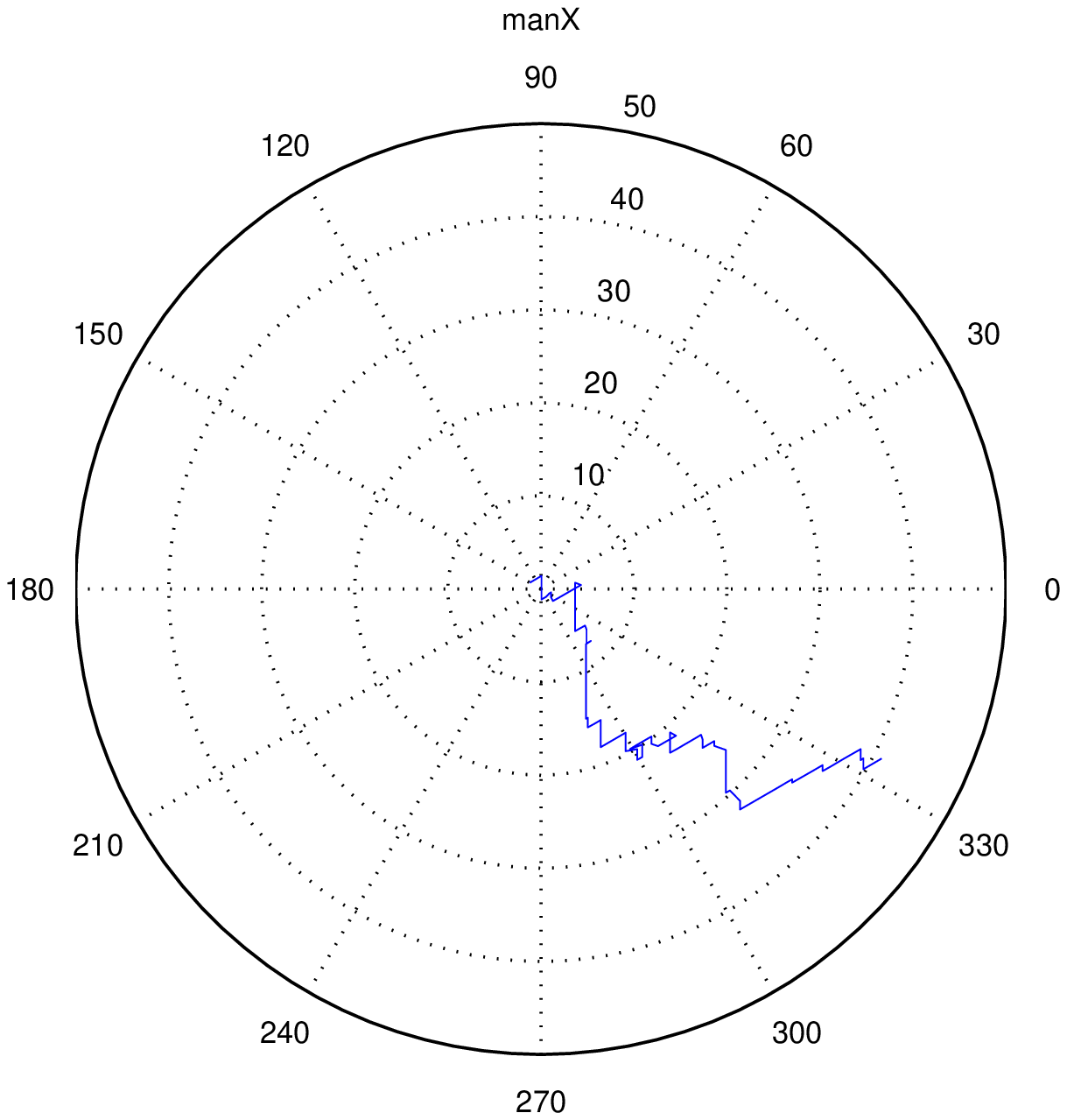, scale=0.5}
      \caption{Polar plot}
    \end{center}
  \end{minipage}
  \hfill
  \begin{minipage}[t]{.45\textwidth}
    \begin{center}  
      \epsfig{file=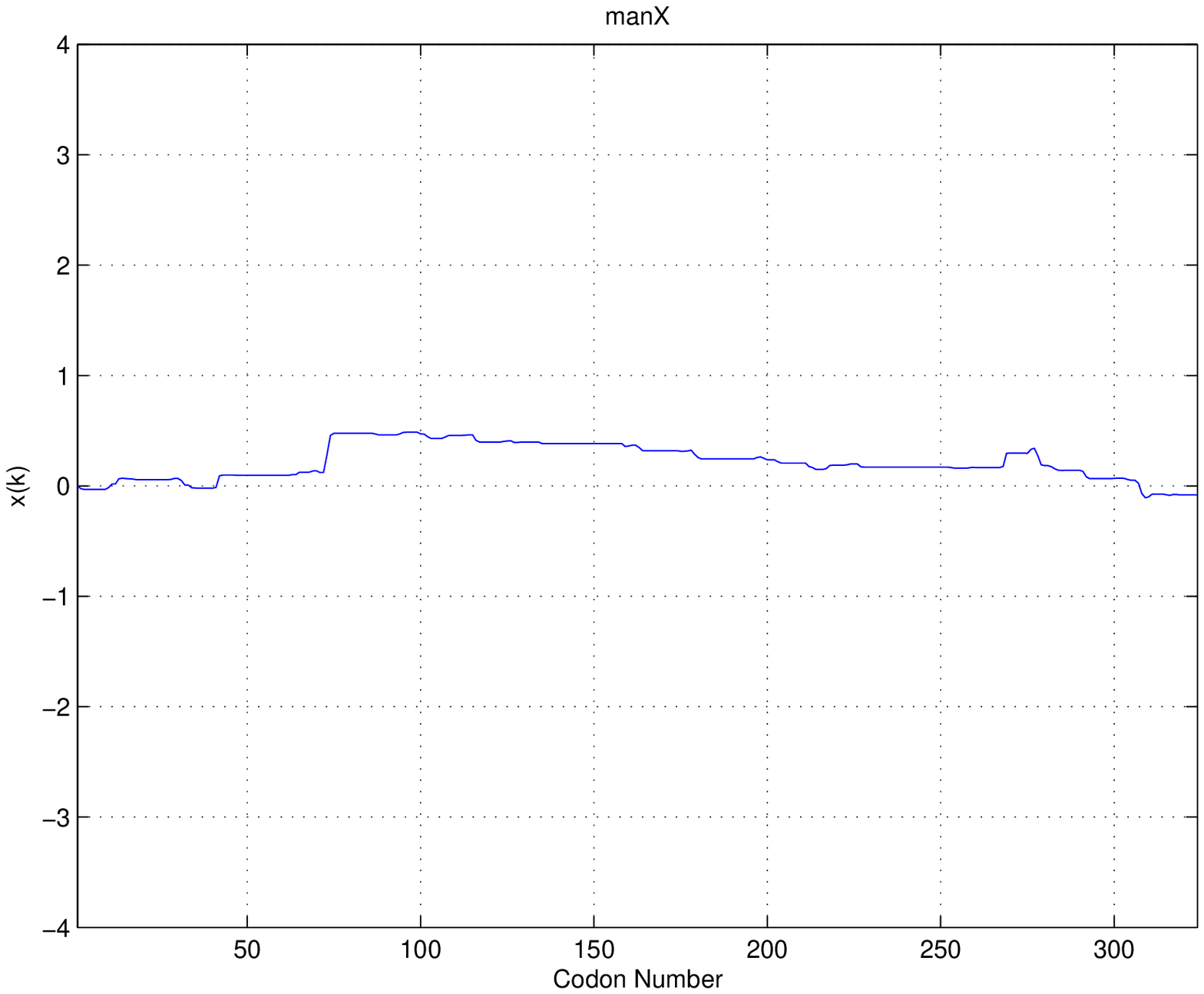, scale=0.5}
      \caption{Displacement plot}
    \end{center}
  \end{minipage}
  \hfill
\end{figure}

\clearpage
\textbf{mglB}
\begin{figure}[h]
  \hfill
  \begin{minipage}[t]{.45\textwidth}
    \begin{center}  
      \epsfig{file=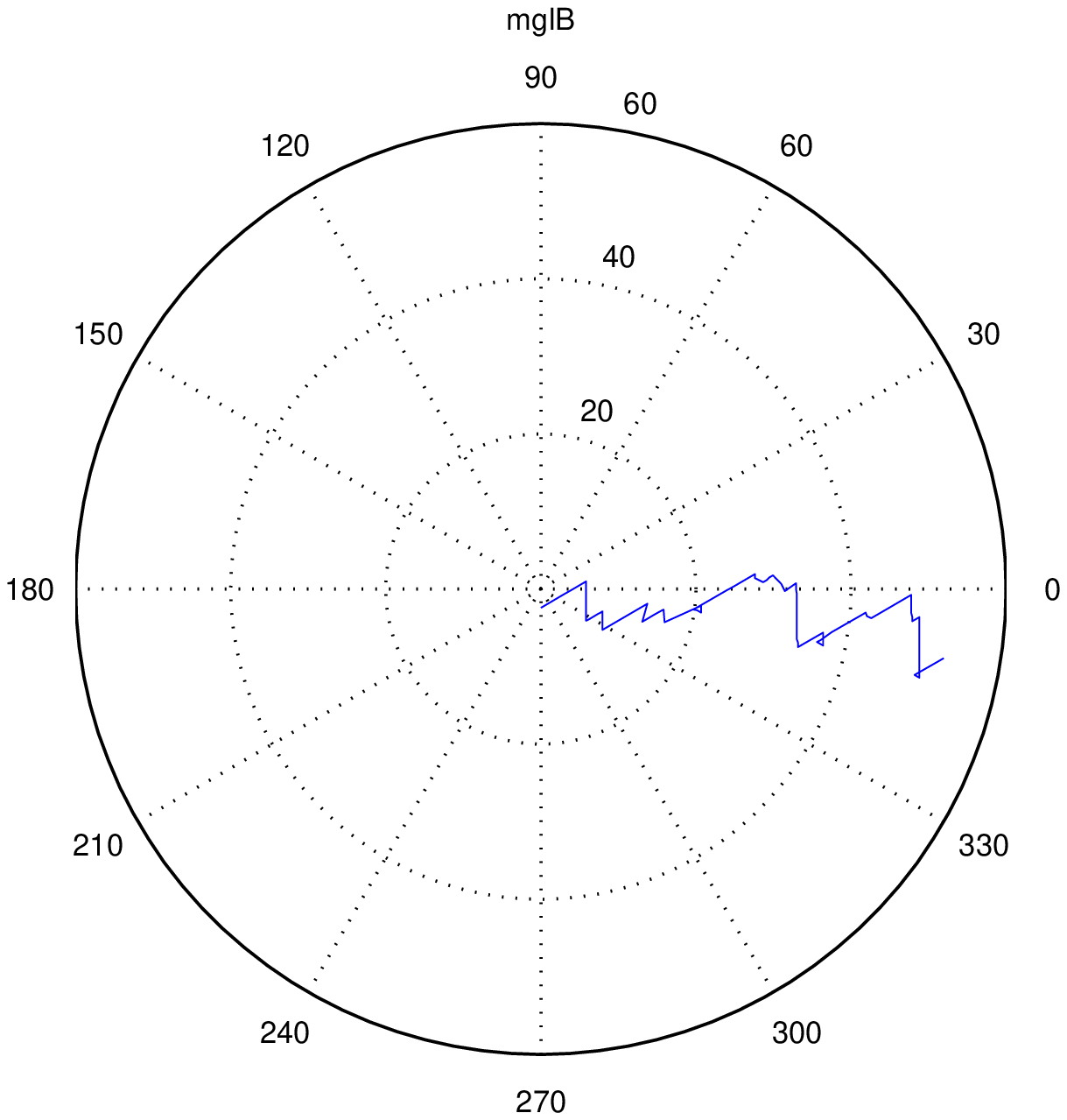, scale=0.5}
      \caption{Polar plot}
    \end{center}
  \end{minipage}
  \hfill
  \begin{minipage}[t]{.45\textwidth}
    \begin{center}  
      \epsfig{file=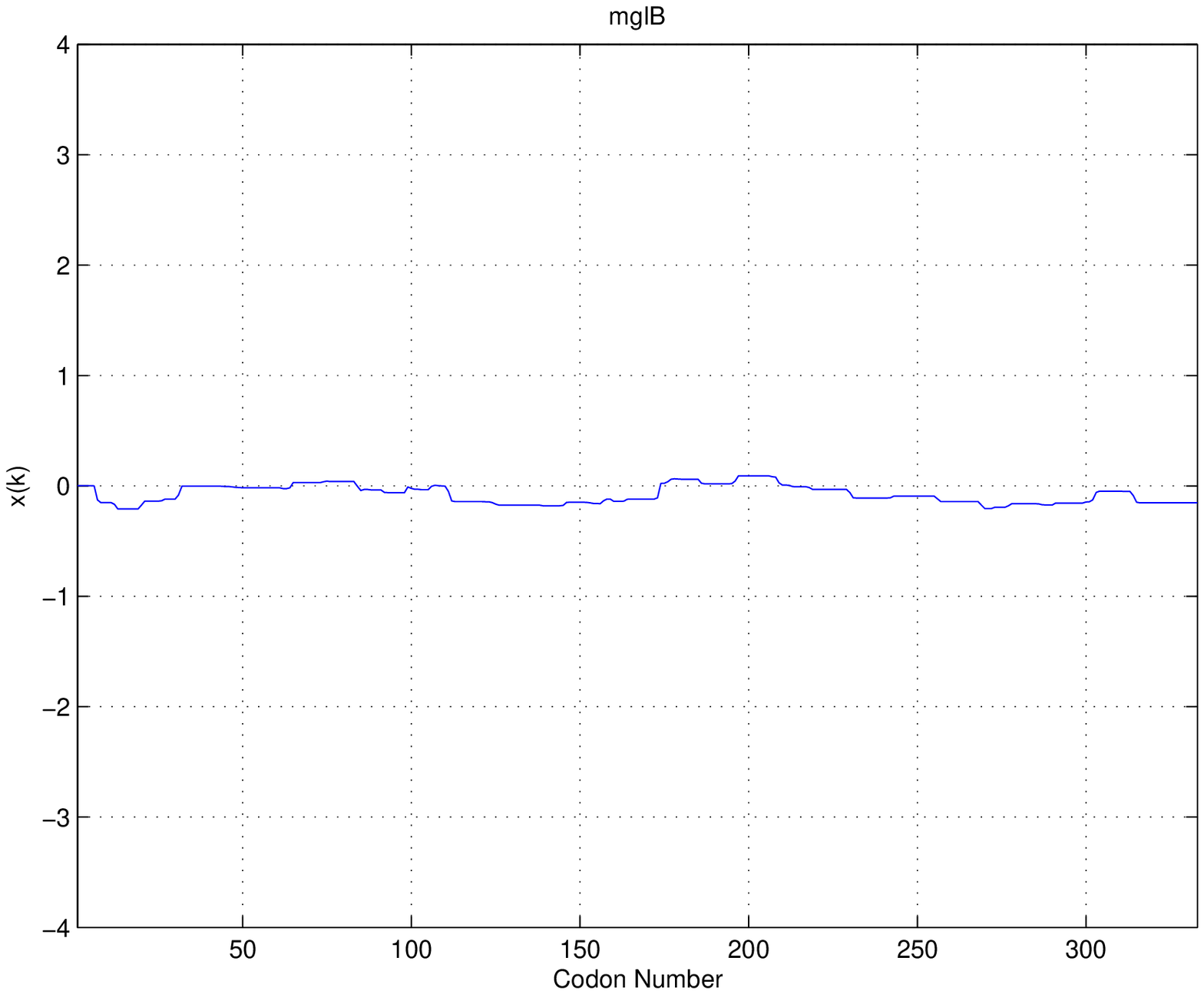, scale=0.5}
      \caption{Displacement plot}
    \end{center}
  \end{minipage}
  \hfill
\end{figure}

\clearpage
\textbf{osmC}
\begin{figure}[h]
  \hfill
  \begin{minipage}[t]{.45\textwidth}
    \begin{center}  
      \epsfig{file=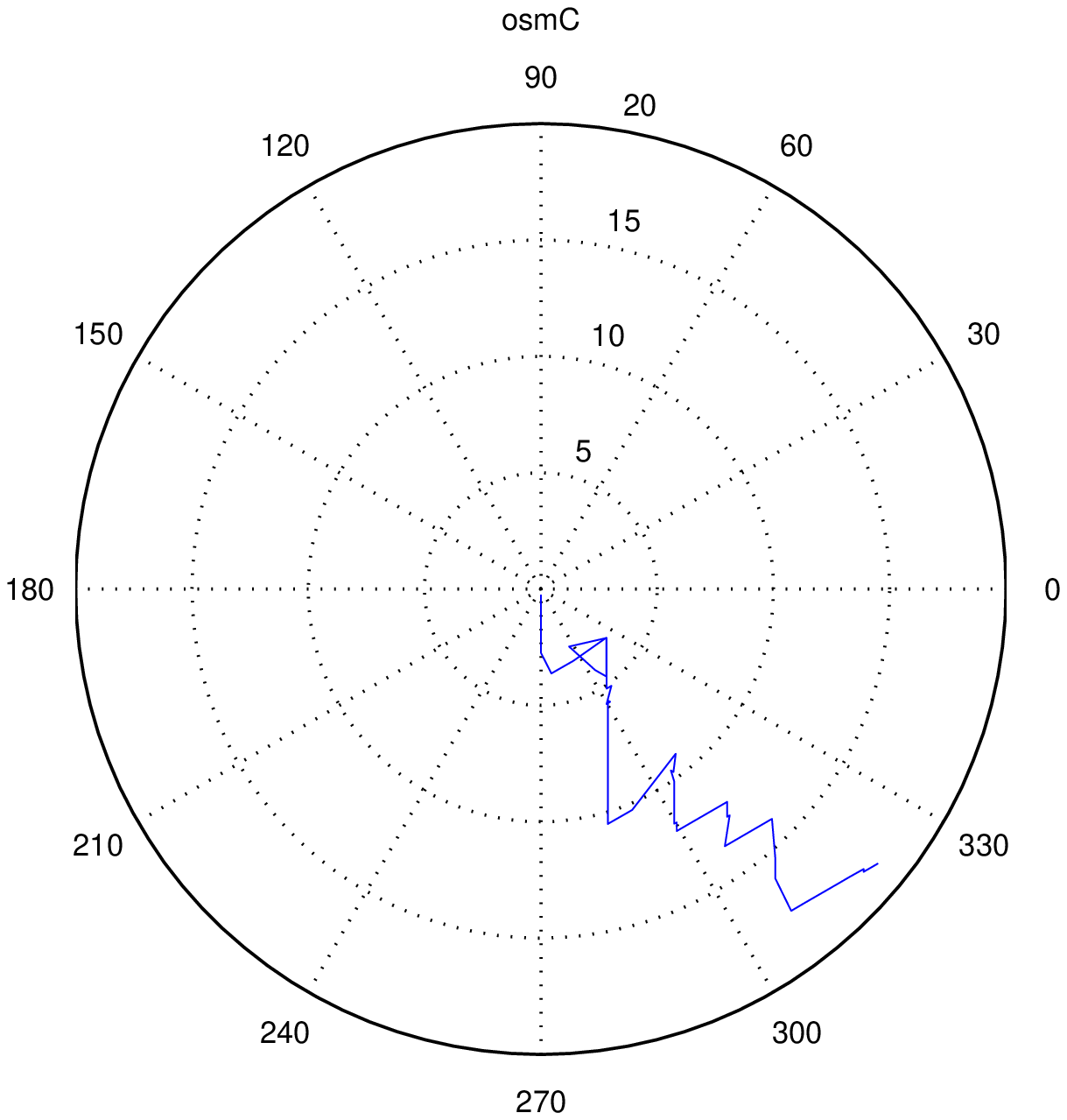, scale=0.5}
      \caption{Polar plot}
    \end{center}
  \end{minipage}
  \hfill
  \begin{minipage}[t]{.45\textwidth}
    \begin{center}  
      \epsfig{file=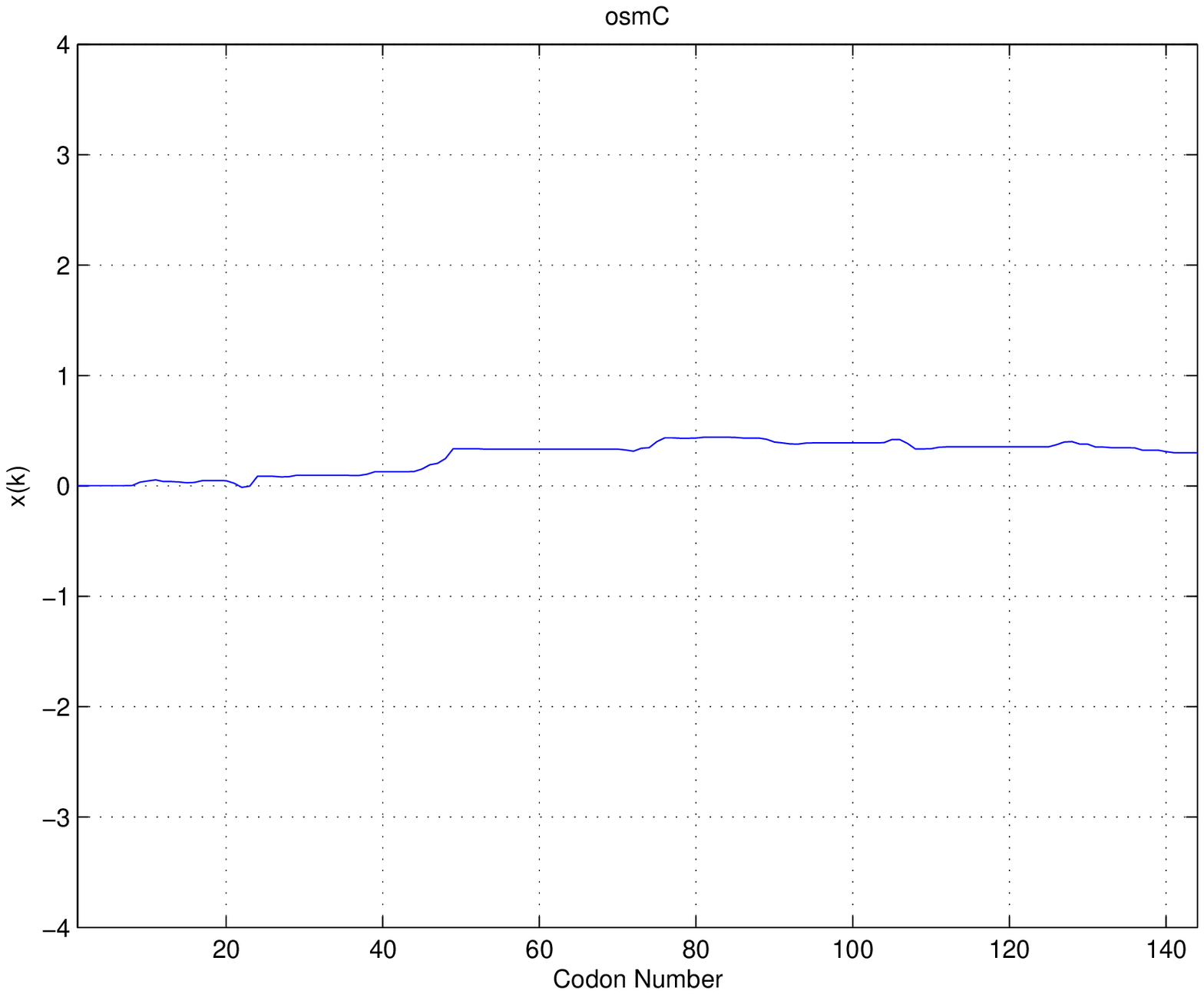, scale=0.5}
      \caption{Displacement plot}
    \end{center}
  \end{minipage}
  \hfill
\end{figure}

\clearpage
\textbf{rbsB}
\begin{figure}[h]
  \hfill
  \begin{minipage}[t]{.45\textwidth}
    \begin{center}  
      \epsfig{file=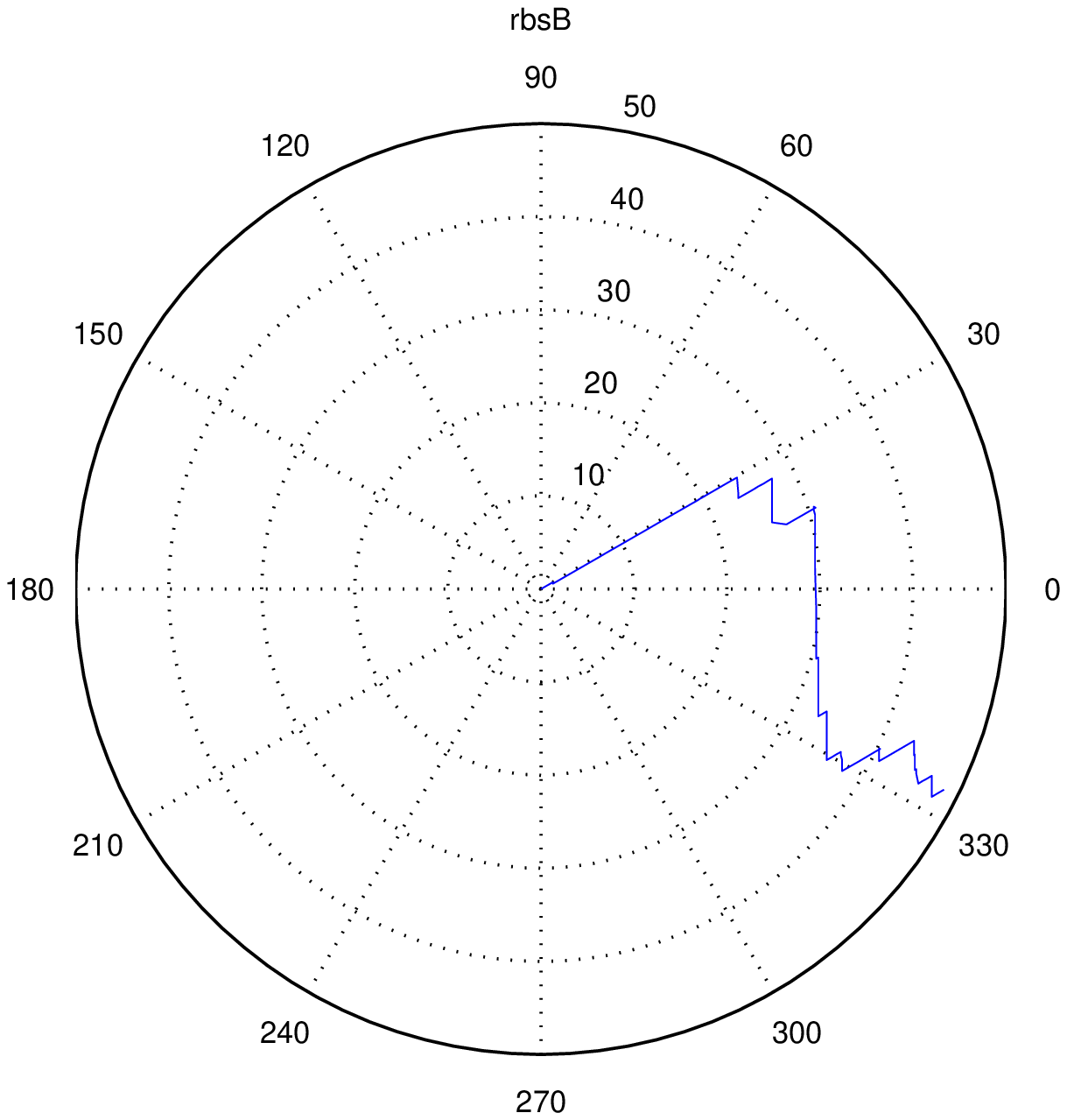, scale=0.5}
      \caption{Polar plot}
    \end{center}
  \end{minipage}
  \hfill
  \begin{minipage}[t]{.45\textwidth}
    \begin{center}  
      \epsfig{file=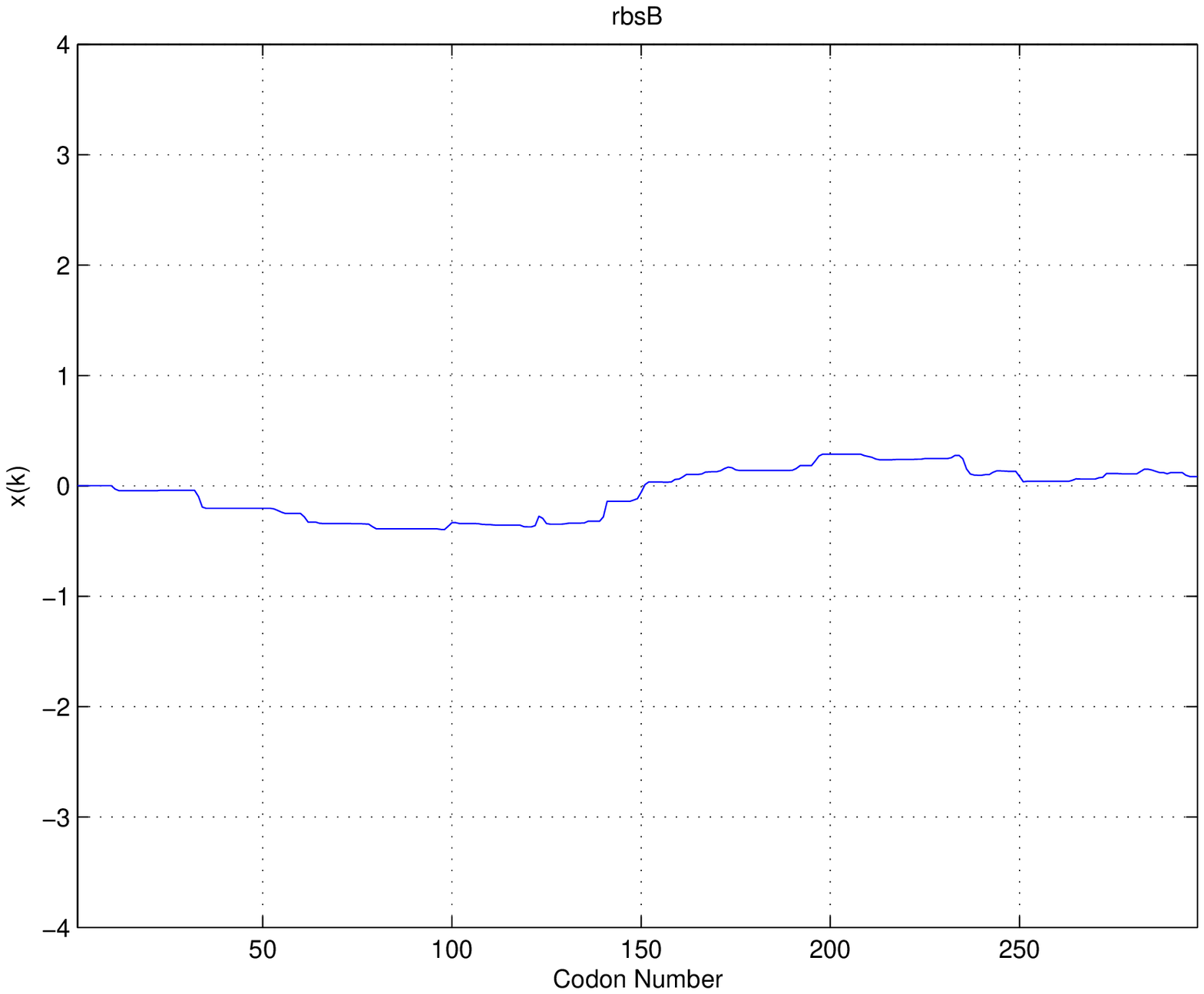, scale=0.5}
      \caption{Displacement plot}
    \end{center}
  \end{minipage}
  \hfill
\end{figure}

\clearpage
\textbf{rplF}
\begin{figure}[h]
  \hfill
  \begin{minipage}[t]{.45\textwidth}
    \begin{center}  
      \epsfig{file=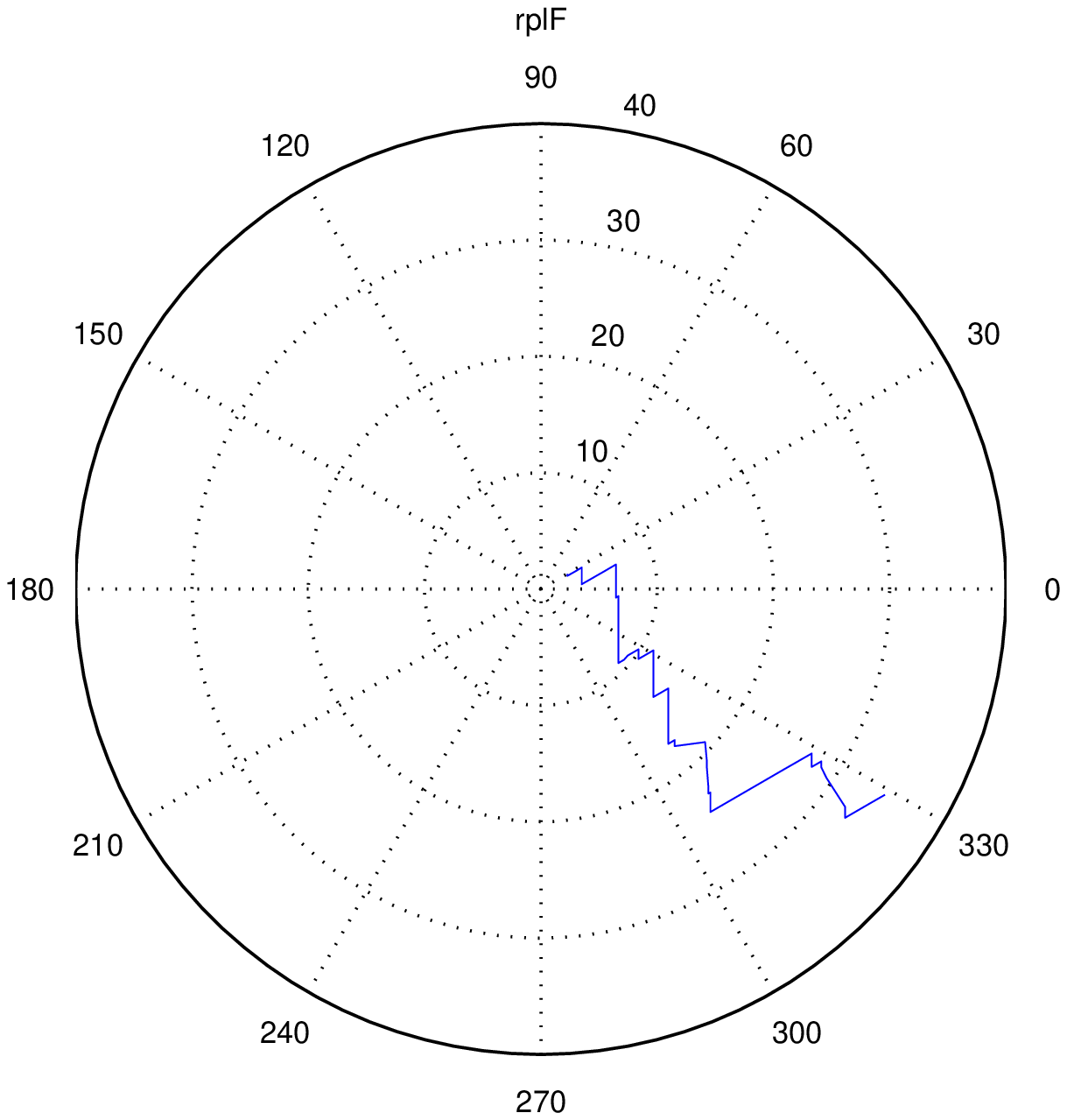, scale=0.5}
      \caption{Polar plot}
    \end{center}
  \end{minipage}
  \hfill
  \begin{minipage}[t]{.45\textwidth}
    \begin{center}  
      \epsfig{file=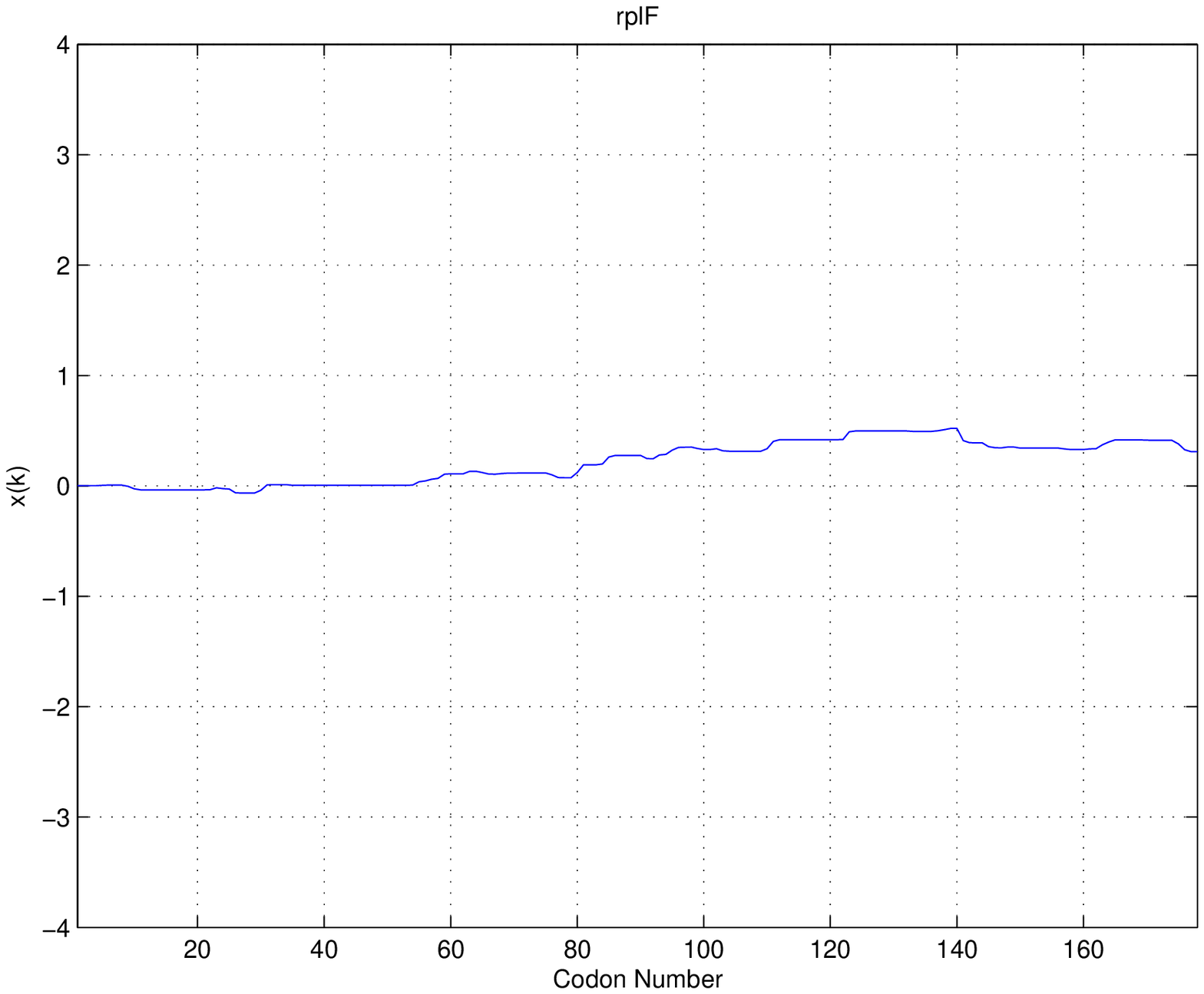, scale=0.5}
      \caption{Displacement plot}
    \end{center}
  \end{minipage}
  \hfill
\end{figure}

\clearpage
\textbf{rpoA}
\begin{figure}[h]
  \hfill
  \begin{minipage}[t]{.45\textwidth}
    \begin{center}  
      \epsfig{file=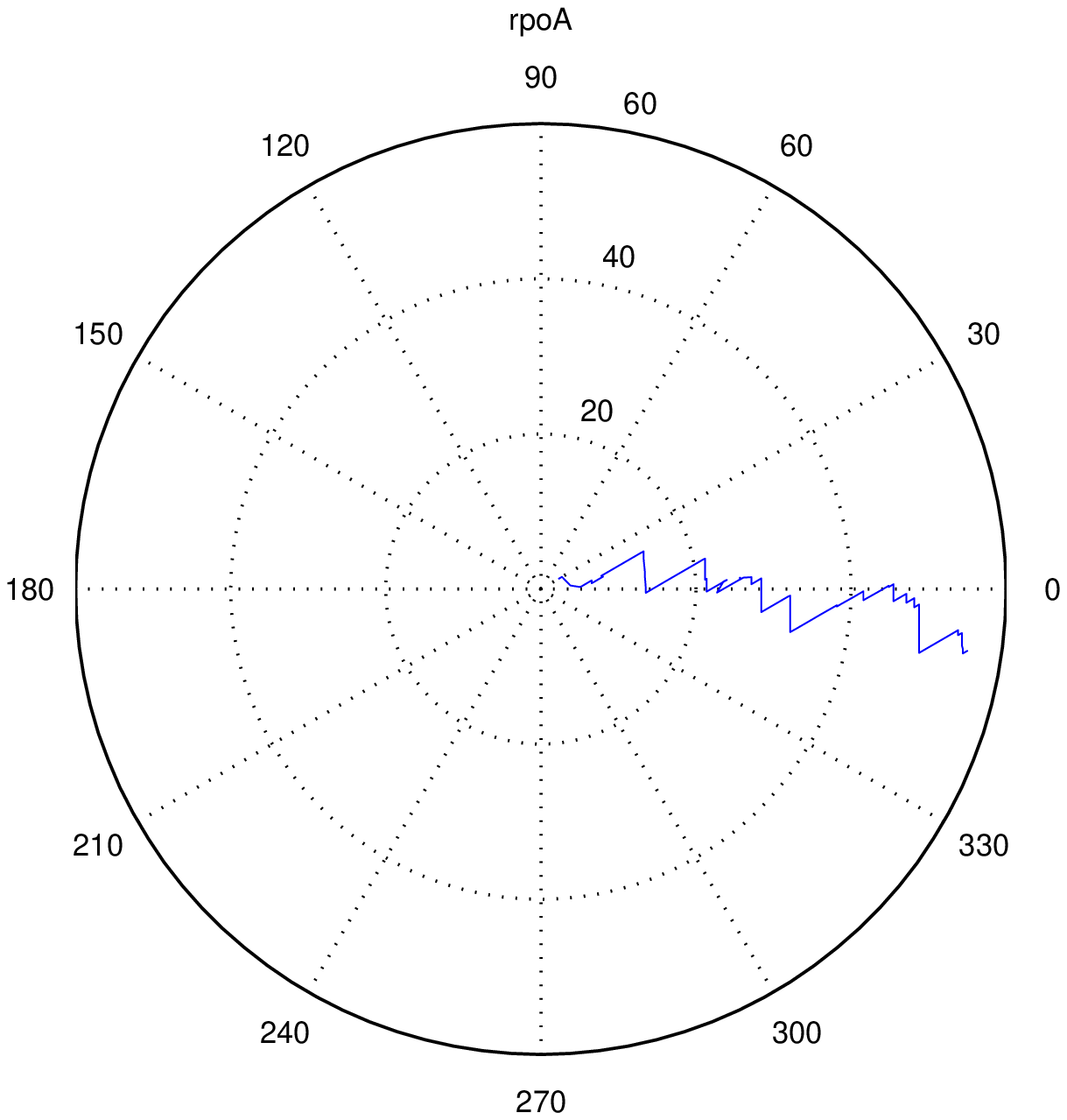, scale=0.5}
      \caption{Polar plot}
    \end{center}
  \end{minipage}
  \hfill
  \begin{minipage}[t]{.45\textwidth}
    \begin{center}  
      \epsfig{file=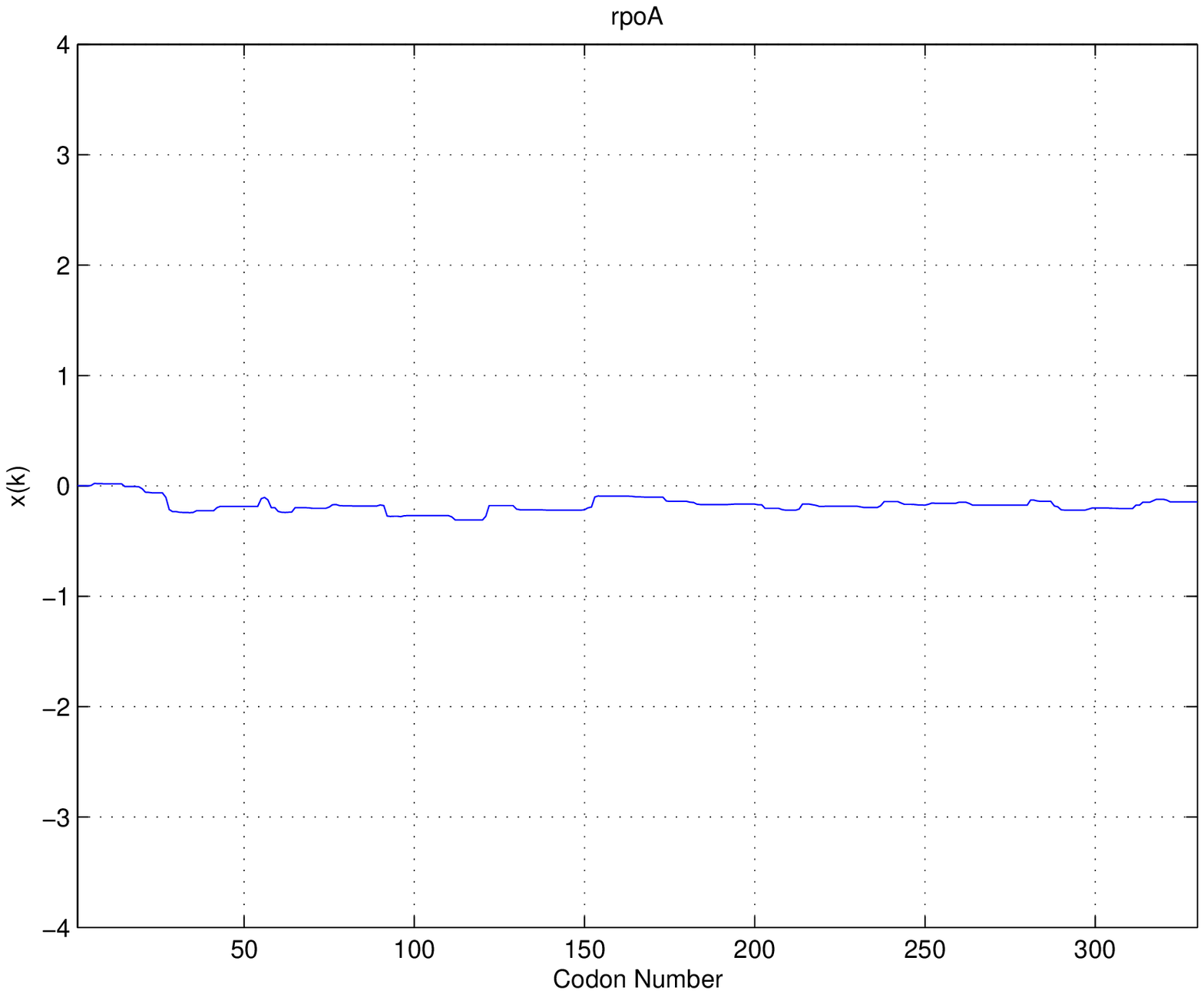, scale=0.5}
      \caption{Displacement plot}
    \end{center}
  \end{minipage}
  \hfill
\end{figure}

\clearpage
\textbf{rpsJ}
\begin{figure}[h]
  \hfill
  \begin{minipage}[t]{.45\textwidth}
    \begin{center}  
      \epsfig{file=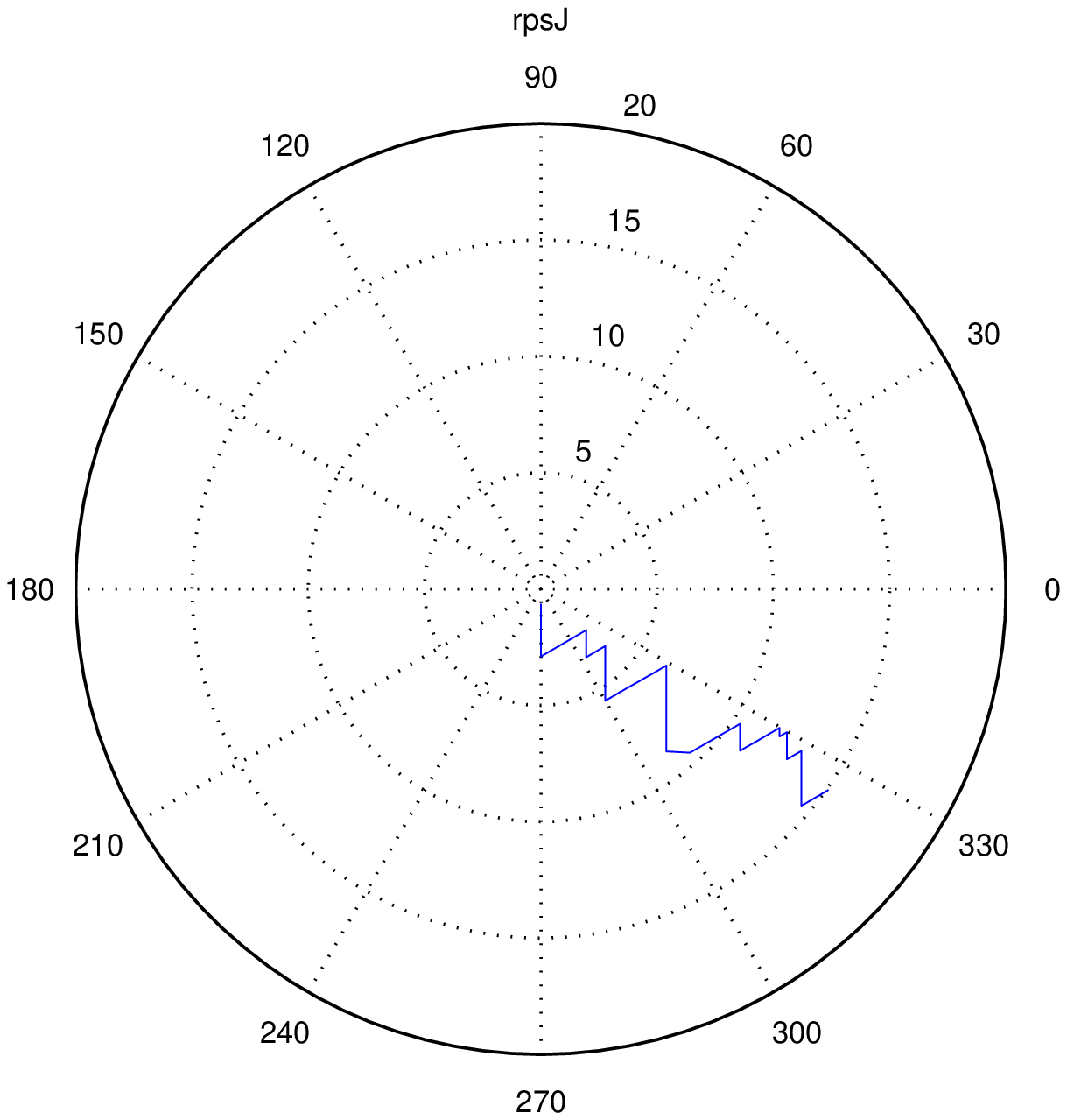, scale=0.5}
      \caption{Polar plot}
    \end{center}
  \end{minipage}
  \hfill
  \begin{minipage}[t]{.45\textwidth}
    \begin{center}  
      \epsfig{file=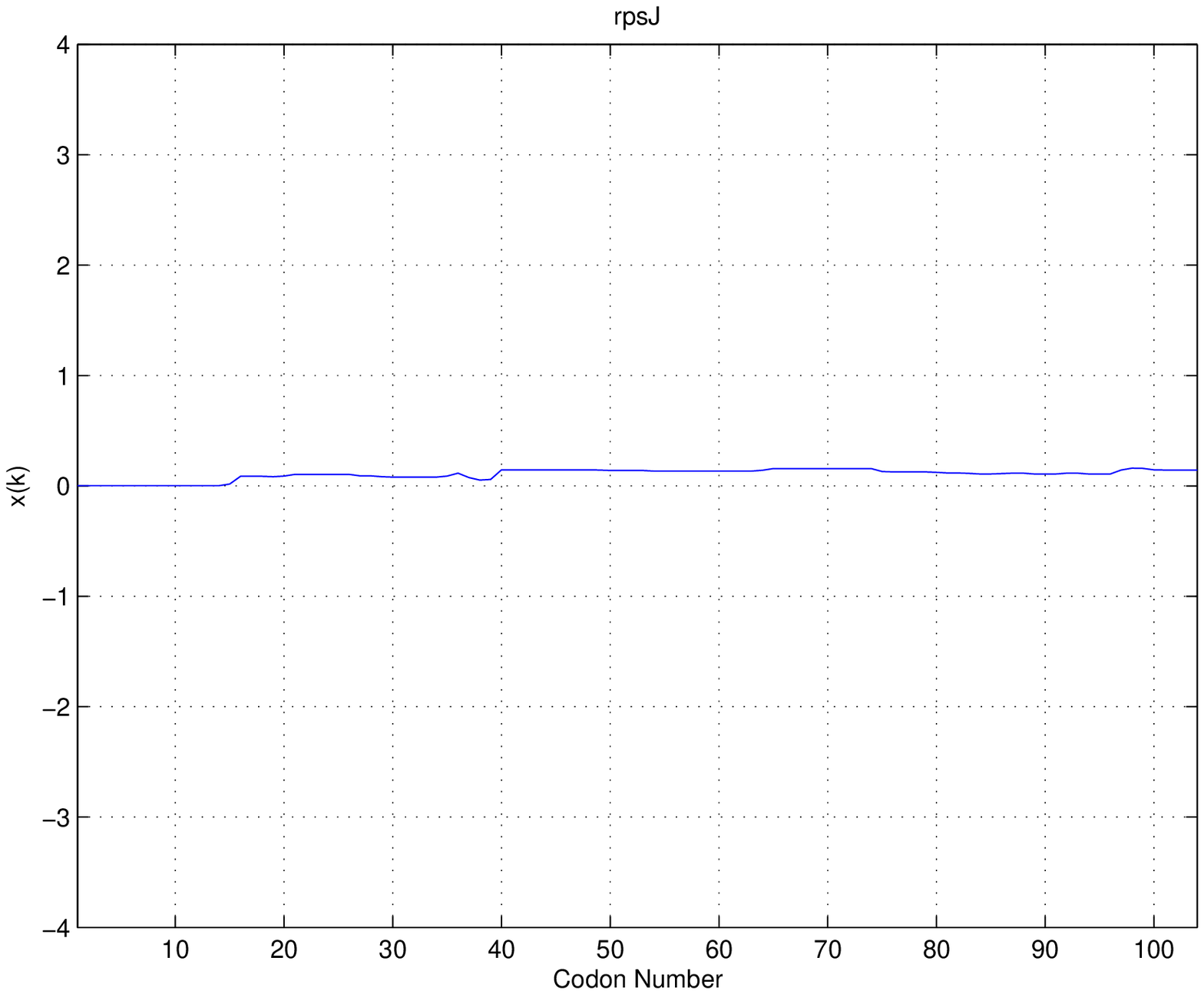, scale=0.5}
      \caption{Displacement plot}
    \end{center}
  \end{minipage}
  \hfill
\end{figure}

\clearpage
\textbf{sdhA}
\begin{figure}[h]
  \hfill
  \begin{minipage}[t]{.45\textwidth}
    \begin{center}  
      \epsfig{file=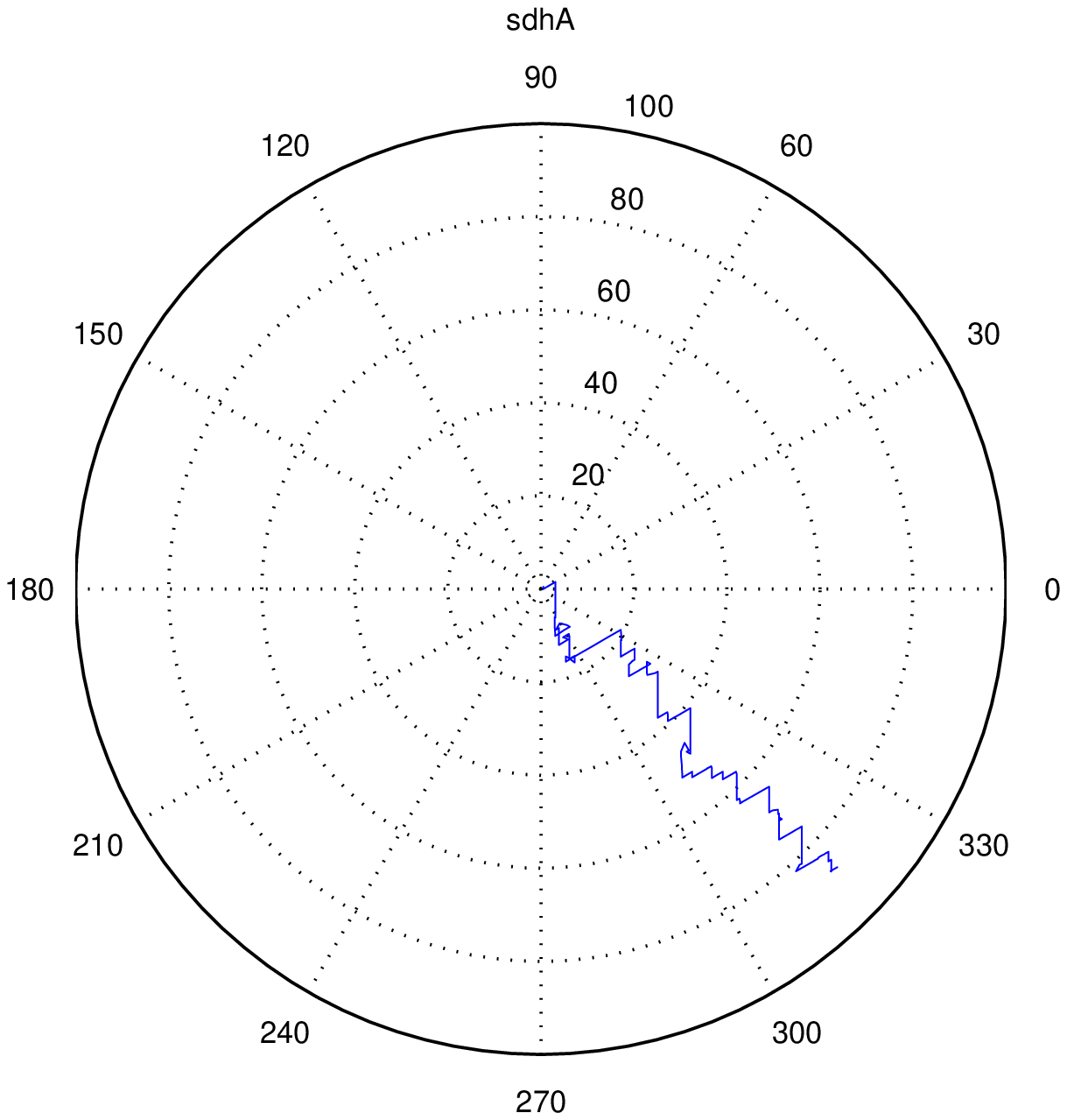, scale=0.5}
      \caption{Polar plot}
    \end{center}
  \end{minipage}
  \hfill
  \begin{minipage}[t]{.45\textwidth}
    \begin{center}  
      \epsfig{file=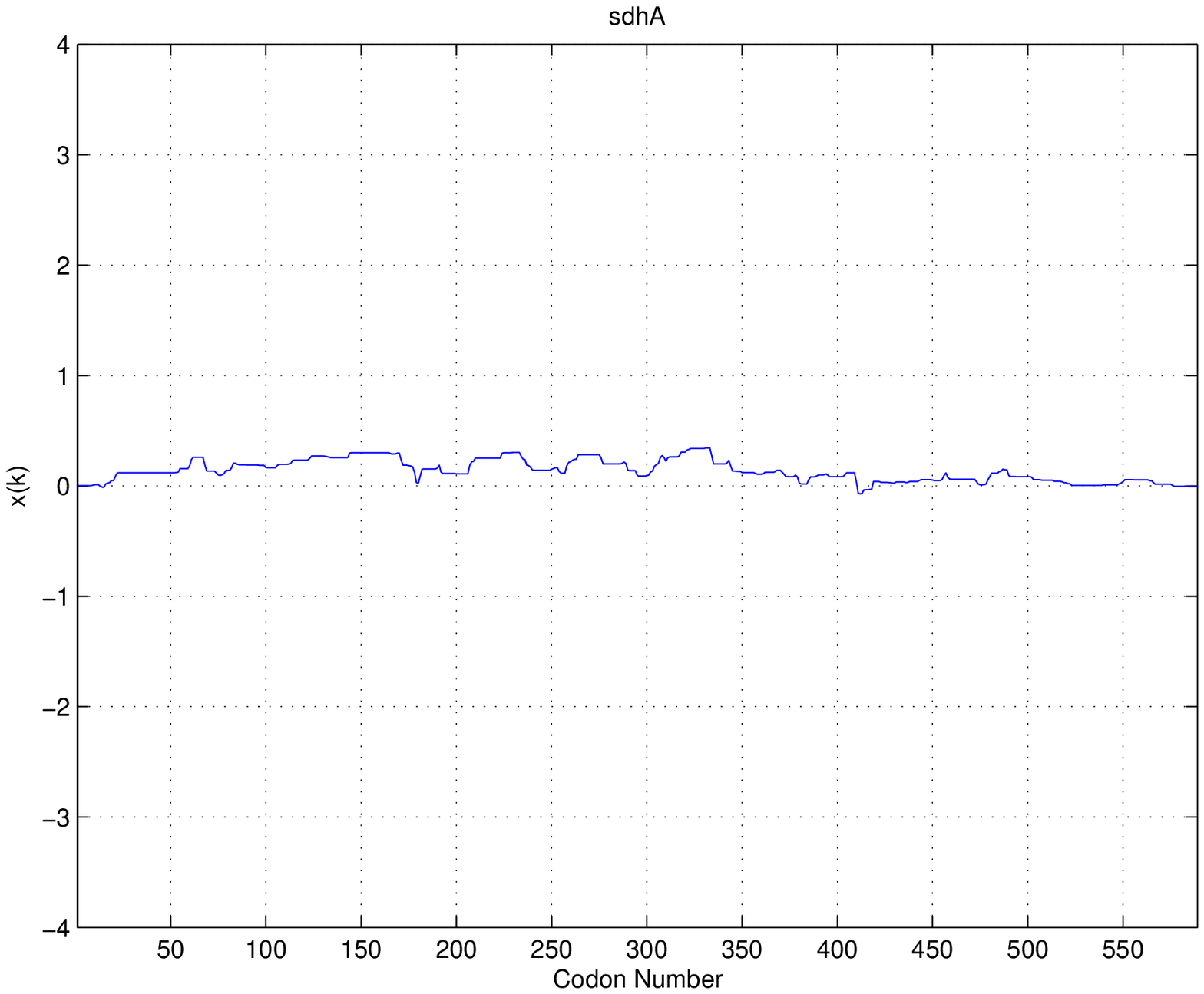, scale=0.5}
      \caption{Displacement plot}
    \end{center}
  \end{minipage}
  \hfill
\end{figure}

\end{document}